\documentclass[%
reprint,
nofootinbib,
amsmath,amssymb,
aps,
prb,
]{revtex4-2}

\usepackage{graphicx}% Include figure files
\usepackage{dcolumn}% Align table columns on decimal point
\usepackage{bm}% bold math
\usepackage[breaklinks=true]{hyperref}% add hypertext capabilities
\hypersetup{colorlinks=true, linkcolor=red!50!black, citecolor=green!50!black, urlcolor=blue!80!black}

\DeclareSymbolFont{usualmathcal}{OMS}{cmsy}{m}{n}
\DeclareSymbolFontAlphabet{\mathcal}{usualmathcal}

\usepackage[utf8]{inputenc}

\usepackage{adjustbox}
\usepackage{float}

\usepackage{mathtools}
\usepackage{stackrel}
\usepackage{scalerel}
\usepackage{leftidx}
\usepackage{xfrac}

\usepackage{euscript}
\let\mathbb\undefined
\usepackage{bbold}

\usepackage[shortlabels]{enumitem}

\usepackage{tcolorbox}
\usepackage{subfigure}

\usepackage[all,2cell,arrow,matrix]{xy} \UseAllTwocells \SilentMatrices
\usepackage{tikz-cd}
\usepackage{tikz}
\usetikzlibrary{arrows,arrows.meta,%
decorations.markings,decorations.pathreplacing,%
decorations.pathmorphing,calc}

\usepackage{verbatim}
\usepackage{comment}
\usepackage{diagbox}
\usepackage{makecell}
\usepackage{colortbl}

\tikzset{->-/.style={decoration={markings,mark=at position #1 with {\arrow{Stealth}}},postaction={decorate}},->-/.default=0.55}

\usepackage[amsmath,amsthm,thmmarks]{ntheorem}
\theoremstyle{definition}

\newtheorem{thm}{Theorem}[section]
\newtheorem{prop}[thm]{Proposition}
\newtheorem{pthm}[thm]{Theorem$^{\text{ph}}$}

\newtheorem{lem}[thm]{Lemma}

{
\theoremsymbol{\mbox{${\blacksquare}$}}
\newtheorem{defn}[thm]{Definition}
}
{
\theoremsymbol{\mbox{$\heartsuit$}}
\newtheorem{expl}[thm]{Example}
}
{
\theoremsymbol{\mbox{$\diamondsuit$}}
\newtheorem{rem}[thm]{Remark}
}
\qedsymbol{\mbox{$\square$}}

\numberwithin{equation}{section}
\numberwithin{thm}{section}

\newcommand\be            {\begin{equation}}
\newcommand\ee            {\end{equation}}
\newcommand\bea           {\begin{eqnarray}}
\newcommand\eea         {\end{eqnarray}}
\newcommand\bnu          {\begin{enumerate}}
\newcommand\enu          {\end{enumerate}}
\newcommand\bit          {\begin{itemize}}
\newcommand\eit          {\end{itemize}}

\newcommand{\pf}{\begin{proof}}
\newcommand{\epf}{\qed\end{proof}}

\newcommand\Cb			{\mathbb{C}}

\newcommand\Zb			{\mathbb{Z}}
\newcommand\Z			{\mathfrak{Z}}

\newcommand\CA			{\EuScript{A}}
\newcommand\CB			{\EuScript{B}}
\newcommand\CC			{\EuScript{C}}

\newcommand\CG			{\EuScript{G}}
\newcommand\CH			{\EuScript{H}}

\newcommand\CM			{\EuScript{M}}
\newcommand\CN			{\EuScript{N}}

\newcommand\CS			{\EuScript{S}}
\newcommand\CT			{\EuScript{T}}

\newcommand\CV			{\EuScript{V}}

\newcommand\CX			{\EuScript{X}}

\newcommand{\FZ}			{\text{\usefont{U}{euf}{m}{n}Z}}

\newcommand\bfe             {\mathbf{e}}
\newcommand\bff             {\mathbf{f}}

\newcommand\bfm             {\mathbf{m}}
\newcommand\bfE             {\mathbf{E}}

\newcommand\bfM             {\mathbf{M}}
\newcommand\bfone             {\mathbf{1}}

\newcommand\SC			{\mathsf{C}}

\DeclareMathOperator{\Hom}{Hom}

\DeclareMathOperator{\Aut}{Aut}

\DeclareMathOperator{\id}{id}

\DeclareMathOperator{\Ind}{Ind}

\DeclareMathOperator{\fpdim}{FPdim}

\DeclareMathOperator{\ob}{ob}
\newcommand{\br}			{\mathrm{br}}

\DeclareMathOperator{\fun}{Fun}
\DeclareMathOperator{\Fun}{Fun}

\DeclareMathOperator{\BMod}{BMod}

\DeclareMathOperator{\Supp}{Supp}
\DeclareMathOperator*{\medotimes}{\vcenter{\hbox{{\scalebox{0.8}{$\bigotimes$}}}}}

\newcommand{\op}			{\mathrm{op}}
\newcommand{\rev}			{\mathrm{rev}}

\newcommand{\one}			{\mathbb{1}}

\newcommand\vect			{\mathrm{Vec}}

\newcommand\rep			{\mathrm{Rep}}

\newcommand{\bscale}	{0.7}
\makeatletter
\newcommand{\ec}[2][]	{{\@ec{#1 |}{#2}}}
\newcommand{\bc}[2][]	{{\@ec{#1}{#2}}}
\newcommand{\@ec}[2]	{\mathchoice
{\displaystyle \raise.9ex\hbox{$\scaleobj{\bscale}{#1}$} {#2}}%
{\textstyle \raise.9ex\hbox{$\scaleobj{\bscale}{#1}$} {#2}}%
{\scriptstyle \raise.55ex\hbox{$\scriptstyle \scaleobj{\bscale}{#1}$} {#2}}%
{\scriptscriptstyle \raise.38ex\hbox{$\scriptscriptstyle \scaleobj{\bscale}{#1}$} {#2}}%
}
\makeatother

\newcommand{\ot}			{\otimes}

\begin{document}

\preprint{APS/123-QED}

\title{Categorical descriptions of 1-dimensional gapped phases \\ with abelian onsite symmetries}% Force line breaks with \\
%\thanks{A footnote to the article title}%

\author{Rongge Xu}
\email{xurongge@westlake.edu.cn}
\affiliation{Department of Physics, Fudan University, Shanghai 200433, China}
\affiliation{School of Science, Westlake University, Hangzhou, 310024, China}
\affiliation{Institute of Natural Sciences, Westlake Institute of Advanced Study, Hangzhou 310024, China}

%  \altaffiliation[Also at ]{Physics Department, XYZ University.}%Lines break automatically or can be forced with \\
\author{Zhi-Hao Zhang}
\email{zhangzhihao@bimsa.cn}
\affiliation{Beijing Institute of Mathematical Sciences and Applications, Beijing, 101408, China}
\affiliation{Wu Wen-Tsun Key Laboratory of Mathematics of Chinese Academy of Sciences, \\
School of Mathematical Sciences, University of Science and Technology of China, Hefei, 230026, China}
\affiliation{Shenzhen Institute for Quantum Science and Engineering, \\
Southern University of Science and Technology, Shenzhen, 518055, China}

% \author{Charlie Author}
%  \homepage{http://www.Second.institution.edu/~Charlie.Author}
% \affiliation{
%  Second institution and/or address\\
%  This line break forced% with \\
% }%
% \affiliation{
%  Third institution, the second for Charlie Author
% }%
% \author{Delta Author}
% \affiliation{%
%  Authors' institution and/or address\\
%  This line break forced with \textbackslash\textbackslash
% }%

% \collaboration{CLEO Collaboration}%\noaffiliation

\date{\today}% It is always \today, today,
            %  but any date may be explicitly specified

\begin{abstract}
%   In \href{https://arxiv.org/abs/2108.08835}{arXiv:2108.08835}, Kong, Wen and Zheng show that the 
%   Macroscopic observables in gapped phases of Ising chain form enriched fusion categories. 

In this work,
we carefully analyze the macroscopic observables in the 1+1D gapped phases with abelian onsite symmetries, and show that the spacetime observables for each gapped phase form 
a clear structure
that can be mathematically described by enriched fusion categories, which uncovers the behavior of non-local excitations that were blurry in traditional Landau paradigm.
%  on both the symmetry preserving and symmetry breaking cases.
These categorical descriptions not only generate the known classification results for symmetry preserving/breaking phases, but also unifies lattice dualities in a broader picture. 
After analyzing the general lattice model together with their boundaries, we give explicit examples including non-trivial SPT phase, where non-trivial boundaries can be given directly through our classification. 
% together with some discussions on general dualities and non-invertible symmetries.
%  under this topological skeleton.
% These examples further demonstrate 
Using enriched categorical descriptions,
the lattice dualities and their gapped phases are unified under a holographic duality between an 2d topological order with gapped 1d boundaries and 1+1D gapped quantum liquids with a categorical symmetry, which shed light on a unified definition of all quantum phases.
% \begin{description}
% \item[Usage]
% Secondary publications and information retrieval purposes.
% \item[Structure]
% You may use the \texttt{description} environment to structure your abstract;
% use the optional argument of the \verb+\item+ command to give the category of each item. 
% \end{description}
\end{abstract}

%\keywords{Suggested keywords}%Use showkeys class option if keyword
                            %display desired
\maketitle

\tableofcontents

% \input{introduction}
% !TeX root = main.tex
% !TeX program = pdfLaTeX

\section{Introduction} \label{sec:introduction}

The description of phases is a central question in condensed matter physics. Since a ``phase'' by itself, is a macroscopic notion, it should have a mathematical description which can be obtained by collecting all physical observables in the long wave length limit (LWLL). However, for a long period of time, the pursuing of this structure has been put aside, especially for phases within Landau's paradigm like the symmetry breaking phases.

Recently, in virtue of category theory, a unified description is proposed for all gapped/gapless quantum liquid phases \cite{KZ22b,KZ22}. It states that a quantum liquid phase $\CX$ can be described by a pair ($\CX_{lqs}, \CX_{sk}$), where $\CX_{lqs}$ is the \emph{local quantum symmetry} and $\CX_{sk}$ is the \emph{topological skeleton} consisting of all topological defects. For gapped quantum liquid phases, the information of local quantum symmetries is encoded in the topological skeletons. Therefore, a gapped quantum liquid phase can be described by a topological skeleton, which is mathematically an {\it enriched (higher) category}.
%For gapped quantum liquid phase, it is enough to describe the phases by topological skeletons, which are enriched (higher) categories.

In \cite{KWZ22}, the validity of this proposal has been checked in the 1+1D Ising chain and Kitaev chain, which are the first concrete examples in the lattice models to show the spacetime observables indeed form an enriched categories. 
% However, for such an important proposal, it is necessary to check it in more general lattice models. 
So in this work, we generalize this result to 1+1D\footnote{Throughout this work, we use $n$d to represent the spatial dimension and $n$D to represent the spacetime dimension.} bosonic gapped quantum phases with a finite abelian onsite symmetry $G$. Our categorical descriptions also give the classification of these bosonic gapped phases manifestly. 
%These categorical descriptions of anomaly-free (i.e., the one-dimensional higher bulk is trivial) 1d systems naturally contain the holographic duality with certain 1d boundaries of 2d topological orders. 

So what are the macroscopic observables in a quantum phase and why do they form an enriched category? It follows from \cite{KWZ22} that, for a 1d lattice model, there are two kinds of observables in LWLL: topological sectors of operators 
% (also called the categorical symmetry) 
and topological sectors of states.
\begin{enumerate}
\item The topological sectors of operators are the spaces of non-local 
% unconfined (tensionless under all admitted perturbations) 
operators that are invariant under the action of local operators. 
% Two topological sectors of operators can be connected by some operators that commutes with the action of local operators, we can such an operator a {\bf morphism}.
An operator between two topological sectors of operators that intertwines the action of local operators is called a morphism. 
We can say the topological sectors of operators form a category, denoted by $\CB$. 
Moreover, these operators can be fused and braided in 1+1D spacetime. In particular, the sector consisting of local operators is the tensor unit of $\CB$, denoted by $\mathbb{1}_{\CB}$. In consequence, $\CB$ is expected to be a braided fusion category $\CB$, which is often called the \emph{categorical symmetry} \cite{JW20,KLWZZ20a}.

\item The topological sectors of states are the topological defect lines (also known as topological excitations) in 1+1D, which are the subspaces of the total Hilbert space that are invariant under the action of local operators. The sector generated by the vacuum state (i.e. the ground state) is called the vacuum sector or the trivial sector, we denote by $\one$. The topological defect lines can be mapped to each other or fused together (see figure \ref{fig:fusion_topological_defect}), which indicates that the topological sectors of states can be characterized by a fusion category, denoted by $\CS$.
\end{enumerate}
Under categorical notations, the space of (possibly non-local) operators which maps a topological sector of states $a$ to another one $b$ is denoted by $\Hom (a, b)$. Since the topological sectors of operators act on those of states, $\Hom (a, b)$ should be an object in the category $\CB$ of topological sectors of operators. 
% From the canonical construction of enriched categories \cite{Kel69}, 
%% arXiv version
Then the topological sectors of states and the hom spaces between them should form a $\CB$-enriched category $\bc[\CB]{\CS}$ ($\CB$ is also called {\it background category} and $\CS$ is also called {\it underlying category}. For a detailed explanation, see appendix \ref{appendix:enriched_categories}).
%% SB version
%Then the topological sectors of states and the hom spaces between them should form a $\CB$-enriched category (for a detailed explanation, see appendix C in Supplementary material).

Furthermore, the fusion behavior of non-local operators should be compatible with the fusion of topological sectors of states. This fusion compatibility can be summarized mathematically by the condition that $\CS$ is equipped with a braided functor $\phi \colon \CB \to \FZ_1(\CS)$, where $\FZ_1(\CS)$ is the Drinfeld center (also called the monoidal center) of $\CS$
%  The functor $\phi$ provides a canonical construction of a $\CB$-enriched fusion category $\bc[\CB]{\CS}$ 
\cite{MP17,KYZZ21}.\footnote{Indeed, the canonical construction requires a braided functor $\overline{\CB} \to \FZ_1(\CS)$ where $\overline{\CB}$ is the time-reversal of $\CB$ obtained by reversing the braiding. In this work we only consider the case that $\CB = \FZ_1(\rep(G))$ for some finite group $G$ and thus $\CB \simeq \overline{\CB}$. So for simplicity we just take a braided functor $\phi \colon \CB \to \FZ_1(\CS)$ and consider its ``canonical construction''.}
%  which is just the topological skeleton of the given phase. 
In addition, 
since all excitations on a 1+1D anomaly-free lattice model can be created from the ground state by non-local operators, and these excitations should be detectable by the double braiding of the non-local operators that create them from the vacuum \cite{KWZ22}, $\phi$ need to be an equivalence $\phi \colon \CB \simeq \FZ_1(\CS)$. 

In other words, the topological sectors of operators form the monoidal center of the category of the topological sectors of states. And the topological skeleton that captures the macroscopic observables of a gapped quantum phase is $\bc[\CB]{\CS} \simeq \bc[\FZ_1(\CS)]{\CS}$.

% Moreover, a topological skeleton can be associated to different phases depending on what local quantum symmetry we assign. A recent understanding indicates that an onsite symmetry should be a special case of local quantum symmetries \cite{KZ22b,KZ22}.
%However, if we do not impose any symmetry on the lattice model, then all non-local operators will be confined by introducing small perturbations. In this case $\CB = \vect$ (the effective operators contains only vacuum sector). Since the system is anomaly-free, then $\CS$ is also $\vect$ and $\bc[\CB]{\CS} = {\vect}$ is the topological skeleton of the trivial $1$d topological order.
%But once we impose an onsite symmetry, then arbitrary perturbations are not allowed, what remains are only the perturbations that respect the symmetry. 
%A non-local operator is unconfined as long as it commutes with symmetric local operators. So in the presence of an onsite symmetry, we only consider the sectors that are invariant under the action of symmetric local operators. %Thus, when we apply a symmetry on the lattice model, the topological sectors of operators becomes the topological sectors of symmetric operators.
%These symmetric topological sectors of operators form the categorical symmetry.

\begin{rem}
% Non-local operators are not allowed to create excitations with infinitely large energy after introducing perturbations. 

Note that non-local operators will be confined due to arbitrary perturbations.
% , since they can create infinitely large energy.
For example, if we do not impose any symmetry, then all non-local operators are confined because arbitrary perturbations lead to infinitely large energy. In this case $\CB$ can only be $\vect$,
%  the category of finite dimensional $\mathbb{C}$-vector spaces.
i.e. there is only one sector consisting of local operators. Since the system is anomaly-free, then $\CS$ is also $\vect$ and $\bc[\CB]{\CS} = {\vect}$ is the topological skeleton of the trivial $1+1$D topological order.
%But once we impose an onsite symmetry, then arbitrary perturbations are not allowed, what remains are only the perturbations that respect the symmetry. 
%A non-local operator is unconfined as long as it commutes with symmetric local operators. In the presence of an onsite symmetry, we only consider the sectors that are invariant under the action of symmetric local operators. %Thus, when we apply a symmetry on the lattice model, the topological sectors of operators becomes the topological sectors of symmetric operators.
%These symmetric topological sectors of operators form the categorical symmetry.
So we only consider non-local operators that will not be confined by any symmetry-allowed perturbations. 
\end{rem}

\begin{rem}
Here we offer a beginner's guide tailored for readers with a physics background to understand categorical language in the context of topological phases of matter. In his seminal papers \cite{Kit03,Kit06}, Kitaev introduces the modular tensor category (MTC) descriptions of anyonic excitations, establishing a foundational link between category theory and quantum models. Levin and Wen \cite{LW05} investigate "string-net condensation" as a physical mechanism underlying topological phases, demonstrating how tensor category theory provides a unifying framework for various quantum states. Reference \cite{Wan10}, along with \cite{NSSFS08}, serves as an accessible introduction to MTCs and their applications in quantum computing. Further developments discussed in \cite{CGW10, KK12, Kap14, LKW16, PWW18, LFHSV21, Sha23} cover topics such as gapped boundaries and domain walls, ground-state degeneracy, symmetry-protected topological phases, fermionic topological orders, tensor network applications, and dualities in topological phases related to fusion categories. Recently, \cite{KZ22} offers a modern and comprehensive guide on applying categorical language to topological orders, which is valuable for building a foundational understanding of how enriched fusion categories can be used to describe various phases.
\end{rem}

\subsection{Holographic duality and topological Wick rotation}

% \mynote{Morita equivalent}
% \mynote{G is vec G}

Though finding the observables in the 1d quantum phases does not need to appeal to higher dimension, it would be more clear to understand the relation between the category $\CB$ of topological sectors of operators and the category $\CS$ of topological sectors of states via a \emph{holographic duality}\footnote{Note the difference between holographic duality and lattice duality based on lattice symmetries.}. \cite{KLWZZ20a,KZ22b}. 

%In literature, a ``holographic duality'' sometimes means a 			``correspondence'' between a boundary and its one-dimensional-higher bulk. 
In this work, by a \emph{holographic duality} (or \emph{topological holographic duality}) we mean a duality between an ($n$+1)d topological order with an $n$d gapped boundary and $n+1$D quantum liquids \cite{KLWZZ20a,KZ22b}. This holographic duality can be realized by the so-called \emph{topological Wick rotation} \cite{KZ20,KZ22b}: given an ($n$+1)d topological order with an $n$d gapped boundary, we can intuitively `rotate' the ($n$+1)d bulk phase via its $n$d boundary to the time direction to get an anomaly-free $n+1$D quantum liquid phase (see figure \ref{fig:holographic_duality}). 
This process allow us to modify an $n$d boundary phase $\CS$ of an $n+1$d topological order $\CB$ into an $n+1$D quantum liquid $\bc[\CB]{\CS}$.  

\begin{figure}[H]
\centering
\begin{tikzpicture}[scale=0.9]
% \useasboundingbox (-3,-0.5) rectangle (1,1.5) ;
\draw[->-,very thick] (-2,0)--(-4,0) node[midway,above] {\footnotesize $\CB = \FZ_1(\CS)$} ;
\draw[fill=white] (-2.1,-0.1) rectangle (-1.9,0.1) node[midway,above] {\footnotesize $\CS$} ;
\node[scale=0.9] at (-3.5,-0.5) {\scriptsize spatial direction} ;
\draw[-latex] (-1.5,0)--(1.5,0) node[midway,above]{\tiny Topological Wick rotation};
\draw[->-,very thick] (3,-1)--(3,1);
%  node[midway,right] {\footnotesize $\CB = \FZ_1(\CS)$} ;
\draw[fill=white] (2.9,-1.1) rectangle (3.1,-0.9) node[midway,right] {\footnotesize $\bc[\CB]{\CS}=\bc[\FZ_1(\CS)]{\CS}$} ;
\node[scale=0.9] at (2,1) {\scriptsize time direction} ;
\end{tikzpicture}
\caption{The left hand side depicts an ($n$+1)d topological order with an $n$d gapped boundary. %The boundary is described by a fusion $n$-category $\CS$ and the bulk is described by the Drinfeld center $\FZ_1(\CS)$.
After the topological Wick rotation, we get its holographic dual as depicted in the right hand side, which is an anomaly-free $n+1$D quantum liquid phase that can be described by $\bc[\CB]{\CS}$.} %described by the enriched fusion $n$-category $\bc[\CB]{\CS} \simeq \bc[\FZ_1(\CS)]{\CS}$.}
\label{fig:holographic_duality}
\end{figure}

We mainly focus on the case that $n = 1$. In this case, the topological excitations in the 2d bulk form a unitary modular tensor category (UMTC) $\CB$ and the topological excitations on the $1$d boundary form a unitary fusion category $\CS$. By the boundary-bulk relation \cite{KWZ15,KWZ17} we have $\CB \simeq \FZ_1(\CS)$. After the topological Wick rotation, we get an anomaly-free $1+1$D quantum liquid phase. The bulk excitations in $\CB \simeq \FZ_1(\CS)$ before the rotation are replaced by the topological sectors of operators in the 1+1D spacetime after the rotation, and the topological excitations in $\CS$ become topological sectors of states. Then we have an enriched fusion category $\bc[\CB]{\CS}\simeq \bc[\FZ_1(\CS)]{\CS}$, which is just the topological skeleton of the obtained anomaly-free $1+1$D phase.

\begin{rem}
The anomaly-free property for the $1+1$D phase after the rotation means $\bc[\CB]{\CS}$ is not a boundary of a one dimensional higher bulk. Mathematically, it is to say enriched fusion category $\bc[\CB]{\CS}$ has trivial Drinfeld center $\FZ_1 \bigl( \bc[\CB]{\CS} \bigr) \simeq \vect$.
% and this condition is equivalent to the boundary-bulk relation before the rotation. More precisely, the boundary-bulk relation $\CB \simeq \FZ_1(\CS)$ is equivalent to $\FZ_1 \bigl( \bc[\CB]{\CS} \bigr) \simeq \vect$ \cite{KYZZ21}.
\end{rem}

%To get the exact categorical description of 1+1D quantum phases, the 2d topological order needs to passes through a "Wick" rotation that rotates the defects in the 2d bulk into the time direction. There are also some works implicitly use this idea based on lattice models with categorical symmetries from various perspectives \cite{JW20, CW22, MMT22, LDOV21, LTLSB21}.
%
%To illustrate our viewpoint, let us briefly review the 2d anyon condensation theory \cite{Kon14} and the so-called \emph{topological Wick rotation} \cite{KZ20,KZ22b}.

\begin{rem}
These kind of topological holographic phenomena have been studied by various groups of people and in different contexts. The holographic duality based on topological Wick rotation was proposed by \cite{KLWZZ20a,KZ22b}. The variations of this duality were also appeared in \cite{KZ18b,VBWBHV18,KZ20,KZ21,JW20,KLWZZ20,KLWZZ20a,AFM20,KZ22b,WJX21,KWZ22,AAXJAP21,LDOV21,CW22,CW22a,MMT22,LJ22,LDV22}. 
%Its lower dimension versions were studied in \cite{KZ18b,VBWBHV18,KZ20,KZ21,JW20,AFM20,KZ22b,WJX21,KWZ22,LDOV21,MMT22,LDV22}. 
%A closely related concept is the so-called categorical symmetry, which provides a slightly different perspective on the holographic duality. For example, see \cite{JW20,KLWZZ20,KLWZZ20a,KZ22b,CW22,CW22a,LJ22}.
\end{rem}

%\begin{rem}
%The low energy effective theory of a 1+1D gapped quantum liquid phase is a topological quantum field theory (TQFT) with a fusion category symmetry, which is a special case of a TQFT with defects (valued in a bicategory) introduced by Davydov, Kong and Runkel \cite{DKR11} (see \cite{Car18} for a review). Later in \cite{BT18}, the notion of a TQFT with defects valued in a fusion 1-category (viewed as a bicategory with only one object) revived with a new name: a TQFT with a fusion category symmetry (see also \cite{TW19,HLS21}).
%\end{rem}

\subsection{Classification of gapped phases and anyon condensation} \label{sec_phase_condenstion}

Now if we fix a categorical symmetry $\CB$ in the 1+1D anomaly-free gapped quantum liquid, how can we find different topological sectors of states $\CS_i$ so as to classify these 1+1D gapped phases $\{\bc[\CB]{\CS_i}\}$ with the same categorical symmetry?

Since anyon condensation theory can be used to classify gapped boundaries of a 2d UMTC $\CB$, and $\CS$  
% with categorical symmetry $\CB$ 
correspond to the 1d gapped boundaries of $\CB$, we can use anyon condensation to give a classification of 1+1D gapped phases with categorical symmetry $\CB$ through holographic duality.

So we briefly review the mathematical theory of 2d anyon condensation \cite{Kon14}. Suppose a 2d topological order (with excitations form a UMTC) $\CC$ is obtained from another 2d topological order $\CB$ via a $2$d condensation, and a $1$d gapped domain wall with the wall excitations form a unitary fusion category $\CS$ is produced as a result of this condensation process. 
%% arXiv version
Then the vacuum $\one_\CC$ of $\CC$-phase is naturally a condensable algebra $A$ in $\CB$ (see appendix \ref{appendix:condensable_pointed} for the definition of a condensable algebra). 
%% SB version
%Then the vacuum $\one_\CC$ of $\CC$-phase is naturally a condensable algebra $A$ in $\CB$ (see appendix A.2 in Supplementary material for the definition of a condensable algebra). 
% Moreover, $\CC$ consists of all deconfined anyons and can be identified with the category $\CB_A^{\mathrm{loc}}$ of local right $A$-modules in $\CB$. 
The excitations $\CS$ on the 1d domain wall
%  include all confined and deconfined anyons and thus $\CS$
can be identified with the category $\CB_A$ of all right $A$-modules in $\CB$. The bulk-to-wall map from the original phase $\CB$ to $\CS$ is given by
\begin{align} \label{right_A_module}
L = -\otimes A \colon \CB &\rightarrow \CB_A \simeq \CS , \quad \nonumber \\
\quad a &\mapsto a \otimes A, \qquad \forall a \in \CB .
\end{align}
% Also, $L$ is a central functor, in the sense that it factors through $\jCB \to \FZ_1(\CS) \xrightarrow{\rm forget} \CS$. 
% The condensable algebra $A$ can also be recovered as $L^R(\one_{\CS})$, where $\one_\CS$ is the vacuum sector in $\CS$ and $L^R$ is the right adjoint of $L$. 
% We also have $\FZ_1(\CS) \simeq \CB \boxtimes \overline{\CC}$ \cite{DMNO13}, where $\overline{\CC}$ denotes the time-reversal of $\CC$.

We mainly focus on the special case that the condensed phase $\CC$ is trivial, namely $\CC \simeq \vect$  (figure \ref{fig:topological_Wick_rotation}(a)). In this case the corresponding condensable algebra $A$ is called \emph{Lagrangian}. Also the domain wall $\CS$ becomes a boundary of the $\CB$-phase and we have the boundary-bulk relation $\CB \simeq \FZ_1(\CS)$ \cite{DMNO13}. Hence the gapped boundaries of the $\CB$-phase are classified by the Lagrangian algebras in $\CB$.

Now we apply the topological Wick rotation to the 2d topological order $\CB$ with boundary $\CS$ and get an anomaly-free 1+1D quantum liquid phase. Then we can view the condensation process as happening in spacetime. 
See figure \ref{fig:topological_Wick_rotation}(b). 
% By anyon condensation theory,
The Lagrangian algebra $A \in \CB$ is condensed on the vacuum sector of states $\one_{\CS}$, in the sense that it consists of the operators that act on the topological sector of the ground state invariably. Therefore, we call this Lagrangian algebra $A$ the \emph{ground state algebra}\footnote{We use Lagrangian algebra in the context of 2d anyon condensation theory and use ground state algebra in 1+1D topological sectors of states.} (ground state algebra). Mathematically, this means that the ground state algebra $A$ is the internal hom $\Hom_{\bc[\CB]{\CS}}(\one_{\CS},\one_{\CS})$ \cite{KYZ21}. On the other hand, as we have known in the anyon condensation theory, the fusion category $\CS$ can be recovered as the module category $\CB_A$. Thus the topological skeleton $\bc[\CB]{\CS}$ of the phase can be completely determined by the categorical symmetry $\CB$ and the ground state algebra $A$.
% $\CS$ can be recovered as the module category $\CB_{A}$  
% That is to say, we can actually condense the topological sectors of operators to those of the states, and the sectors of operators and the corresponding states also satisfy the boundary-bulk relation $\CB \simeq \FZ_1(\CS)$.

\begin{figure}[H]
\centering
\begin{tikzpicture}[scale=0.7]
            \filldraw[fill=gray!20, draw=none](-6.8,0)--(-5.3,0)--(-3.3,1.5)--(-4.8,1.5)--(-6.8,0);
            \filldraw[fill=gray!10, draw=none](2.1,0)--(2.1,1.5)--(4.1,3)--(4.1,1.5)--(2.1,0);
            \filldraw[fill=black, draw=black] (-5.7,0.25)node[left]{\text{$a$}} circle (0.05);
            \draw[thick,-latex](-5.6,0.25) --(-5.05,0.25) ;
            \filldraw[fill=black, draw=black] (-4.96,0.25) circle (0.05);
            
            \node[scale=0.9]at(-4.5, 0.25){\scriptsize $L(a)$};
            % \draw[](-6.8,0)--(-5.3,0);
            \draw[dashed](-5.3,0)--(-3.7,0);
            % \draw[](-4.8,1.5)--(-3.3,1.5);
            \draw[dashed](-3.3,1.5)--(-1.7,1.5);
            % \draw[](-6.8,0)--(-4.8,1.5);
            \draw[dashed](-3.7,0)--(-1.7,1.5);
            \draw[thick](-5.3,0)--(-3.3,1.5);
            \draw[thick](-3.3,1.5)--(-4.3,0.75);
            \node[rotate=38]at(-4.1,0.7){\scriptsize $\CS$};
            \node[]at(-5,-0.8){\text{$(a)$}};
            \node[]at(3,-0.8){\text{$(b)$}};
            \draw[-latex](-6.3,1.4) --(-6.3,2);
            \node[]at(-6.1, 1.7){$t$};
            \draw[-latex](-1,0.5)--(1,0.5) ;
            \node[scale=0.8]at(0, 0.75){\scriptsize \text{topological wick rotation}};
            \draw[dashed](2.1,0)--(3.6,0);
            \draw[dashed](4.1,1.5)--(5.6,1.5);
            % \node[]at(4, 0.8){\text{$\vect$}};
            \node[scale=0.9]at(-3, 1.1){\scriptsize $\CC \simeq \vect$};
            \node[rotate=0]at(3.1, 1.3){\footnotesize $\bc[\FZ_1(\CS)]{\CS}$};
            \node[scale=0.9]at(-4.7, 1.1){\scriptsize $\CB = \FZ_1(\CS)$};

            \draw[dashed](3.6,0)--(5.6,1.5);
            % \draw[](2.1,0)--(2.1,1.5);
            % \draw[](4.1,1.5)--(4.1,3);
            % \filldraw[fill=black, draw=black] (2.5,1.1)node[right]{\text{$a$}} circle (0.05);
            % \draw[thick,-latex](2.5,1) --(2.5,0.4) ;
            % \filldraw[fill=black, draw=black] (2.5,0.3) circle (0.05);
            % \node[]at(2.9,0.3){\text{$L(a)$}};

            \draw[thick](2.1,0)--(4.1,1.5);
            % \draw[](2.1,1.5)--(4.1,3);
            \draw[thick](4.1,1.5)--(3.1,0.75);
            % \node[right, rotate=38]at(3.1, 0.5){\text{$\CS \simeq \CB_A$}};
\end{tikzpicture}
\caption{(a) depicts a normal 2d anyon condensation process from a $2$d topological order $\CB$ to vacuum $\vect$, which generates a boundary phase described by $\CS \simeq \CB_A$. After topological Wick rotation, we get (b), an anomaly-free 1+1D phase described by $\bc[\CB]{\CS} \simeq \bc[\FZ_1(\CS)]{\CS}$. The original boundary phase $\CS$ now becomes the topological sector of sates, and the original 2d bulk phase $\CB$ becomes the categorical symmetry that acts on those states.} 
\label{fig:topological_Wick_rotation}
\end{figure}

% From the definition of the right adjoint functor, i.e., $\Hom_{\CS}(L(a),\one_{\CS}) = \Hom_{\CB}(a, L^R(\one_{\CS}))$, we get $\Hom_{\CS}(a \odot \one_{\CS}, \one_{\CS}) = \Hom_{\CB}(a,[\one_{\CS},\one_{\CS}]_{\CB})$. Thus the Lagrangian algebra $L^R(\one_{\CS})$ can be written as the internal hom $[\one_{\CS},\one_{\CS}]_{\CB}$.

% In Levin-Wen type of lattice models \cite{LW05,KK12}, a bulk lattice is defined by a spherical fusion category $\CS$ and the boundary lattice is defined by a $\CS$-module $\CM$.
% The $1$d boundary phase determined by an ${}_{\CS}\CM$-boundary lattice model can be obtained by condensing the algebra $L^{\CR}_{\CM}(\id_{\CM})$ in the $\FZ_1(\CS)$-bulk.
% Kitaev quantum double models \cite{Kit03} cover a subset of phases defined by Levin-Wen models. 

We mainly focus on the case that the $2$d topological order is realized by the Kitaev quantum double model \cite{Kit03}, which is the topological order that exhibits gauge symmetry associated with a finite group $G$. Let $\rep(G)$ denote the category of $\mathbb{C}$-linear representation of $G$ (or $G$-charges, physically). The topological excitations of this $2$d topological order form the UMTC $\FZ_1(\rep(G))$ \cite{BK01}. The Lagrangian algebras in $\FZ_1(\rep(G))$ are classified by pairs $(H, \omega)$, where $H$ is a subgroup of $G$ and $\omega \in \mathrm{H}^2(H, U(1))$ is a 2-cohomology class \cite{Dav10a}. We use $A(H,\omega)$ to denote such a Lagrangian algebra. Readers may check appendix \ref{appendix:condensable_pointed} for the case that $G$ is abelian. Equivalently, the gapped boundaries of the Kitaev quantum double model is also classified by the pair $(H,\omega)$ \cite{BSW11}, and the topological excitations on the boundary is $\FZ_1(\rep(G))_{A(H,\omega)}$. These classification results exactly meet with a known classification of 1d bosonic gapped phases with an onsite $G$-symmetry \cite{CGW11,SPGC11}. 

% Then we can use these 1d boundary phases to classify the anomaly free 1+1D phases after topological Wick rotation.
After topological Wick rotation, $\FZ_1(\rep(G))_{A(H,\omega)}$ becomes the topological sector of states, and the topological features of the quantum double model captured by $\FZ_1(\rep(G))$ becomes the categorical symmetry that acts on those states (see figure \ref{fig:enriched_category} (b)), and the ground state algebra is exactly $A(H,\omega)$.

\begin{expl}
In particular, when $H = G$, the 1+1D phases after the rotation are 1d symmetry protected topological (SPT) orders with symmetry $G$. The topological sectors of states are all given by the symmetry charges which form the fusion category $\rep(G)$, and the 2d bulk $\FZ_1(\rep(G))$ before the rotation is the gauged phase of the 2d trivial $G$-SPT phase \cite{LKW16a,LKW17,KLWZZ20}. The 2-cohomology classes $\omega \in H^2(G;U(1))$ determine different actions of the categorical symmetry $\FZ_1(\rep(G))$ on the category $\rep(G)$ of topological sectors of states. Therefore, the SPT orders can be distinguished by the actions of the topological sectors of operators on the topological sectors of states. In particular, it suffices to use the ground state algebra $A(G,\omega)$ consisting of the operators that act on the ground state invariably to distinguish the SPT orders. We discuss the example with $G = \Zb_2 \times \Zb_2$ in Section \ref{sec_Z2Z2_SPT}. This gives a method to distinguish different 1d SPT orders without seeking a boundary or gauging the bulk. We believe that this method can be generalized to higher dimensions.
\end{expl}

Lattice models constructed in this work are mainly for trivial 2-cohomology. We expect that, with different choices of $H \subseteq G$, we obtain different symmetry breaking phases with the global symmetry $G$ on the 1+1D gapped phase breaking to its subgroup $H$. And the topological features on these different symmetry breaking phases should be captured by the enriched fusion category $\bc[\FZ_1(\rep(G))]{\FZ_1(\rep(G))_{A(H)}}$. Also, in this case $A(H)$ is the full center \cite{Dav10} of the function algebra $F_H \coloneqq \fun(G/H) \in \rep(G)$, and $\FZ_1(\rep(G))_{A(H)}$ is equivalent to the bimodule category ${}_{F_H} \rep(G)_{F_H} \coloneqq \BMod_{F_H|F_H}(\rep(G))$. Or to say a boundary phase of $\FZ_1(\rep(G))$ can be described by $\FZ_1(\rep(G))_{A(H)}$ \cite{Dav10} from the 2d condensation perspective, or as ${}_{F_H} \rep(G)_{F_H}$ form the 1d condensation perspective \cite{XY24}. Thus, in this work we check the following physical theorem, which was originally proposed in \cite{KZ22b,KWZ22}, in concrete lattice models:

% For every subgroup $H \subseteq G$, there is a gapped boundary of the Kitaev quantum double model described by the bimodule category ${}_{F_H} \rep(G)_{F_H} \coloneqq \BMod_{F_H|F_H}(\rep(G))$,
%  of $(F_H,F_H)$-bimodules in $\rep(G)$,

\begin{pthm} \label{pthm:symmetry_breaking_phase}
A 1+1D bosonic gapped phase with an onsite symmetry $G$ can be described by the enriched fusion category $\bc[\FZ_1(\rep(G))]{\FZ_1(\rep(G))_{A(H)}} \simeq \bc[\FZ_1(\rep(G))]{{}_{F_H} \rep(G)_{F_H}}$ if the symmetry spontaneously breaks to a subgroup $H \subseteq G$.
\end{pthm}

%        So there is another way to state Theorem$^{\text{ph}}$ \ref{pthm:symmetry_breaking_phase}:
%
%\begin{pthm} \label{pthm:topological_skeleton}
%%    The topological skeleton of the 1d bosonic gapped phase with a finite onsite symmetry $G$ is given by the enriched fusion category $\bc[\FZ_1(\rep(G))]{\FZ_1(\rep(G))_{A})}$, where $A$ is a Lagrangian algebra in $\FZ_1(\rep(G))$.
%%    The set of Lagrangian algebras in $\FZ_1(\rep(G))$ gives the classification result of the 1d bosonic gapped quantum phases with an onsite symmetry.
%The topological skeleton of a 1d bosonic gapped phase with an onsite symmetry $G$ is the enriched fusion category $\bc[\FZ_1(\rep(G))]{\FZ_1(\rep(G))_{A(H)}}$ if the symmetry spontaneously breaks to a subgroup $H \subseteq G$.
%\end{pthm}

The simplest example to illustrate theorem \ref{pthm:symmetry_breaking_phase} is for $G = \Zb_2$, namely, we have the holographic duality between two 1d gapped boundaries of 2d toric code model \cite{Kit03} and the two 1+1D gapped phases of transverse Ising chain, see figure \ref{fig:enriched_category} (a). 
The process of demonstrating that the $\Zb_2$ SPT phase can be describe by ${}^{\FZ_1(\rep(\mathbb{Z}_2))}\rep(\mathbb{Z}_2)$ and the spontaneous symmetry-breaking phase can be described by ${}^{\FZ_1(\rep(\mathbb{Z}_2))}\vect_{\mathbb{Z}_2}$, is given in \cite{KWZ22}.
We would retell this story in section \ref{section:Ising}. 
Conceptually, we generalize the 2d topological order $\CB$ from the toric code model ${\FZ_1(\rep(\mathbb{Z}_2))}$ to the quantum double model ${\FZ_1(\rep(G))}$ . 
In order to show that the gapped boundaries of 2d quantum double model is one to one correspond to the 1+1D gapped phases after topological Wick rotation,
we explicitly construct a 1+1D lattice model with onsite symmetry $G$ (see Hamiltonian \ref{eq:Hamiltonian}) and read off its observables. 
It turns out that, for each symmetry breaking phase where $G$ breaks to its subgroup $H$, the topological sectors of operators and states together form enriched fusion category $\bc[\FZ_1(\rep(G))]{\FZ_1(\rep(G))_{A(H)}} \simeq \bc[\FZ_1(\rep(G))]{{}_{F_H} \rep(G)_{F_H}}$, as the holographic duality indicates. 
We depict this correspondence in figure \ref{fig:enriched_category} (b), in which $H_1,H_2,H_3\ldots$ label different subgroups of $G$.

Note that though model \ref{eq:Hamiltonian} is constructed artificially as a concrete example to check Theorem$^{\text{ph}}$ \ref{pthm:symmetry_breaking_phase}, 
what matters here is indeed the categorical symmetry that defines the macroscopic observables (even lattices with different global symmetries as long as they are Morita equivalent as fusion categories).
Namely, we can also have other
models to get the same categorical description as long as
they realize the same phases.
We invent a general method in finding the descriptions of topological sectors
of states with onsite $G$-symmetry in the next section.

% \footnote{This kind of intuition may be generalized to a more general symmetry, which should provide more general 1d gapped phases after topological Wick rotation (think of replacing the bulk phase by the Levin-Wen model). We can also start another way around, to first pick a special lattice model (like the Haldane model) then give the categorical descriptions of its phases. Note that the topological Wick rotation approach is also valid in higher dimensions.}.
%The underlying category $\CS$ should be given by $\FZ_1(\rep(G))_{A(H, \omega)} $, or ${}_{F_H} \rep(G)_{F_H} \simeq \FZ_1(\rep(G))_{A_{(H)}}$ if we set $\omega$ trivially (figure \ref{fig:enriched_category}, $H_a,H_b,H_c...$ labels different subgroups of $G$).

\begin{figure*}
\centering
\begin{tikzpicture}
        \filldraw[fill=gray!20, draw=white] (-8,0) rectangle (-4,2);
        \draw[very thick](-8, 0)--(-6,0);
        \draw[very thick, dashed](-6, 0)--(-4,0);
        \filldraw[fill=white, draw=black] (-6.1,-0.1) rectangle (-5.9,0.1);
        \node[] at(-6,0.8){$\FZ_1(\rep(\mathbb{Z}_2))$};
        \node at(-4.3,-0.4){$\vect_{\mathbb{Z}_2}$};
        \node at(-7.5, -0.4){$\rep(\mathbb{Z}_2)$};

        \draw[very thick](0,0)--(4,0);
        \draw[draw=none](-8,0)--(-7,0);
        \filldraw[fill=white] (1.9,-0.1) rectangle (2.1,0.1);
        \node at(0,0.5){${}^{\FZ_1(\rep(\mathbb{Z}_2))}\rep(\mathbb{Z}_2)$};
        \node at(4,0.5){${}^{\FZ_1(\rep(\mathbb{Z}_2))}\vect_{\mathbb{Z}_2}$};

        \node at(-7,1.7){\scriptsize \text{2d Toric code}};
        \node at(0.9,1.7){\scriptsize \text{1+1D Ising chain}};
        \draw[-latex](-3,1)--(-2.4,1);

	\node at (-2,-0.7) {(a)} ;

\begin{scope}[yshift=-3.5cm]
        \filldraw[fill=gray!20, draw=white] (-8,0) rectangle (-4,2);
        \draw[very thick](-8, 0)--(-4,0);
        \filldraw[fill=white, draw=black] (-7.1,-0.1) rectangle (-6.9,0.1);
        \filldraw[fill=white, draw=black] (-5.1,-0.1) rectangle (-4.9,0.1);

        \node at(-6.6,1.7){\scriptsize \text{2d Quantum double}};
        \node at(1.6,1.7){\scriptsize \text{1+1d gapped phases with $G$}};
        \draw[-latex](-3,1)--(-2.4,1);
        \node[] at(-6,0.8){$\FZ_1(\rep(G))$};
        
        \node at(-3.5,-0.4){\tiny $\FZ_1(\rep(G))_{A(H_3)}\quad \cdots$};
        \node at(-6,-0.4){\tiny$ \FZ_1(\rep(G))_{A(H_2)}$};
        \node at(-8.4, -0.4){\tiny$\cdots \quad \FZ_1(\rep(G))_{A(H_1)}$};
        \node at(-3.5,0){$\cdots$};
        \node at(-8.5,0){$\cdots$};
        \node at(4.5,0){$\cdots$};
        \node at(-0.5,0){$\cdots$};

        \draw[very thick](0,0)--(4,0);
        \filldraw[fill=white] (0.9,-0.1) rectangle (1.1,0.1);
        \filldraw[fill=white] (2.9,-0.1) rectangle (3.1,0.1);
        \node[]at(1.9,-0.5){\tiny${}^{\FZ_1(\rep(G))}\FZ_1(\rep(G))_{A(H_2)}$};

        \node at(-0.8,0.5){\tiny $\cdots \quad {}^{\FZ_1(\rep(G))}\FZ_1(\rep(G))_{A(H_1)}$};
        \node at(4.6,0.5){\tiny ${}^{\FZ_1(\rep(G))}\FZ_1(\rep(G))_{A(H_3)}\quad \cdots$};
        \filldraw[draw=none](6.7,0)--(6.84,0);

	\node at (-2,-0.7) {(b)} ;
\end{scope}
\end{tikzpicture}
\caption{A picturesque explanation of the holographic duality between 2d topological orders with gapped boundaries and 1d gapped quantum phases.
For the left part of (a), we have 2d toric code mode with the smooth boundary $\rep(\Zb_2)$ and rough boundary $\vect_{\Zb_2}$. After topological Wick rotation, these two gapped boundaries correspond to the 
two gapped phases of transverse Ising chain, namely, they become the $\Zb_2$ SPT phase ${}^{\FZ_1(\rep(\mathbb{Z}_2))}\rep(\mathbb{Z}_2)$ and the symmetry-breaking phase ${}^{\FZ_1(\rep(\mathbb{Z}_2))}\vect_{\mathbb{Z}_2}$.
More generally, we have (b), in which the gapped boundaries of 2d quantum double model classified by Lagrangian algebra $A(H)$, one to one correspond to $1$d anomaly-free gapped phases $\bc[\FZ_1(\rep(G))]{\FZ_1(\rep(G))_{A_{(H)}}}$ with onsite symmetries after topological Wick rotation.} \label{fig:enriched_category}
\end{figure*}

% \begin{rem}
% Though we construct a 1d lattice model with abelian onsite symmetry %in section \ref{section:lattice_model} 
% to check the predicted enriched categories, our description is not limited to this model. What matters here is the symmetry that defines the macroscopic observables, namely we can use different models to get the same categorical description as long as they realize the same phase.
% \end{rem}

% \begin{rem}
%     Duality
% \end{rem}

% Now we are about to check the enriched category description $\bc[\FZ_1(\rep(G))]{\FZ_1(\rep(G))_{A_{(H)}}}$ or $\bc[\FZ_1(\rep(G))]{{}_{F_H} \rep(G)_{F_H}}$ works in concrete lattice models.
%  In order to check the observables in each phase, we would give and analyze the lattice model in the next section.
% the observables in the lattice models,
Here is the layout of this paper: In section \ref{section:G_symmetry}, we analyze the categorical structure of the topological defects with onsite $G$-symmetry and show the topological sectors of states $\CS$ respect symmetry can be obtained from the so-called equivariantization.
In section \ref{section:lattice_model}, we construct lattice models with a finite abelian onsite symmetry $G$ and demonstrate that the categorical symmetry is $\FZ_1(\rep(G))$. We then analyze topological skeletons of each phase and prove our main result Theorem$^{\text{ph}}$ \ref{pthm:symmetry_breaking_phase}. %and \ref{pthm:topological_skeleton}. 
In section \ref{section:boundaries}, we further check the enriched category descriptions of the $0$d boundaries and domain walls of these 1d gapped phases and show they also agree with the predictions of holographic duality. In section \ref{section:example}, we perform some physical examples including transverse field Ising model, its Kramers-Wanniers dual, clock model and cluster model with non-trivial SPT order. 
In section \ref{sec:GS_algebra}, we discuss the physical perspectives in regard of ground state algebra.
We also discuss the unification power of enriched categories in lattice dualities in section \ref{section:conclusion}.
%% arXiv version
Some background knowledge and calculations are performed in the \hyperref[appendix]{appendices}.

\section{Topological sectors of states with onsite \texorpdfstring{$G$}{G}-symmetry} \label{section:G_symmetry}

% A topological order with an onsite $G$-symmetry is described by its observables in LWLL as a quantum phase. 
% Recall that for a topological order without symmetry, its observables in LWLL form a (higher) category. 
In this section, we translate the behavior of excitations in a topological order with an onsite $G$-symmetry into categorical language, in which we invent a general method to find the topological sectors of states 
with onsite symmetry $G$
% we argue that the category of $G$-symmetric topological sectors of states is directly related to 
% the category of observables 
through considering those states without symmetry. 
Analyses throughout this section do not depend on  $G$ abelian and 2-cocycle trivial or not.
This technique is also valid in higher dimensions.
For readability, we hide some of the categorical details, which can be found in appendix \ref{appendix:equivariantization}.

\subsection{The category of topological sectors of states and group actions} \label{section:topological_sectors of states}

First we consider observables in topological orders without symmetry. 
% There is a class of observables in LWLL of a topological order, called \emph{topological sectors of states} or \emph{topological excitations}.
%Microscopically, a \emph{defect} in a lattice model can be realized as a local modification of the lattice or the Hamiltonian. In the long wave length limit, a defect may break the uniformity of the phase and becomes an observable in LWLL. A defect in LWLL is called a \emph{topological defect}.
%
%The simplest way to obtain a defect is to add a local term in the Hamiltonian. Suppose $\delta \mathcal H_\xi$ is an operator acting around a site $\xi$. Then the ground state of the new Hamiltonian $(\mathcal H + \delta \mathcal H_\xi)$ is different from that of the original Hamiltonian $H$ in general. With respect to the original Hamiltonian $H$, the new ground state looks like an excitation around the site $\xi$, but coincides with the original ground state far from the site $\xi$. Intuitively, the local term $\delta \mathcal H_\xi$ traps an excitation located at the site $\xi$. Conversely, any excitation located at a site $\xi$ can be trapped by some operators $\delta \mathcal H_\xi$. Therefore, such a defect is characterized by an excitation of the Hamiltonian $H$.
%Microscopically, a state $\lvert \psi \rangle$ is a local observable. 
In LWLL, the coarse-graining process integrates and averages out the microscopic degrees of freedom by the action of local operators. Thus given a local operator $A$, the states $A \lvert \psi \rangle$ and $\lvert \psi \rangle$ should be viewed as the same after coarse-graining. These states form a subspace of the total Hilbert space:
\[
\{A \lvert \psi \rangle \mid A \text{ is a local operator}\} .
\]
This subspace is called the \emph{topological sector of states} (or topological excitation) generated by $\lvert \psi \rangle$, which can be viewed as a macroscopic observable \cite{KZ22a}. 
% In general, a topological sector of states is a subspace of the total Hilbert space that is invariant under the action of local operators.

Given an $n$d topological order $\SC_n$, all (particle-like, i.e. 0+1D) topological sectors of states form a category $\CC$. The hom space between two topological sectors of states consists of non-local operators that map one topological sector of states to another \cite{KW14}
% (such operators are also called \emph{instantons}  because they are localized in the spacetime and behave like a $0$D defect). 
Since topological sectors of states are invariant under the action of local operators, the hom spaces between them are also invariant under the action of local operators. 
A space of non-local operators that is closed under the action of local operators is called a \emph{topological sector of operators}.
It is clear that each hom space fall into a topological sector of operators.
% \begin{rem}
% More precisely, the hom spaces consist of topological sectors of unconfined operators that map one topological sector of states to another one. An operator is called unconfined if it is tensionless under all admitted perturbations. When there is no symmetry, all non-local operators are confined by introducing arbitrary perturbations.
% \end{rem}

\begin{rem}
When $n \geq 1$, the topological sectors of states in $\SC_n$ can be fused together in space: if two topological sectors of states $x$ and $y$ are closed enough to each other, they can be viewed as a single topological sector of states. 
The non-local operators can also be fused together because they are localized in the spacetime and behave like a $0$D defect.
Figure \ref{fig:fusion_topological_defect} depicts the fusion of topological sectors of states in a 1+1D topological order. Then $\CC$ is a (multi-)fusion category with the tensor product given by the fusion of topological sectors of states.
% When $n \geq 2$, 0+1D topological sectors of states in $\SC_n$ can also be braided in space and hence $\CC$ is a braided fusion category.
\end{rem}

\begin{figure}[htbp]
\centering
\[
\begin{array}{c}
        \begin{tikzpicture}[scale=0.8]
            \useasboundingbox (-0.3,0) rectangle (4.3,2) ;
            % \fill[color=gray!30,opacity=0.5] (0,0) rectangle (4,2);
            \draw[dashed] (1.5,0)--(1.5,2);
            \draw[dashed] (2.5,0)--(2.5,2) ;
            \filldraw[color=black] (1.5,0) circle (0.07) node[below] {$x$};
            \filldraw[color=black] (2.5,0) circle (0.07) node[below] {$y$};
            \draw[very thick] (0,0)--(4,0) node[above right] {$\SC$} ;
            \draw[-latex] (-0.5,1)--(-0.5,1.6) node[left]{$t$};
        \end{tikzpicture}
    \end{array}
    \xrightarrow{\text{fusion}}
    \begin{array}{c}
        \begin{tikzpicture}[scale=0.8]
            \useasboundingbox (0,0) rectangle (3.3,2) ;
            % \fill[color=gray!30,opacity=0.5] (0,0) rectangle (3,2);
            \draw[dashed] (1.5,0)--(1.5,2);
            \draw[very thick] (0,0)--(3,0) node[above right] {$\SC$} ;
            \filldraw[color=black] (1.5,0) circle (0.07) node[below] {$x \otimes y$};
        \end{tikzpicture}
\end{array}
\]
\caption{The fusion of topological defect in a 1+1D topological order $\SC$.}
\label{fig:fusion_topological_defect}
\end{figure}

% \mynote{Maybe add fusion of morphisms}

%\subsection{Group actions}\label{section:topological_sectors of states_onsite}

% To find the categorical description of the topological sectors of states in this lattice model, we need to look into details of the topological sectors of states in a topological order imposed with symmetry. So consider an $n$d topological order $\SC_n$ without symmetry, where the topological sectors of states and their morphisms form a category, denoted by $\CC$. 

Now we assume that the $n$d topological order $\SC_n$ is equipped with a unitary onsite symmetry given by a finite group $G$. Microscopically, the $G$-symmetry is usually realized by a collection of operators $\{U(g)\}_{g \in G}$. As a symmetry on a quantum system, these operators should satisfy the following `projective' relations:
%  because quantum states are defined up to a phases factor:
\[
U(g) U(h) \lvert \psi_x \rangle = \omega_x(g,h) U(gh) \lvert \psi_x \rangle , \quad g,h \in G ,
\]
where $\omega_x(g,h)$ is a phase factor, and $x \in \CC$ is the 
% simple 
topological sector of states generated by a state $\lvert \psi_x \rangle$.

For an onsite symmetry, the operator $U(g)$ is the tensor product of local operators on every site. Hence for any local operator $A$, the operator $U(g) A U(g)^{-1}$ is still a local operator. Therefore, given a topological sector of states $x \in \CC$, there is a well-defined topological sector of states $T_g(x) \in \CC$ defined by
\[
T_g(x) \coloneqq \{U(g) \lvert \psi \rangle \mid \lvert \psi \rangle \in x\} .
\]
Moreover, if $B$ is an operator that maps a topological sector of states $x$ to another one $y$, then $U(g) B U(g)^{-1}$ is an operator that maps $T_g(x)$ to $T_g(y)$. Hence we find the symmetry induces a functor $T_g \colon \CC \to \CC$ for every $g \in G$.

Note that the functors $T_g \circ T_h$ and $T_{gh}$ are not necessarily equal on the nose because the states $U(g) U(h) \lvert \psi_x \rangle$ and $U(gh) \lvert \psi_x \rangle$ may be differed by a phase factor $\omega_x(g,h)$. 
Such phase factors induce a natural isomorphism $\gamma_{g,h} \colon T_g \circ T_h \Rightarrow T_{gh}$, in which the component $(\gamma_{g,h})_x$ is given by the phase factor $\omega_x(g,h)^{-1}$ for all $x \in \CC$. By computing $U(g)U(h)U(k) \lvert \psi_x \rangle$ in two different ways we obtain the following equation:
\[
\omega_{T_k(x)}(g,h) \omega_x(gh,k) = \omega_x(h,k) \omega_x(g,hk) , \quad g,h,k \in G .
\] 
This equation translates to commutative diagram \ref{diag:G_action_appendix}.
% \[
%     \xymatrix{
%     (T_g T_h) T_k \ar@{=}[rr] \ar@{=>}[d]_{\gamma_{g,h} 1} & & T_g (T_h T_k) \ar@{=>}[d]^{1 \gamma_{h,k}} \\
%     T_{gh} T_k \ar@{=>}[r]^{\gamma_{gh,k}} & T_{ghk} & T_g T_{hk} \ar@{=>}[l]_{\gamma_{g,hk}}
%     }
% \]
% Mathematically, this condition means that the functors $\{T_g\}_{g \in G}$ and natural isomorphisms 
% $\{\gamma_{g,h}\}_{g,h \in G}$ 
% form a monoidal functor $T \colon G \to \Aut(\CC)$.
%% arXiv version
In other words, the $G$-symmetry on the topological order $\SC_n$ equips the category $\CC$ with a $G$-action.
%  (see appendix \ref{appedix:actions_equivariantization} for details).
%% SB version
%In other words, the $G$-symmetry on the topological order $\SC_n$ equips the category $\CC$ with a $G$-action (see appendix B.1 in Supplementary material for details).

\begin{rem}
%% arXiv version
Recall that $\CC$ is a (multi-)fusion category when $n \geq 1$. Since $G$ is an onsite symmetry, the symmetry action operators $\{U(g)\}_{g \in G}$ preserve the fusion of topological sectors of states. Thus the $G$-action on $\CC$ is indeed a monoidal action, i.e., $T$ is a monoidal functor $T \colon G \to \Aut_\otimes(\CC)$ (see appendix \ref{appedix:actions_equivariantization} for details).
%% SB version
% Recall that $\CC$ is a (multi-)fusion category when $n \geq 1$. Since $G$ is an onsite symmetry, the symmetry action operators $\{U(g)\}_{g \in G}$ preserve the fusion of topological sectors of states. Thus the $G$-action on $\CC$ is indeed a monoidal action, i.e., $T$ is a monoidal functor $T \colon G \to \Aut_\otimes(\CC)$ (see appendix B.1 in Supplementary material for details).
\end{rem}

\subsection{Equivariantization and \texorpdfstring{$G$}{G}-symmetric topological sectors of states} \label{section:equivariantization}

% For a topological order with symmetry, 
Not all topological sector of states are observables in LWLL after imposing a symmetry $\{U(g)\}_{g\in G}$. 
% We want to find all topological sector of states that respect $G$-symmetry.
% since there might be some topological sector of states $x$ are not invariant under the action of the symmetry.
% if it is not related to the symmetry.
An operator is $G$-symmetric if it commutes with each $U(g)$.
An observable in LWLL for a topological order with $G$-symmetry should be a subspace $x$ of the total Hilbert space that is invariant under both the action of $G$-symmetric local operators and the $G$-action.
% We define such sector $x$ as a \emph{$G$-symmetric topological sector of states}. 
%  to be a subspace of the total Hilbert space that is invariant under the action of $G$-symmetric local operators and also invariant under the $G$-action. 
% Then $G$-symmetric topological sectors of states are observables in LWLL for a topological order with $G$-symmetry.
%A topological sector of states $x$ is called \emph{$G$-symmetric} if it is invariant under the $G$-action, i.e., $T_g(x) = \{U(g) \lvert \psi \rangle \mid \lvert \psi \rangle \in x\}$ is still $x$ for every $g \in G$. For a topological order with $G$-symmetry, if we want to add a local term $\delta \mathcal H_\xi$ to the Hamiltonian, the operator $\delta \mathcal H_\xi$ itself must by $G$-symmetric, i.e., $[\delta \mathcal H_\xi,U(g)] = 0$. Equivalently, the topological sector of states trapped by $\delta \mathcal H_\xi$ is $G$-symmetric.
% In general, a $G$-symmetric topological sector of states 

Since we only know how to express topological sectors of states without symmetry in $\CC$, so in order to formulate $x$,
% Since $x$ may not be invariant under action of all local operators,
we first apply all local operators on $x$ to obtain a larger subspace $\bar x$ that is an object in $\CC$.
Note that $x$ is a $G$-invariant subspace (i.e., subrepresentation) of $\bar x$, and this invariant subspace is characterized by its behavior under the $G$-action. 
To restore the $G$-action of $x$ on $\bar x$, we act the operator $U(g)$, the subspace $x$ is invariant, but the states in $x$ may be changed by a (possibly non-abelian) phase. Categorically, such a phase under the $U(g)$-action is described by an isomorphism
\[
u_g \colon T_g(\bar x) \xrightarrow{\sim} \bar x , \quad g \in G .
\]
By comparing the action of $U(g) U(h)$ and $U(gh)$ for $g,h \in G$, we get commutative diagram \ref{diag:equivariant_object_appendix} for $\bar x$.
% \[
% %\be \label{diag:equivariantization_object}
% \begin{array}{c}
% \xymatrix{
% T_g T_h (\bar{x}) \ar[r]^{(\gamma_{g,h})_{\bar{x}}} \ar[d]^{T_g(u_h)} & T_{gh}(\bar{x}) \ar[d]^{u_{gh}} \\
% T_g(\bar{x}) \ar[r]^{u_g} & \bar x
% }
% \end{array}
% %\ee
% \]
Then the $G$-symmetric topological sector of states $x$ is characterized by the pair $(\bar x,\{u_g\}_{g \in G})$.  
% satisfying commutative diagram \ref{diag:equivariant_object_appendix}. 
% It is possible that $\bar x$ contains different $G$-symmetric topological sector of states, which are characterized by different $\{u_g\}$'s.

In addition, an operator between two such topological sectors of states $(\bar{x},\{u_g\}_{g \in G})$ and $(\bar{y},\{v_g\}_{g \in G})$ should also be symmetric, i.e., commutes with operators $U(g)$ for all $g \in G$. Categorically, it means diagram \ref{diag:equivariant_operator_appendix} commutes.
% \[
% \begin{array}{c}
% \xymatrix{
% T_g (\bar{x}) \ar[r]^{T_g(f)} \ar[d]^{u_g} & T_g(\bar{y}) \ar[d]^{v_g} \\
% \bar{x} \ar[r]^{f} & \bar{y}
% }
% \end{array}
% \]
%% arXiv version
Hence, the category of $G$-symmetric topological sector of states is an \emph{equivariantization} (see Definition \ref{defn:equivariantization}). Thus we have the following physical theorem.
%% SB version
%Hence, the category of $G$-symmetric topological sector of states is an \emph{equivariantization} (see Definition B.2 in Supplementary material). Thus we have the following physical theorem.

\begin{pthm} \label{pthm:symmetry_equivariantization}
Let $\SC_n$ be an $n$d topological order with a unitary onsite $G$-symmetry. Suppose all particle-like topological sectors of states (after ignoring the symmetry) form a category $\CC$.
\bnu[(1)]
\item The $G$-symmetry induces a $G$-action on the category $\CC$.
\item The category of (particle-like) $G$-symmetric topological sectors of states is equivalent to the equivariantization $\CC^G$.
\enu
\end{pthm}

% \begin{rem}
% Note that the above analysis does not depend on the dimension of $\SC_n$ nor $G$ symmetry abelian or not.
% When $n \geq 1$, the $G$-action should preserve the fusion of topological sectors of states, 
% thus each $T_g$ is a monoidal functor and each $\gamma_{g,h}$ is a monoidal natural isomorphism. It follows that $T$ is a monoidal functor from $G$ to the monoidal category $\Aut^{\otimes}(\CC)$ of monoidal autoequivalences of $\CC$.
% when $n \geq 2$, the $G$-action should also preserve the braiding. 
% thus $T$ is a monoidal functor from $G$ to the monoidal category $\Aut^{\mathrm{br}}(\CC)$ of braided autoequivalences of $\CC$.
% If $\CC$ is a (braided) monoidal category equipped with a $G$-action, the equivariantization $\CC^G$ is also a (braided) monoidal category.
% \end{rem}
As a result, 
for a 1+1D gapped phase with onsite symmetry $G$,
$\CC^G$ is the underlying category of its topological skeleton. We will demonstrate these structures in a 1+1D lattice model in the next section.

%For the $1$d lattice model with onsite abelian $G$ symmetry, $\CC^G$ is indeed equivalent to $\FZ_1(\rep(G))_{A_{(H)}} \simeq {}_{F_H} \rep(G)_{F_H}$, as we will show in section \ref{section:enriched_category}.
% In the next section, we would see the equivariantization $\CC^G$ in our lattice model is indeed ${}_{F_H} \rep(G)_{F_H} \simeq \FZ (\rep(G))_{A(H)}$.

% Now we are about to check the enriched category description ${}^{\FZ_1(\rep(G))}\FZ_1(\rep(G))_{A_{(H)}} \simeq {}^{\FZ_1(\rep(G))}{}_{F_H} \rep(G)_{F_H}$ works in real $1$d lattice model with finite abelian onsite symmetries. 
% In order to check the observables in each phase, we would give and analyze the lattice model in the next section.

% \begin{rem}
% Here the equivariantization $\CC^G$ is not the same as the topological skeleton of the topological order with $G$-symmetry, as introduced in section \ref{sec:introduction}. Actually, $\CC^G$ is the underlying category of the topological skeleton with the morphisms being sectors of local operators.
% \end{rem}

% \input{lattice}
% !TeX root = main.tex
% !TeX program = pdfLaTeX

\section{General analysis for lattice model with abelian onsite \texorpdfstring{$G$}{G}-symmetry} \label{section:lattice_model}

% Let $G$ be a finite abelian group. 
In this section, we construct a lattice model exhibits an onsite abelian $G$-symmetry to analyze its macroscopic observables in both symmetric and symmetry breaking cases. % From which we would see the topological sector of operators form a braided fusion category $\CB \simeq \FZ(\rep(G))$. 
This model construction can be generalized to any 1+1D gapped phases with fusion categorical symmetries \cite{Ina22, LDOV21}.

\subsection{The lattice model and ground states}\label{sec:model}

% In order to construct the lattice model with an onsite abelian $G$-symmetry, \
Consider the 1+1D lattice (defined on an infinitely long open interval) with the total Hilbert space $\CH_{\text{tot}} = \bigotimes_{i \in \mathbb{Z}} \CH_i$, where each local Hilbert space $\CH_i$ is the group algebra $\Cb[G]$ spanned by an orthonormal basis $\{\lvert g \rangle_i \mid g \in G\}$.
For each site $i$ and $g \in G$, we define an operator $L_g^i$ acting on $\CH_i$ as follows:
\[
L_g^i \lvert h \rangle_i \coloneqq \lvert gh \rangle_i , \quad \forall h \in G .
\]
%Then each local Hilbert space $\mathcal H_i$ equipped with the $G$-action $\{L_g^i\}_{g \in G}$ is a $G$-representation.
The global $G$-symmetry action is defined by
\[
U(g) \coloneqq \bigotimes_i L_g^i , \quad \forall g \in G .
\]
%We need to find (local and non-local) operators commute with $ U(g)$ (i.e., $[P, U(g)] = 0$),  these operators are called $G$-symmetric operators. A topological sector of operators should be invariant under the action of $G$-symmetric local operators and the symmetries.

In order to construct the Hamiltonian, we need to define some operators. Recall that all group homomorphisms $G \to \mathrm U(1)$ form a finite abelian group $\hat G$, called the \emph{dual group} of $G$. Equivalently, $\hat G$ is also the group of equivalence classes of irreducible $G$-representations. For each site $i$ and $\rho \in \hat G$, we define an operator $Z_\rho^i$ acting on $\mathcal H_i$ by
\[
Z_\rho^i \lvert g \rangle_i \coloneqq \rho(g) \lvert g \rangle_i .
\]
It is clear that the following equations hold for all $g,k \in G$ and $\rho,\sigma \in \hat G$:
\begin{gather}
L_g^i L_k^i = L_{gk}^i , \qquad (L_g^i)^\dagger = (L_g^i)^{-1} = L_{g^{-1}}^i , \nonumber \\
Z_\rho^i Z_\sigma^i = Z_{\rho \sigma}^i , \qquad (Z_\rho^i)^\dagger = (Z_\rho^i)^{-1} = Z_{\rho^{-1}}^i , \nonumber \\
L_g^i Z_\rho^i (L_g^i)^{-1} = \rho(g)^{-1} Z_\rho^i . \label{eq:Lg_Zrho}
\end{gather}

Let $H \subseteq G$ be a subgroup of $G$. Note that $\widehat{G/H}$ can be naturally viewed as a subgroup of $\hat G$: every group homomorphism $\lambda \colon G/H \to \mathrm U(1)$ induces a group homomorphism $G \twoheadrightarrow G/H \xrightarrow{\lambda} \mathrm U(1)$, which is nontrivial if $\lambda$ is nontrivial. Then for each site $i$ we define two Hermitian operators:
\[
X_H^i \coloneqq \frac{1}{\lvert H \rvert} \sum_{h \in H} L_h^i , \quad \quad Z_H^{i,i+1} \coloneqq \frac{\lvert H \rvert}{\lvert G \rvert} \sum_{\rho \in \widehat{G/H}} Z_\rho^i (Z_\rho^{i+1})^\dagger .
\]
It is easy to verify that these operators are mutually commuting projectors and $G$-symmetric (i.e. commute with $U(g)$ for all $g \in G$).

\medskip
The lattice model we considered depends on the symmetry group $G$ and a chosen subgroup $H \subseteq G$. The Hamiltonian is defined as follows:
\be \label{eq:Hamiltonian}
\mathcal{H} \coloneqq \sum_i (1 - X_H^i) + \sum_i (1 - Z_H^{i,i+1}) .
\ee
%where $B, J$  are coupling constants, since they are insignificant for our result, we set them to be $1$ in the following paper. \mynote{(why coupling constant?)}

%Note that when we pick $H$ to be $G$, the Hamiltonian realizes a SPT phase, which has only one term $ \mathcal{H} = \sum_i (1 - X_G^i) $ left. On the other hand, when we set all subgroups $H$ to be $\{e\}$, 
%% where $e$ is the unit element of $G$, 
%$\mathcal{H} = \sum_i (1 - E_{\{e\}}^{i,i+1})$ realizes a symmetry completely broken phase (SCB).

\begin{rem}
When $G = \Zb_n$, the Hamiltonian equivalently realizes the bosonic quantum clock model \cite{Fen12},
where $L^i$ and $Z^i$ are $n\times n$ generalizations of the Pauli matrices $\sigma_x$ and $\sigma_z$.
In particular, when $G = \Zb_2$, it recovers the well-known 1d quantum Ising model. We illustrate both cases explicitly in section \ref{section:example}.
\end{rem}

% \begin{rem}
% After our paper appeared on arXiv, we noticed that there were constructions of 1d lattice models with general fusion category symmetries (also called algebraic higher symmetries) in \cite{Ina22,LDOV21}.  The lattice models we used in this work can be viewed as special cases of their constructions with $\vect_G$-symmetries. However, they did not find the enriched category of all macroscopic observables.
% \end{rem}

Since the Hamiltonian \eqref{eq:Hamiltonian} is the sum of commuting projectors, it is not hard to verify that the ground state subspace has dimension $\lvert G \rvert / \lvert H \rvert$ and there is an orthonormal basis labeled by coset $x \in G/H$:
\be \label{eq:ground_state_H}
\lvert \psi_x \rangle \coloneqq \bigotimes_i \frac{1}{\sqrt{\lvert H \rvert}} \sum_{g \in x} \lvert g \rangle_i .
\ee
Moreover, we have $U(g) \lvert \psi_x \rangle = \lvert \psi_{gx} \rangle$ for $g \in G$ and $x \in G/H$. Thus the ground state $\lvert \psi_H \rangle$ (indeed, every ground state $\lvert \psi_x \rangle$) is stable under the action of $U(h)$ for $h \in H$. Hence the lattice model \eqref{eq:Hamiltonian} spontaneously breaks the $G$-symmetry to the subgroup $H$.

%\subsubsection{The ground state subspaces}
%
%Fix a subgroup $H \subseteq G$, as
%\be \label{eq:Lg_Zrho}
%L_g^i Z_\rho^i (L_g^i)^{-1} = \rho(g)^{-1} Z_\rho^i ,
%\ee
%both $X_H^i$ and $Z_H^{j,j+1}$ are $U$-symmetric operators and they are mutually commutative for all $g \in G$ and sites $i,j$, i.e.
%\[
%    [U(g),X_H^i] = [U(g),Z_H^{j,j+1}] = [X_H^i,X_H^j] = [Z_H^{i,i+1},Z_H^{j,j+1}] = [X_H^i,Z_H^{j,j+1}] = 0 ,
%\]
%Hence, the ground state subspace is the common eigenspace of all $X_H^i$'s and $Z_H^{j,j+1}$'s such that the eigenvalues of them get $1$. As the Hamiltonian realizes a phase which breaks the symmetry from $G$ to $H$, it is not hard to see that the ground state subspace has dimension $\lvert G \rvert / \lvert H \rvert$, and there is an orthonormal basis labeled by $x \in G/H$:
%\[
%    \lvert \psi_x \rangle \coloneqq \bigotimes_i \frac{1}{\sqrt{\lvert H \rvert}} \sum_{g \in x} \lvert g \rangle_i .
%\]
%\begin{rem}
%    As a $G$-representation, the ground state subspace is isomorphic to $\fun(G/H) =: F_H $, which is a condensable algebra in $\rep(G)$, with $\rep(G)_{F_H} \simeq \rep(H)$.
%\end{rem}

\subsection{Topological sectors of operators}\label{sec:topo_operator}

%Through the above analysis, we can easily see two kinds of $G$-symmetric non-local operators, which are one-to-one correspond to irreducible representations of the quantum double (sometimes called fluxion and chargeon \cite{BSW11}):

Now let us analyse the topological sector of operators within this model.
There are two kinds of $G$-symmetric non-local operators:
\[
M_g^i \coloneqq \prod_{j \leq i} L_g^j , \, \forall g \in G , \quad \quad E_\rho^i = \prod_{j\geq i} Z_\rho^j ( Z_\rho^{j+1})^\dagger , \, \forall \rho \in \hat G .
\]
% \[
%     M_g^{i} \coloneqq \prod_{j \leq i} L_g^j , \, \forall g \in G , \qquad \qquad Z_\rho^{ i} = \prod_{j \leq i} ( Z_\rho^{j-1})^\dagger Z_\rho^j , \, \forall \rho \in \hat G .
% \]
Since $E_\rho^i$ is the product of infinitely many $G$-symmetric operators, it should be viewed as a $G$-symmetric non-local operator. On the other hand, the single operator $Z_\rho^i$ is a non-symmetric local operator. Note that $E_\rho^i = Z_\rho^i$ as an operator that catches all the corrected properties near site $i$.
%  We use different notations to distinguish them.

\begin{rem}
These two operators are also known as fluxions and chargeons in a quantum double model \cite{BSW11}. In some literatures \cite{JW20,CW22}, they are called ``patch operators''.
\end{rem}

Apparently, the product $M_g^{ i} E_\rho^j$ is also a symmetric non-local operator. We denote the topological sector of operators generated by $M_g^{ i} E_\rho^j$ by $\mathcal O_{(g,\rho)}$.
They form the simple objects of the background category.

By \eqref{eq:Lg_Zrho} we have
\[
M_g^{ i} E_\rho^j = \begin{cases} E_\rho^j M_g^{ i} , & i < j , \\ \rho(g)^{-1} E_\rho^j M_g^i , & i \geq j . \end{cases}
\]
Thus, $M_g^{ i} E_\rho^j M_h^{ k} E_\sigma^{l}$ and $M_g^{ i} M_h^{ k} E_\rho^j E_\sigma^{l}$ differ by a coefficient at most. It follows that the fusion products of these topological sectors of operators are given by
\be \label{eq:operator_fusion}
\mathcal O_{(g,\rho)} \otimes \mathcal O_{(h,\sigma)} = \mathcal O_{(gh,\rho \sigma)} .
\ee
This recovers the fusion structure in $\FZ_1(\rep(G))$. 

Moreover, for $i<k<j$ we have
\[
M_g^{ i} M_{g^{-1}}^{ j} E_\rho^k M_g^{ j} M_{g^{-1}}^{ i} = \rho(g) \cdot E_\rho^k .
\]
This means that the double braiding of two topological sectors $\mathcal O_{(g,1)}$ and $\mathcal O_{(e,\rho)}$ is
\[
\mathcal{O}_{(g,1)} \otimes \mathcal{O}_{(e,\rho)} \xrightarrow{\rho(g) \cdot 1(e)} \mathcal{O}_{(g,1)} \otimes \mathcal{O}_{(e,\rho)} .
\]
Conceptually, we can regard this process as first creating a pair of fluxions at sites $i$ and $j$, then winding a chargeon around one of the fluxions (where the chargeon would cross in between the fluxions on site $k$), after this process, we annihilate the pair of fluxions and obtain a phase $\rho(g)$. For a general braiding process, we have
\begin{align*}
M^{ i}_g E^{ i}_{\rho} E^{ j}_{\rho^{-1}} M^{ j}_{g^{-1}} M^{ k}_h E^{ k}_{\sigma} M^{ j}_g E^{ j}_{\rho} E^{ i}_{\rho^{-1}} M^{ i}_{g^{-1}} \\
= \rho(h)\sigma(g) M^{ k}_h E^{ k}_{\sigma} ,
\end{align*}

and this equation gives the double braiding
\begin{equation} \label{eq:operator_braiding}
\mathcal{O}_{(h,\sigma)} \otimes \mathcal{O}_{(g,\rho)} \xrightarrow{\sigma(g) \rho(h)} \mathcal{O}_{(h,\sigma)} \otimes \mathcal{O}_{(g,\rho)}.
\end{equation}
%\begin{equation} \label{eq:operator_braiding}
%    \mathcal{O}_{(h,\sigma)} \otimes \mathcal{O}_{(g,\rho)}
%    \xrightarrow{\rho(h)} \mathcal{O}_{(g,\rho)} \otimes \mathcal{O}_{(h,\sigma)}
%    \xrightarrow{\sigma(g)}
%    \mathcal{O}_{(h,\sigma)} \otimes \mathcal{O}_{(g,\rho)}.
%\end{equation}

%% arXiv version
The topological sectors of operators $\{\mathcal O_{(g,\rho)}\}_{g \in G,\rho \in \hat G}$, together with the fusion and braiding properties \eqref{eq:operator_fusion} \eqref{eq:operator_braiding}, form a braided fusion category equivalent to $\FZ_1(\rep(G))$ (see appendix \ref{appendix:pointed_metric_group} and example \ref{expl:abelian_center_metric} for details).
%% SB version
%This recovers the braiding structure in $\FZ_1(\rep(G))$. Thus we have the following physical theorem (see appendix A.1 and example A.3 in Supplementary material for details).

\begin{pthm}
In a 1d lattice model with an onsite abelian $G$-symmetry, the topological sectors of operators $\{\mathcal O_{(g,\rho)}\}_{g \in G,\rho \in \hat G}$ , together with the fusion and braiding properties \eqref{eq:operator_fusion} \eqref{eq:operator_braiding}, form a braided fusion category equivalent to $\FZ_1(\rep(G))$.
\end{pthm}

%Since we focus on the case that $G$ is a finite abelian group. In this case, $\FZ_1(\rep(G))$ is a pointed braided fusion category and can be described by the metric group $(G \oplus \hat G,q)$, where $q(g,\rho) \coloneqq \rho(g)$ (appendix \ref{appendix:pointed_metric_group}). The simple objects of the metric group are just $\{\mathcal{O}_{(g,\rho)}\}_{g \in G,\rho \in \hat G}$. The forgetful functor $L_G: \FZ_1(\rep(G)) \to \rep(G)$ is given by $\mathcal{O}_{(g,\rho)} \mapsto \rho$, and the double braiding is given by
%\begin{equation} \label{eq:braiding}
%    \mathcal{O}_{(h,\sigma)} \otimes \mathcal{O}_{(g,\rho)}
%    \xrightarrow{\rho(h)} \mathcal{O}_{(g,\rho)} \otimes \mathcal{O}_{(h,\sigma)}
%    \xrightarrow{\sigma(g)}
%    \mathcal{O}_{(h,\sigma)} \otimes \mathcal{O}_{(g,\rho)}.
%\end{equation}
%
%In other words, the topological sector of operators indeed form a braided fusion category $\FZ_1(\rep(G))$.

\subsection{The topological skeletons}\label{section:enriched_category}

In order to convince the readers of the categorical structure stated in Theorem$^{\text{ph}}$ \ref{pthm:symmetry_breaking_phase}, we would like to give the exact categorical description for each phase step by step. It would be beneficial to first check the symmetry preserving case and the symmetry completely broken case.

\subsubsection{Symmetry preserving case (the trivial SPT)}

Consider the phase realized by \eqref{eq:Hamiltonian} with $H=G$. In other words, the Hamiltonian is
\be \label{eq:Hamiltonian_symmetry_preserving}
\mathcal H = \sum_i (1 - X_G^i) .
\ee
In this case there is no symmetry breaking and the unique ground state is
%\be \label{eq:ground_state_symmetry_preserving}
\[
\lvert \Omega_G \rangle \coloneqq \bigotimes_i \frac{1}{\sqrt{\lvert G \rvert}} \sum_{g \in G} \lvert g \rangle_i .
\]
%\ee
Since the ground state %\eqref{eq:ground_state_symmetry_preserving}
$\lvert \Omega_G \rangle$ is a product state, the Hamiltonian \eqref{eq:Hamiltonian_symmetry_preserving} realizes the trivial $G$-SPT order.

We use $\mathcal S_1$ to denote the topological sector of states generated by the ground state $\lvert \Omega_G \rangle$ (i.e. the trivial topological sector of states).
Note that $M_g^i \lvert \Omega_G \rangle = \lvert \Omega_G \rangle$ for all sites $i$ and $g \in G$, so the non-local operator $M_g^i$ does not generate a new topological sector of states.

Then we consider the topological sector of states generated by $E_{\rho}^i$.
For each $\rho \in \hat G$ and site $i$, we define
\[
\lvert \rho,i \rangle \coloneqq E_\rho^i \lvert \Omega_G \rangle. 
% ( = Z_\rho^i \lvert \Omega_G \rangle ) .
\]
In particular, $\lvert 1,i \rangle = \lvert \Omega_G \rangle$ where $1 \in \hat G$ is the trivial $G$-representation. By the orthogonality of characters, one can verify that $X_G^i \lvert \rho,i \rangle = 0$ if $\rho \neq 1$ and $X_G^i \lvert 1,i \rangle = X_G^i \lvert \Omega_G \rangle = \lvert \Omega_G \rangle$. Moreover, for two sites $i$ and $j$, the states $\lvert \rho,i \rangle$ and $\lvert \rho,j \rangle$ generate the same $G$-symmetric topological sector of states because they are differed by a symmetric local operator $Z_\rho^j (Z_\rho^i)^\dagger$:
\[
\lvert \rho,j \rangle = Z_\rho^j (Z_\rho^i)^\dagger \lvert \rho,i \rangle. 
% = \biggl( \prod_{i \leq k < j} E_\rho^{k+1} (E_\rho^k)^\dagger \biggr) \lvert \rho,i \rangle.
\]
Therefore, we use $\mathcal S_\rho$ to denote the $G$-symmetric topological sector of states generated by $\lvert \rho,i \rangle$, they form simple objects of the underlying category.

\begin{rem}\label{SPT_rem_without_sym}
    The state $\lvert \rho,i \rangle$ can also be regarded as created from the ground state by a non-symmetric local operator $Z_\rho^i$.
    So if we ignore the $G$-symmetry, the state $\lvert \rho,i \rangle$ belongs to the trivial topological sector of states $\mathcal{S}_1$, in this situation the underlying category is just $\vect$.
\end{rem}

Intuitively, the $G$-symmetric topological sector of states $\mathcal S_\rho \otimes \mathcal S_\sigma$ (i.e. the fusion of $\mathcal S_\rho$ and $\mathcal S_\sigma$) should be generated by two non-local operators $E_{\rho}^i$, $E^j_{\sigma}$, i.e. consider the following state:
\[
\lvert \rho,i;\sigma,j \rangle \coloneqq E_\rho^i E_\sigma^{j} \lvert \Omega_G \rangle \in \mathcal{S}_{\rho}\ot\mathcal{S}_{\sigma},
\]
where $\rho,\sigma \in \hat G$ and $i<j$. On the other hand, by acting the symmetric local operator $Z_\rho^j (Z_\rho^i)^\dagger$ on this state we find that $\lvert \rho,i;\sigma,j \rangle$ generates the same $G$-symmetric topological sector of states as $\lvert \rho\sigma,j \rangle$:
\[
Z_\rho^j (Z_\rho^i)^\dagger \lvert \rho,i;\sigma,j \rangle = E_\rho^{j} E_\sigma^{j} \lvert \Omega_G \rangle = E_{\rho \sigma}^{j} \lvert \Omega_G \rangle = \lvert \rho\sigma,j \rangle.
\]
It follows that the fusion rules of topological sectors of states of the phase realized by \eqref{eq:Hamiltonian_symmetry_preserving} are given by
\[
\mathcal S_\rho \otimes \mathcal S_\sigma = \mathcal S_{\rho \sigma} .
\]
Therefore, the fusion category of topological sectors of states is equivalent to the fusion category $\rep(G)$ of finite-dimensional $G$-representations.

\smallskip
From another perspective, if we view \eqref{eq:Hamiltonian_symmetry_preserving} as a system without symmetry, it realizes the trivial 1d topological order (without symmetry) because the ground state $\lvert \Omega_G \rangle$ is a product state. Thus the topological sectors of states form a fusion category equivalent to $\vect$ (see also remark \ref{SPT_rem_without_sym}). 
%% arXiv version
By Theorem$^{\text{ph}}$ \ref{pthm:symmetry_equivariantization}, if we view \eqref{eq:Hamiltonian_symmetry_preserving} as a system with $G$-symmetry, the $G$-symmetric topological sectors of states form a fusion category equivalent to the equivariantization $\vect^G \simeq \rep(G)$ (see example \ref{expl:vect_equivariantization_rep_G}), which coincide with the upper analysis.
%% SB version
%By Theorem$^{\text{ph}}$ \ref{pthm:symmetry_equivariantization}, if we view \eqref{eq:Hamiltonian_symmetry_preserving} as a system with $G$-symmetry, the $G$-symmetric topological sectors of states form a fusion category equivalent to the equivariantization $\vect^G$, which is just $\rep(G)$ (see example B.4 in Supplementary material).

\smallskip

Now we have the topological sectors of operators $\FZ_1(\rep(G))$ and the topological sectors of states $\rep(G)$. 
% Since non-local operators can naturally act on topological excitations, this will reflect the action of background category on the underlying category.
We want to figure out the action of $\FZ_1(\rep(G))$ on $\rep(G)$ on the macroscopic level. 

Since $M_g^{i}$ only has trivial actions on the ground state $\lvert \Omega_G \rangle$, the operator $M_g^{i} E_\rho^{j}$ also maps the sector $\mathcal S_\sigma$ to $\mathcal S_{\rho \sigma}$ for all $g \in G$ and $\rho,\sigma \in \hat G$. 
Equivalently, $M_g^{i} E_{\sigma \rho^{-1}}^{j}$ maps the topological sector $\mathcal S_\rho$ to $\mathcal S_{\sigma}$. 
Hence, the action of $\FZ_1(\rep(G))$ on $\rep(G)$ can be expressed by
\be
\mathcal{O}_{(g,\sigma \rho^{-1})}\odot \mathcal{S}_{\rho}=S_{\sigma}
\ee

This action can also be characterized by the hom spaces between topological sectors of states:
\be \label{SPT_hom}
\Hom(\mathcal S_\rho,\mathcal S_\sigma) = \bigoplus_{g \in G} \mathcal O_{(g,\sigma \rho^{-1})} .
\ee
This is indeed the internal hom of $\bc[\FZ_1(\rep(G))]{\rep(G)}$.

Equivalently, this action can be determined by the ground state algebra consisting of the operators that act on the topological sector of the ground state invariably. In this case, the ground state algebra is
\[
A(G)\coloneqq \Hom(\mathcal S_1,\mathcal S_1) = \bigoplus_{g \in G} \mathcal O_{(g, 1)} \in \FZ_1(\rep(G)) .
\]
In the view of holographic duality, the trivial action of $A(G)$ or the operators $M_g^{i}$ on the ground state can be interpreted as the condensation of fluxions.

Hence we have proved the following physical theorem within a concrete model step by step:

\begin{pthm}\label{thm_SPT}
The macroscopic observables of the $1+1$D trivial SPT order with the bosonic onsite abelian symmetry $G$ form enriched fusion category $\bc[\FZ_1(\rep(G))]{\rep(G)} \simeq  \bc[\FZ_1(\rep(G))]{\FZ_1(\rep(G))_{A(G)}}$.
\end{pthm}

\subsubsection{Symmetry completely broken case}
\label{sec:SCB}

The procedures of finding the underlying category when $G$ is completely broken (i.e., $H=\{e\}$ is the trivial group) follows a similar path. The Hamiltonian now is
\be \label{eq:Hamiltonian_symmetry_broken}
    \mathcal{H} = \sum_i (1 - Z_{\{e\}}^{i,i+1}) .
\ee
In this case, the ground state subspace has dimension $\lvert G \rvert$ and there is an orthonormal basis labeled by group elements $g \in G$:
\[
\lvert \psi_g \rangle \coloneqq \bigotimes_i \lvert g \rangle_i .
\]
Clearly we have $U(g) \lvert \psi_h \rangle = \lvert \psi_{gh} \rangle$ for $g,h \in G$. %Thus the ground state subspace, as a $G$-representation, is isomorphic to the function algebra $\fun(G)$.

\smallskip

First we view \eqref{eq:Hamiltonian_symmetry_broken} as a system without symmetry. Since each ground state $\lvert \psi_g \rangle$ is a product state, \eqref{eq:Hamiltonian_symmetry_broken} realizes the direct sum of $\lvert G \rvert$ copies of the trivial 1d topological order (physically we need a fine tuning to realize this phase). There are $\lvert G \rvert^2$ simple topological sectors of states (domain walls) $\{\mathcal T_{g,h}\}_{g,h \in G}$ generated by the following states:
\[
\lvert \psi_{g,h,i} \rangle \coloneqq \bigl( \bigotimes_{j \leq i} \lvert g \rangle_j \bigr) \otimes \bigl( \bigotimes_{j > i} \lvert h \rangle_j \bigr) , \quad g,h \in G .
\]

\begin{figure}[H]
\centering
\begin{tikzpicture}[scale=0.8]
        \draw[thick](-4,0)--(5, 0);
        \draw[thick](0, 0.1)--(0, -0.1) node[above] at (0, 0.2) {$|g\rangle$} node[below] at (0, -0.2) {$i$};
        \draw[thick](-1, 0.1)--(-1, -0.1)node[above] at (-1, 0.2) {$|g\rangle$} node[below] at (-1, -0.2) {$i-1$};
        \draw[thick](-2, 0.1)--(-2, -0.1) node[above] at (-2, 0.2) {$|g\rangle$} node[below] at (-2, -0.2) {$i-2$};
        \draw[thick](-3, 0.1)--(-3, -0.1) node[above] at (-3, 0.2) {$|g\rangle$} node[below] at (-3, -0.2) {$\cdots$};
        \draw[thick](1, 0.1)--(1, -0.1) node[above] at (1, 0.2) {$|h\rangle$} node[below] at (1, -0.2) {$i+1$};
        \draw[thick](2, 0.1)--(2, -0.1)node[above] at (2, 0.2) {$|h\rangle$} node[below] at (2, -0.2) {$i+2$};
        \draw[thick](3, 0.1)--(3, -0.1)node[above] at (3, 0.2) {$|h\rangle$} node[below] at (3, -0.2) {$\cdots$};
        \draw[thick](4, 0.1)--(4, -0.1) node[above] at (4, 0.2) {$|h\rangle$} node[below] at (4, -0.2) {$\cdots$};
        \node at(5, 0.2){$\cdots$};
        \node at(-4, 0.2){$\cdots$};
\end{tikzpicture}
\caption{A topological sector of states without symmetry $\mathcal T_{g,h}$ can be generated by the state $\lvert\psi_{g,h,i}\rangle$, which can be interpreted as a 0+1D domain wall on 1+1D lattice model.}
\end{figure}

Clearly the fusion rules of these topological sectors of states are given by
\[
\mathcal T_{g,h} \otimes \mathcal T_{k,l} = \delta_{h,k} \mathcal T_{g,l} , \quad g,h,k,l \in G .
\]
Therefore, these topological sectors of states form a multi-fusion category equivalent to $\mathrm{Mat}_{\lvert G \rvert}(\vect)$ of $\lvert G \rvert$-by-$\lvert G \rvert$ matrices valued in $\vect$, or equivalently, $\Fun(\vect_G,\vect_G)$ of linear functors.

\smallskip
If we view \eqref{eq:Hamiltonian_symmetry_broken} as a system with $G$-symmetry, the symmetry group $G$ acts on the multi-fusion category $\Fun(\vect_G,\vect_G)$ of topological sectors of states, and the fusion category of $G$-symmetric topological sector of states is equivalent to the equivariantization $\Fun(\vect_G,\vect_G)^G$ by Theorem$^{\text{ph}}$ \ref{pthm:symmetry_equivariantization}. 
The $G$-action on states is as follows:
\[
U(g) \lvert \psi_{h,k,i} \rangle = \lvert \psi_{gh,gk,i} \rangle , \quad g,h,k \in G .
\]
Thus the $G$-action on $\Fun(\vect_G,\vect_G) \simeq \mathrm{Mat}_{\lvert G \rvert}(\vect)$ is given by
\[
g(\mathcal T_{h,k}) = \mathcal T_{gh,gk} .
\]
Mathematically, the equivariantization $\Fun(\vect_G,\vect_G)^G$ is equivalent to $\vect_G$ as fusion categories. 
%% arXiv version
This is a special case of Proposition \ref{prop:bimodule=equivariantization} below (see also remark \ref{rem:equivariantization_special_case}). 
%% SB version
%This is a special case of Proposition \ref{prop:bimodule=equivariantization} below (see also remark B.11 in Supplementary material). 

\begin{figure}[H]
\centering
\begin{tikzpicture}[scale=0.8]
        \draw[thick](-4,0)--(5, 0);
        \draw[thick](0, 0.1)--(0, -0.1) node[above] at (0, 0.2) {$|gh\rangle$} node[below] at (0, -0.2) {$i$};
        \draw[thick](-1, 0.1)--(-1, -0.1)node[above] at (-1, 0.2) {$|gh\rangle$} node[below] at (-1, -0.2) {$i-1$};
        \draw[thick](-2, 0.1)--(-2, -0.1) node[above] at (-2, 0.2) {$|gh\rangle$} node[below] at (-2, -0.2) {$i-2$};
        \draw[thick](-3, 0.1)--(-3, -0.1) node[above] at (-3, 0.2) {$|gh\rangle$} node[below] at (-3, -0.2) {$\cdots$};
        \draw[thick](1, 0.1)--(1, -0.1) node[above] at (1, 0.2) {$|gk\rangle$} node[below] at (1, -0.2) {$i+1$};
        \draw[thick](2, 0.1)--(2, -0.1)node[above] at (2, 0.2) {$|gk\rangle$} node[below] at (2, -0.2) {$i+2$};
        \draw[thick](3, 0.1)--(3, -0.1)node[above] at (3, 0.2) {$|gk\rangle$} node[below] at (3, -0.2) {$\cdots$};
        \draw[thick](4, 0.1)--(4, -0.1) node[above] at (4, 0.2) {$|gk\rangle$} node[below] at (4, -0.2) {$\cdots$};
        \node at(5, 0.2){$\cdots$};
        \node at(-4, 0.2){$\cdots$};
\end{tikzpicture}
\caption{When symmetry $G$ is imposed on the system, the global symmetry $U(g)$ does not generate a new symmetric sector, e.g. $\mathcal T_{gh,gk}$ and $\mathcal T_{h,k}$ both belong to the $G$-symmetric topological sector of states $\mathcal{S}_{hk^{-1}}$.}
\end{figure}

The mathematical proof can also be reformulated in physical language by explicitly finding the $G$-symmetric topological sectors of states in this lattice model. 
% In this way we prove the equivalence $\Fun(\vect_G,\vect_G)^G \simeq \vect_G$ in this concrete lattice model.
We consider the ground states first. Since the $G$-action permutes the ground states $\lvert \psi_g \rangle$, each ground state (and hence the ground state subspace) generates the same $G$-symmetric topological sector of states:
\[
\mathcal S_e \coloneqq \bigoplus_{g \in G} \mathcal T_{g,g} .
\]
Similarly, given $g \in G$, the states $M_g^{i} \lvert \psi_h \rangle = \lvert \psi_{gh,h,i} \rangle$ for all $h \in G$ and sites $i$ generate the same $G$-symmetric topological sector of state (note that $E_\rho^i \lvert \psi_h \rangle = \rho(h) \lvert \psi_h \rangle$ does not create a new topological sector of states):
\[
\mathcal S_g \coloneqq \bigoplus_{h \in G} \mathcal T_{gh,h} .
\]
% \mynote{The topological sector of states $T_{g,h}$ belongs to which symmetric sector $S_{k}$ is determined by the difference $k=gh^{-1}$.}
These $\mathcal{S}_g$'s form the simple objects of the underlying category.
It is not hard to see that the fusion products of $G$-symmetric topological sectors of states are given by
\[
\mathcal S_g \otimes \mathcal S_h = \mathcal S_{gh} .
\]
Hence the fusion category of $G$-symmetric topological sectors of states is equivalent to $\vect_{G}$.

Now we have the topological sectors of operators $\FZ_1(\rep(G))$ and the topological sectors of states $\vect_{G}$. 
Since $M_g^i$ maps the sector $\mathcal{S}_h$ to $\mathcal{S}_{gh}$, and $E_\rho^j$ has only trivial actions on these sectors, the operator $M_g^{i} E_\rho^{j}$ maps the sector $\mathcal S_h$ to $\mathcal S_{gh}$ for all $g,h \in G$ and $\rho \in \hat G$. 
The action of $\FZ_1(\rep(G))$ on $\vect_{G}$ on the macroscopic level is given by 
\be
\mathcal{O}_{(hg^{-1},\rho)}\odot \mathcal{S}_g=\mathcal{S}_h
\ee

This action can also be characterized by the hom  spaces between $G$-symmetric topological sectors of states:
\be\label{SCB_hom}
\Hom(\mathcal S_g,\mathcal S_h) = \bigoplus_{\rho \in \hat G} \mathcal O_{(h g^{-1},\rho)} .
\ee
This is indeed the internal hom of $\bc[\FZ_1(\rep(G))]{\vect_G}$. In particular, the ground state algebra is
\[
A(\{e\}) \coloneqq \Hom(\mathcal S_e,\mathcal S_e) = \bigoplus_{\rho \in \hat G} \mathcal O_{(e,\rho)} \in \FZ_1(\rep(G)) .
\]
In the view of holographic duality, the trivial action of $A(\{e\})$ or the operators $E_\rho^i$ on the sector of ground state can be interpreted as the condensation of chargeons.

Hence we have proved the following physical theorem:

\begin{pthm}\label{thm_cmplbreak}
The macroscopic observables of the $1+1$D completely symmetry-breaking order with the bosonic onsite abelian symmetry $G$ form enriched fusion category $\bc[\FZ_1(\rep(G))]{\vect_G} \simeq \bc[\FZ_1(\rep(G))]{\FZ_1(\rep(G))_{A(\{e\})}}$.
\end{pthm}

\subsubsection{General cases (symmetry partially broken)}\label{section:general_cases}

For the general symmetry breaking cases, we fix a subgroup $H \subseteq G$ in Hamiltonian \eqref{eq:Hamiltonian}. Recall that the ground state subspace has dimension $\lvert G \rvert / \lvert H \rvert$ and there is an orthonormal basis $\{\lvert \psi_x \rangle\}_{x \in G/H}$ \eqref{eq:ground_state_H} satisfying $U(g) \lvert \psi_x \rangle = \lvert \psi_{gx} \rangle$ for $g \in G$ and $x \in G/H$. Thus the ground state subspace is isomorphic to the function algebra $F_H \coloneqq \fun(G/H)$ as a $G$-representation.

\smallskip
Again we view \eqref{eq:Hamiltonian} as a system without symmetry. Since each ground state $\lvert \psi_x \rangle$ \eqref{eq:ground_state_H} is a product state, \eqref{eq:Hamiltonian} realizes the direct sum of $n \coloneqq \lvert G \rvert / \lvert H \rvert$ copies of the trivial 1d topological order. There are $n^2$ simple topological sectors of states (domain walls) $\{\mathcal T_{x,y}\}_{x,y \in G/H}$ generated by
\[
\lvert \psi_{x,y,i} \rangle \coloneqq \bigl( \bigotimes_{j \leq i} \frac{1}{\sqrt{\lvert H \rvert}} \sum_{g \in x} \lvert g \rangle_j \bigr) \otimes \bigl( \bigotimes_{j > i} \frac{1}{\sqrt{\lvert H \rvert}} \sum_{h \in y} \lvert h \rangle_j \bigr) ,
\]
for $x,y \in G/H$, 
and the fusion rules are given by
\[
\mathcal T_{x,y} \otimes \mathcal T_{z,w} = \delta_{y,z} \mathcal T_{x,w} , \quad x,y,z,w \in G/H .
\]

\begin{figure}[H]
\centering
\begin{tikzpicture}[scale=0.8]
        \draw[thick](-4,0)--(5, 0);
        \draw[thick](0, 0.1)--(0, -0.1) node[above] at (0, 0.2) {$\lvert\psi_x\rangle$} node[below] at (0, -0.2) {$i$};
        \draw[thick](-1, 0.1)--(-1, -0.1)node[above] at (-1, 0.2) {$\lvert\psi_x\rangle$} node[below] at (-1, -0.2) {$i-1$};
        \draw[thick](-2, 0.1)--(-2, -0.1) node[above] at (-2, 0.2) {$\lvert\psi_x\rangle$} node[below] at (-2, -0.2) {$i-2$};
        \draw[thick](-3, 0.1)--(-3, -0.1) node[above] at (-3, 0.2) {$\lvert\psi_x\rangle$} node[below] at (-3, -0.2) {$\cdots$};
        \draw[thick](1, 0.1)--(1, -0.1) node[above] at (1, 0.2) {$\lvert\psi_y\rangle$} node[below] at (1, -0.2) {$i+1$};
        \draw[thick](2, 0.1)--(2, -0.1)node[above] at (2, 0.2) {$\lvert\psi_y\rangle$} node[below] at (2, -0.2) {$i+2$};
        \draw[thick](3, 0.1)--(3, -0.1)node[above] at (3, 0.2) {$\lvert\psi_y\rangle$} node[below] at (3, -0.2) {$\cdots$};
        \draw[thick](4, 0.1)--(4, -0.1) node[above] at (4, 0.2) {$\lvert\psi_y\rangle$} node[below] at (4, -0.2) {$\cdots$};
        \node at(5, 0.2){$\cdots$};
        \node at(-4, 0.2){$\cdots$};
\end{tikzpicture}
\caption{A topological sector of states without symmetry $\mathcal T_{x,y}$ can be generated by the state $\lvert\psi_{x,y,i}\rangle$, which can be interpreted as a 0+1D domain wall on 1+1D lattice model.}
\end{figure}

The multi-fusion category of topological sectors of states is equivalent to $\mathrm{Mat}_n(\vect)$, or the functor category $\Fun(\vect_{G/H},\vect_{G/H})$.

% \begin{rem}
% The fact that the 1d bulk is anomaly-free can be seen from the Drinfeld center of $\Fun(\vect_{G/H},\vect_{G/H})$, i.e., $\FZ_1(\Fun(\vect_{G/H},\vect_{G/H})) \simeq \vect$.
% \end{rem}

Now we view \eqref{eq:Hamiltonian} as a system with $G$-symmetry. The $G$-action on $\Fun(\vect_{G/H},\vect_{G/H}) \simeq \mathrm{Mat}_n(\vect)$ is induced by the $G$-action on states:
\[
U(g) \lvert \psi_{x,y,i} \rangle = \lvert \psi_{gx,gy,i} \rangle , \quad g \in G , x,y \in G/H .
\]
In other words, we have
\[
g(\mathcal T_{x,y}) = \mathcal T_{gx,gy} .
\]
The fusion category of $G$-symmetric topological sectors of states is equivalent to the equivariantization $\Fun(\vect_{G/H},\vect_{G/H})^G$. Also we have the following mathematical result:

\begin{prop} \label{prop:bimodule=equivariantization}
Let $G$ be a finite group and $H \subseteq G$ be a subgroup. Then the fusion category ${}_{F_H} \rep(G)_{F_H}$ is monoidally equivalent to the equivariantization $\Fun(\vect_{G/H},\vect_{G/H})^G$.
\end{prop}

%% arXiv version
Thus the fusion category of $G$-symmetric topological sectors of states is also equivalent to the fusion category ${}_{F_H} \rep(G)_{F_H}$ of $(F_H,F_H)$-bimodules in $\rep(G)$. As in the symmetry completely broken case, we translate the mathematical proof of Proposition \ref{prop:bimodule=equivariantization} in appendix \ref{appendix:equivariantization_proof} to physical language by explicitly finding the $G$-symmetric topological sectors of states.
%% SB version
%Thus the fusion category of $G$-symmetric topological sectors of states is also equivalent to the fusion category ${}_{F_H} \rep(G)_{F_H}$ of $(F_H,F_H)$-bimodules in $\rep(G)$. As in the symmetry completely broken case, we translate the mathematical proof of Proposition \ref{prop:bimodule=equivariantization} in appendix B.2 in Supplementary material to physical language by explicitly finding the $G$-symmetric topological sectors of states.

First note that $M_g^j\lvert \psi_{x,x,j} \rangle$ generates the topological sector of states $\mathcal{T}_{gx,x}$ without symmetry, for each $g \in G$, $x\in G/H$.
Then for each $\rho \in \hat G$, the following states generate a simple $G$-symmetric topological sector of states, denoted by $\mathcal S_{g,\rho}$:
\[
\bigl\{E_\rho^i M_g^j\lvert \psi_{x,x,j} \rangle \bigr\}_{x \in G/H}
\]

\begin{figure}[H]
\centering
\begin{tikzpicture}[scale=0.8]
        \draw[thick](-4,0)--(5, 0);
        \draw[thick](0, 0.1)--(0, -0.1) node[above] at (0, 0.2) {$\cdots$} node[below] at (0, -0.2) {$\cdots$};
        \draw[thick](-1.1, 0.1)--(-1.1, -0.1)node[above] at (-1.1, 0.2) {$Z_{\rho}^i\lvert\psi_{gx}\rangle$} node[below] at (-1.1, -0.2) {$i$};
        \draw[thick](-2.2, 0.1)--(-2.2, -0.1) node[above] at (-2.2, 0.2) {$\lvert\psi_{gx}\rangle$} node[below] at (-2.2, -0.2) {$i-1$};
        \draw[thick](-3.3, 0.1)--(-3.3, -0.1) node[above] at (-3.3, 0.2) {$\lvert\psi_{gx}\rangle$} node[below] at (-3.3, -0.2) {$\cdots$};
        \draw[thick](1.1, 0.1)--(1.1, -0.1) node[above] at (1.1, 0.2) {$\lvert\psi_{gx}\rangle$} node[below] at (1.1, -0.2) {$j$};
        \draw[thick](2.2, 0.1)--(2.2, -0.1)node[above] at (2.2, 0.2) {$\lvert\psi_{x}\rangle$} node[below] at (2.2, -0.2) {$j+1$};
        \draw[thick](3.3, 0.1)--(3.3, -0.1)node[above] at (3.3, 0.2) {$\lvert\psi_x\rangle$} node[below] at (3.3, -0.2) {$j+2$};
        \draw[thick](4.4, 0.1)--(4.4, -0.1) node[above] at (4.4, 0.2) {$\lvert\psi_x\rangle$} node[below] at (4.4, -0.2) {$\cdots$};
        \node at(5.5, 0.2){$\cdots$};
        \node at(-4.4, 0.2){$\cdots$};
\end{tikzpicture}
\caption{When symmetry $G$ is imposed on the system, the global symmetry $U(g)$ does not generate a new symmetric sector, e.g. $\mathcal T_{gx,gy}$ and $\mathcal T_{x,y}$ both belong to the $G$-symmetric topological sector of states.
    Futhermore, the $G$-action would also be distinguished by the action of $E_\rho^i$, which leads to another structure $\rho$ on the symmetric topological sector.}
\end{figure}

Or we can write a $G$-symmetric topological sector of states as
\[
\mathcal{S}_{g,\rho}=(\bigoplus_{x\in G/H}\mathcal{T}_{gx,x},\rho)
\]

But there are some redundancy in counting of $\mathcal{S}_{g,\rho}$ for each $g \in G$ and $\rho \in \hat G$.
Note that
\bnu[(1)]
\item for $g,g' \in G$ satisfying $g^{-1} g' \in H$, we have $ M_g^j\lvert \psi_{x,x,j} \rangle =  M_{g'}^j\lvert \psi_{x,x,j} \rangle$;
\item for $\rho,\rho' \in \hat G$ satisfying $\rho^{-1} \rho' \in \widehat{G/H}$, the states $E_\rho^i \lvert \psi_{x,x,j} \rangle$ and $E_{\rho'}^{i} \lvert \psi_{x,x,j} \rangle$ are differed by a nonzero factor, thus correspond to the same quantum state.
\enu
Therefore, $\mathcal S_{g,\rho}$ only depends on the equivalence classes (cosets) $[g] \in G/H$ and $[\rho] \in \hat G / \widehat{G/H} \simeq \hat H$. So we denote these $G$-symmetric topological sectors of state by $\{\mathcal S_{[g],[\rho]}\}_{[g] \in G/H , [\rho] \in \hat H}$. 
They form distinguished simple objects of the underlying category.

The fusion products of topological sectors of states are given by
\[
\mathcal S_{[g], [\rho]} \otimes \mathcal S_{[h], [\sigma]} = \mathcal S_{[gh], [\rho\sigma]} , \quad g,h \in G , \rho,\sigma \in \hat G .
\]
Hence the fusion category of $G$-symmetric topological sectors of states is equivalent to ${}_{F_H} \rep(G)_{F_H}$.

\smallskip
Now we have the topological sectors of operators $\FZ_1(\rep(G))$ and the topological sectors of states ${}_{F_H} \rep(G)_{F_H}$. 
It is not hard to see that the operator $M_g^{i} E_\rho^{j}$ maps the sector $\mathcal S_{[h],[\sigma]}$ to $\mathcal S_{[gh],[\rho\sigma]}$ for all $g,h \in G$ and $\rho,\sigma \in \hat G$.
The action of $\FZ_1(\rep(G))$ on ${}_{F_H} \rep(G)_{F_H}$ on the macroscopic level is given by 
\[
\mathcal{O}_{(k,\lambda)}\odot \mathcal{S}_{[g],[\rho]}=\mathcal{S}_{[kg],[\lambda\rho]}
\]

Thus the hom spaces between $G$-symmetric topological sectors of states are given by
\[
\Hom(\mathcal S_{[g], [\rho]},\mathcal S_{[h], [\sigma]}) = \bigoplus_{\substack{k \in [hg^{-1}] \\ \lambda \in [\sigma \rho^{-1}]}} \mathcal{O}_{(k,\lambda)} .
\]
%% arXiv version
By comparing with example \ref{expl:internal_hom_monoidal_module_enriched} we see that this enriched category $\bc[\FZ_1(\rep(G))]{{}_{F_H} \rep(G)_{F_H}}$ is precisely the one obtained from the canonical construction of the obvious action of $\FZ_1(\rep(G))$ on ${}_{F_H} \rep(G)_{F_H}$.
%% arXiv version
%By comparing with example C.5 in Supplementary material we see that this enriched category $\bc[\FZ_1(\rep(G))]{{}_{F_H} \rep(G)_{F_H}}$ is precisely the one obtained from the canonical construction of the obvious action of $\FZ_1(\rep(G))$ on ${}_{F_H} \rep(G)_{F_H}$.

%We see the fusion products of topological sectors of states meet with the fusion structure of ${}_{F_H} \rep(G)_{F_H} $.
%
%\begin{rem}
%    Indeed, $\rep(G)$ and ${\rm Vec}_{G}$ are just special cases of ${}_{F_H} \rep(G)_{F_H}$, with $\rep(G)=\rep(G)_{F_G|F_G}$ and
%    ${\rm Vec}_{G}=\rep(G)_{F_{\{e\}}|F_{\{e\}}}$.
%\end{rem}

Hence we have proved the following through concrete models:
\begin{pthm}
The macroscopic observables of the $1+1$D gapped phase with an onsite abelian symmetry $G$ form enriched fusion category $\bc[\FZ_1(\rep(G))]{{}_{F_H} \rep(G)_{F_H}}$ if the symmetry spontaneously breaks to a subgroup $H \subseteq G$.
\end{pthm}

\begin{rem}
The categorical description $\bc[\FZ_1(\rep(G))]{{}_{F_H} \rep(G)_{F_H}}$ is anomaly-free in the sense that the Drinfeld center of this category is trivial:
$$
\FZ_1 \bigl( \bc[\FZ_1(\rep(G))]{{}_{F_H} \rep(G)_{F_H}} \bigr) \simeq \vect .
$$
See \cite[section 5.3]{KYZZ21} for the definition of the Drinfeld center ($E_1$-center) of an enriched fusion category and the proof of the above equivalence.
\end{rem}

\begin{rem} \label{rem:H_symmetric_Lagrangian_algebra}
    In the view of holographic duality, we may say the operators are `partially condensed', with $E_\rho^i$ acting trivially on the trivial $G$-symmetric topological sector of states $\mathcal S_{[e],[1]}$ for each $\rho \in \widehat {G/H}$, and $M_g^{i}$ acting trivially on $\mathcal S_{[e],[1]}$ for each $g \in H$. More precisely, these operators form the ground state Lagrangian algebra
    \[
    A(H) \coloneqq \Hom(\mathcal S_{[e],[1]},\mathcal S_{[e],[1]}) = \bigoplus_{\substack{g \in H \\ \rho \in \widehat{G/H}}} \mathcal O_{(g,\rho)} ,
    \]
    in a 2d quantum double model $\FZ_1(\rep(G))$.
    
    In this way the underlying bimodule category ${}_{F_H} \rep(G)_{F_H}$ corresponds to the 1d gapped boundary $\FZ_1(\rep(G))_{A(H)}$ of right $A(H)$-modules in $\FZ_1(\rep(G))$.
Or to say, a 1+1D gapped phase with an onsite symmetry $G$ can be equivalently written as $\bc[\FZ_1(\rep(G))]{\FZ_1(\rep(G))_{A(H)}}$ if the symmetry spontaneously breaks to a subgroup $H \subseteq G$.
    \end{rem}

By now we have finished the proof of our main theorem \ref{pthm:symmetry_breaking_phase} in concrete models.
  
% \input{boundary}
% !TeX root = main.tex
% !TeX program = pdfLaTeX

\section{0+1D boundaries and domain walls with abelian onsite symmetry} \label{section:boundaries}

In this section we discuss the 0+1D open boundaries with an onsite abelian symmetry $G$ explicitly broken and domain walls of those 1+1D bosonic gapped phases.

% % Mathematically, the pair (X, x) is a kind of E0-algebra.
% Therefore, the category C of particle-like topological
% defects can also be viewed as that of ‘0d boundary conditions’ at a given site i, which is denoted by Ci
% .
% This point of view is sometimes very useful. We can specify a given ‘0d boundary condition’ x at the
% site i by a pair (Ci
% , x)
% The category of boundary conditions.

% Recall that 0+1D topological defects and instantons in an nd topological order form a category for n ≥ 1
% (see Theoremph 3.2.12). For a 0d topological order x, we cannot talk about 0+1D topological defects of
% x, but 0+1D topological defects (or equivalently, 0d topological orders) living on the world line of x.
% Then Theoremph 3.2.12 still holds for a 0d topological order x in the following sense:

\subsection{Boundaries} \label{sec_boundary}

In principle, for a $0$d topological phase, it is not precise to talk about particle-like excitations, since each excitation can be viewed as a $0$d phase itself, but there is still a 0+1D categorical structure given by the `category of boundary conditions'.

More precisely, the category of boundary conditions is defined as follows:
\bit
\item The objects are 0d boundary phases, which can be identified with the topological sector of states generated by the ground state.
\item The morphisms are topological sectors of operators between objects, living on the worldline of boundary phases.
\eit

In this section we consider 0+1D open boundary that explicitly breaks the bulk symmetry $G$ to its subgroup $H_b \subseteq G$.
That is, 
% each Hamiltonian specifies a boundary condition $x$, and 
all $H_b$-symmetric 0d boundary phases form a category $\CX$.
% That is, each simple object $x$ specifies a boundary condition
% We sometimes write a $0$d boundary $x \in \CX$ as a pair $(\CX, x)$ to emphasize the category $\CX$ of 0d boundaries \cite{KZ22a}.
The topological sectors of operators can be fused along the world line. So they form a fusion category $\CA$. We call this fusion category $\CA$ the categorical symmetry of the boundaries. Therefore, the category of boundary conditions is an $\CA$-enriched category $\bc[\CA]{\CX}$.

%\begin{rem}
%    Another way to describe the `category of boundary conditions' $\CX$ is that, all $0$d topological phases living on the world line of $x$, together with instantons between them, form a category $\CX$. 
%In principle, we can just write down the canonical boundary Hamiltonian that preserves the symmetry we want, this would give us a distinguished element $x \in \CX$. Then we can act the non-local operators on it to get other boundary conditions.
%\end{rem}

In practice, we can just write down a boundary Hamiltonian that preserves the symmetry we want. This would give us a distinguished object $x \in \bc[\CA]{\CX}$. Then we can act the non-local operators on it to get other boundary conditions.

\subsubsection{0+1D open boundaries of 1+1D trivial SPT bulk} \label{sec_boundary_trivial_SPT}

In order to see the categorical structures of the boundaries phases in lattice models,
we again start from the most obvious case, where we choose the 1+1D bulk to be the trivial SPT order, whose topological skeleton is $\bc[\FZ_1(\rep(G))]{\rep(G)}$ as $\rm theorem^{ph}$ \ref{thm_SPT} shows. We pick the boundary phase on the left side, i.e. $\mathcal H_{\text{tot}} = \bigotimes_{i\geq 0}\mathcal{H}_i$ (the right side is similar) and analyze its boundary conditions:

\vspace{1em}

\textbf{$G$-symmetric boundary conditions}: First we consider the boundary conditions that preserve $G$-symmetry. For example, the following boundary Hamiltonian preserves the $G$-symmetry on the boundary:
   \be\label{cano_bdy_preserving}
\mathcal{H} = \sum_{i\geq 0} (1-X^i_G) .
   \ee
Its ground state is simply $\bigotimes_{i \geq 0} \frac{1}{\sqrt{\lvert G \rvert}} \sum_{g \in G} \lvert g \rangle_i$. This Hamiltonian is the most trivial case that preserves $G$-symmetry on the boundary (where we can obtain it by taking $\rho$ to be the trivial representation $1$ in equation \eqref{$G$-symmetric boundary}), we may call this Hamiltonian the canonical choice. Adding a $G$-symmetric operator to this Hamiltonian also gives a $G$-symmetric boundary condition. For example, for every $\rho \in \hat G$ the following Hamiltonian realizes a $G$-symmetric boundary condition:
\be\label{$G$-symmetric boundary}
\mathcal H = \sum_{i>0} (1 - X_G^i) + \bigl( 1 - \frac{1}{\lvert G \rvert} \sum_{g \in G} \rho(g) L_g^0 \bigr) .
\ee
Its ground state is $\frac{1}{\sqrt{\lvert G \rvert}} \sum_{g \in G} \rho(g) \lvert g \rangle_0 \otimes \bigl( \bigotimes_{i > 0} \frac{1}{\sqrt{\lvert G \rvert}} \sum_{g \in G} \lvert g \rangle_i \bigr)$, and the corresponding boundary phase is denoted by $\mathcal S_\rho$. Then the underlying category of the enriched category of $G$-symmetric boundary conditions matches with $\rep(G)$.

\begin{rem}
We can also apply the equivariantization technique introduced in section \ref{section:equivariantization} in 0+1D case. Since the 1d SPT bulk without symmetry is $\vect$ \ref{SPT_rem_without_sym}, its open boundary when view as a 0d topological order can only be $\vect$. After imposing symmetry $G$ on the boundary, the topological sectors of states form $\vect^{G} \simeq \rep(G)$ rightfully so. 
\end{rem}

When acting $E_{\sigma}^{0}$ on the ground state, 
% $\frac{1}{\sqrt{\lvert G \rvert}} \sum_{g \in G} \rho(g) \lvert g \rangle_0$ 
site $0$ becomes $\frac{1}{\sqrt{\lvert G \rvert}} \sum_{g \in G} \sigma(g)\rho(g) \lvert g \rangle_0 \in \mathcal{S}_{\sigma\rho}$.
Or to say, $E_{\sigma}^0$ maps sector $\mathcal{S}_{\rho}$ to sector $\mathcal{S}_{\sigma\rho}$. 
Note that for every $g \in G$, the operator $M^i_g = \prod_{k \leq i} L_g^k$ is the product of symmetric local operators. So it should be viewed as `local' on $G$-symmetric boundaries.
% So the non-local operators just has $E^i_{\rho}$ part left. 
Therefore, the topological sectors of operators mapping one $G$-symmetric boundary conditions to another are invariant under the action of $M^0_g$. 
% This invariance means that the operators $M^0_g$ could be viewed as `local' on $G$-symmetric boundaries.
Mathematically, this means that the topological sectors of operators are the modules over the algebra $A(G) = \bigoplus_{g\in G} \mathcal{O}_{(g,1)} \in \FZ_1(\rep(G))$. Therefore, the categorical symmetry of symmetry preserving boundary conditions is equivalent to the category $\FZ_1(\rep(G))_{A(G)}$. More explicitly,
\begin{align*}
    \Hom(\mathcal{S_{\rho}}, \mathcal{S_{\sigma}})&= \bigoplus_{g\in G} \mathcal{O}_{(g, \sigma \rho^{-1})}\\
    &=  \mathcal{O}_{(e, \sigma \rho^{-1})} \otimes (\bigoplus_{g\in G} \mathcal{O}_{(g,1)}) \\
    &= \mathcal{O}_{(e, \sigma \rho^{-1})} \otimes A(G)\\
    &\mapsto \sigma \rho^{-1} \in \rep(G),
\end{align*}
In other words, the 1+1D bulk topological sector of operators $\FZ_1(\rep(G))$ is condensed to be a 0+1D topological sector of operators $\FZ_1(\rep(G))_{A(G)} \simeq \rep(G)$.
%, because the hom space (topological sector of operators) between $G$-symmetric boundary conditions is invariant under these operators. , and the hom spaces between $G$-symmetric boundary conditions are (right) $A(G)$-modules as explained in \eqref{right_A_module}.  

And the action of categorical symmetry $\rep(G)$ on the underlying category $\rep(G)$ is given by 
\[
    \sigma\rho^{-1}\odot \mathcal{S}_{\rho}=S_{\sigma}
\]
So the topological skeleton of this $G$-symmetric 0+1D boundary is enriched category 
\[\bc[\rep(G)]{\rep(G)}.\]

% \mynote{
\begin{rem}
    $\bc[\rep(G)]{\rep(G)}$ 
    % can not only be understood as $\bc[\FZ_1(\rep(G))_{A(G)}]{\rep(G)}$, but 
    can also be understood as $\bc[\rep(G)]{\rep(G)_{F_G}}$ through 1d condensation, in which $F_G$ is just the tensor unit $1 \in \rep(G)$.
    The 0+1D boundary description $\bc[\rep(G)]{\rep(G)_{F_G}}$ is parallel to the 1+1D bulk description $\bc[\FZ_1(\rep(G))]{\FZ_1(\rep(G))_{A(G)}}$ in the sense that the full center of $F_G$ is $A(G)$.
    % $\FZ_0(\bc[\rep(G)]{\rep(G)_{F_G}}) \simeq \bc[\FZ_1(\rep(G))]{\FZ_1(\rep(G))_{A(G)}}$.
\end{rem}
% }

\vspace{1em}

\textbf{Symmetry completely broken boundary conditions}: Then we consider the boundary conditions that do not preserve any symmetry, i.e. $H_b=\{e\}$. For example, we can choose the boundary Hamiltonian to be
%    \be \label{bdy_completely}
\[    \mathcal{H}  = \sum_{i>0} (1-X^i_G) + (1-Z^{0}_{\{e\}}), \]
%    \ee
where $Z^{0}_{\{e\}} := \frac{1}{\lvert G \rvert} \sum_{\rho\in \hat{G}} Z^0_{\rho}$. Its ground state is $\lvert e \rangle_0 \otimes \bigl(\bigotimes_{i > 0} \frac{1}{\sqrt{\lvert G \rvert}} \sum_{g \in G} \lvert g \rangle_i \bigr)$, which completely breaks the $G$-symmetry on site $0$. 

For each $g\in G$, $M_g^{0}$ is a local operator, but $U(g^{-1})M_g^0$ is a non-local operator and defines a non-trivial topological sector of operators because the $G$-symmetry is broken on the boundary.
Since $E_{\rho}^0$ acts on $\lvert e \rangle_0$ trivially, there is only one simple boundary condition that completely breaks the symmetry, denoted by $\mathcal S$. 
Therefore, the underlying category of the symmetry completely broken boundary conditions matches with $\vect$.

\begin{rem}
    We can also check that the equivariantization technique holds.
    Unlike in section \ref{sec:SCB} where the 1+1D bulk is spontaneously broken, here the 0+1D open boundary is explicitly broken. Or to say, there is no symmetry on this boundary.
    So we have $\vect^{\{e\}} \simeq \vect$. 
    \end{rem}
%More generally, for every $g \in G$ the following Hamiltonian also breaks the $G$-symmetry completely:
%\[
%\mathcal{H}  = \sum_{i>0} (1-X^i_G) + \bigl( 1 - \frac{1}{\lvert G \rvert} \sum_{\rho \in \hat G} \rho(g)^{-1} Z_\rho^0 \bigr) .
%\]
%Its ground state is given by $\lvert g \rangle_0 \otimes \bigl(\bigotimes_{i > 0} \frac{1}{\sqrt{\lvert G \rvert}} \sum_{g \in G} \lvert g \rangle_i \bigr)$, and the corresponding boundary condition is denoted by $\mathcal S_g$.

Note that $Z^0_{\rho}$ is a local operator.
On symmetry completely broken boundaries, the operator $E^0_{\rho}$ for every $\rho \in \hat G$ is also `local' since $Z^0_{\rho}E^0_{\rho}$ is an identity operator.
These $E_{\rho}^0$ operators form the Lagrangian algebra $A(\{e\}) = \bigoplus_{\rho \in \hat{G}} \mathcal{O}_{(e, \rho)}$, and the hom spaces between symmetry completely broken boundary conditions are (right) $A(\{e\})$-modules. 
Therefore, the categorical symmetry of symmetry completely broken boundary conditions is equivalent to the category $\FZ_1(\rep(G))_{A(\{e\})}$. 
More explicitly,
% The structure of the enriched category $\bc[\FZ_1(\rep(G))_{A(\{e\})}]{\vect}$ of symmetry completely broken boundary conditions is determined by
\begin{align*}
\Hom(\mathcal{S}, \mathcal{S})&= \bigoplus_{\substack{g\in G \\ \rho \in \hat{G}}} \mathcal{O}_{(g, \rho)}\\
&= \bigl( \bigoplus_{g\in G} \mathcal{O}_{(g,1)} \bigr) \otimes \bigl( \bigoplus_{\rho \in \hat{G}} \mathcal{O}_{(e, \rho)} \bigr) \\
&= \bigl( \bigoplus_{g\in G} \mathcal{O}_{(g,1)} \bigr) \otimes A(\{e\}) \\
&\mapsto \bigoplus_{g\in G}\mathbb{C}_g \in \vect_{G}.
\end{align*}  
% Moreover, there is a monoidal equivalence
%     \begin{align*}
%         \vect_G & \simeq \FZ_1(\rep(G))_{A(\{e\})} \\
%         \mathbb{C}_g & \mapsto \mathcal{O}_{(g,1)} \otimes A(\{e\}) ,
%     \end{align*}
where $\mathbb{C}_g$ is the complex number field with $g$-grading. 
In other words, the 1+1D bulk topological sector of operators $\FZ_1(\rep(G))$ is condensed to be a 0+1D topological sector of operators $\FZ_1(\rep(G))_{A(\{e\})} \simeq \vect_{G}$.

And the action of categorical symmetry $\vect_G$ on the underlying category $\vect$ is given by 
\[
    \mathbb{C}_g \odot \mathcal{S} = \mathcal{S}
\]

So the topological skeleton of this $G$-symmetry completely broken 0+1D boundary is enriched category 
$$\bc[\vect_{G}]{\vect}$$.

% \begin{rem}
% There are many different boundary Hamiltonians realizing the same boundary condition in LWLL. For example, we can just take the symmetric Hamiltonian
% \[    \mathcal{H} = \sum_{i\geq 0} (1-X^i_G)  \]
% and view it as having no symmetry. Then it realizes a symmetry completely broken boundary condition, which is the same as the above one because their ground state can be connected to a local operator.
% \end{rem}

\vspace{1em}

\textbf{Symmetry partially broken boundary conditions}: Fix a subgroup $H_b \subseteq G$. Now we consider the boundary conditions that only preserve the subgroup $H_b$. 
The canonical choice of Hamiltonian that break the ground
state symmetry on site $0$ from $G$ to $H_b$ can be written as
\[
\mathcal{H} = \sum_{i>0} (1-X^i_G) + (1-Z^{0}_{H_b})+(1-X^{0}_{H_b}), 
\] 
where $Z^{0}_{H_b} = \frac{\lvert H_b \rvert}{\lvert G \rvert} \sum_{\rho\in \widehat{G/{H_b}}} Z^0_{\rho}$.
The ground state is then given by
\[
\frac{1}{\sqrt{\lvert H_b \rvert}} \sum_{g \in H_b} \lvert g \rangle_0 \otimes (\bigotimes_{i > 0} \frac{1}{\sqrt{\lvert G \rvert}} \sum_{g \in G} \lvert g \rangle_i) .
\]
For every $\sigma \in \hat G$, we can generate the upper Hamiltonian to be the Hamiltonian which realizes each $H_b$-symmetric boundary condition, for example:
%    \be \label{SPT_partially}
\begin{align*}  
\mathcal{H} = \sum_{i>0} (1-X^i_G) + ( 1 - Z^{0}_{H_b}) \\
+ \bigl(1 - \frac{1}{\lvert H_b \rvert} \sum_{h \in H_b} \sigma(h) L_h^0 \bigr) . 
\end{align*}
%    \ee 
Its ground state is given by $\frac{1}{\sqrt{\lvert H_b \rvert}} \sum_{g \in H_b} \sigma(g) \lvert g \rangle_0 \otimes \bigl( \bigotimes_{i > 0} \frac{1}{\sqrt{\lvert G \rvert}} \sum_{g \in G} \lvert g \rangle_i \bigr)$. Note that it only depends on the equivalence class (coset) $[\sigma] \in \hat G / \widehat{G/H_b} \simeq \hat H_b$, so we denote the corresponding $H_b$-symmetric boundary condition by $\mathcal S_{[\sigma]}$. Therefore, the underlying category of the enriched category of $H_b$-symmetric boundary conditions matches with $\rep(H_b)$.

\begin{rem}
    The underlying category
    %  of the enriched category $\bc[{}_{F_H} \rep(G)_{F_H}]{\rep(H_b)}$ of $H_b$-symmetric boundary conditions 
     can also be obtained from equivariantization, in which 
    %  By ignoring the symmetry, the 1d bulk is the trivial topological order whose topological skeleton is $\vect$. Thus its boundary conditions also form the category $\vect$. 
    the category of $H_b$-symmetric boundary conditions should be the equivariantization $\vect^{H_b} \simeq \rep(H_b)$.
\end{rem}

Similarly to above two cases, 
% the categorical symmetry of boundary conditions only depends on the symmetry imposed on the boundary. 
for $H_b$-symmetric boundary conditions, clearly the operators $M_g^0$ for all $g \in H_b$ do not change the boundary condition. Dually, since these boundary conditions can be viewed as $H_b$-representations, the operators $E_\rho^0$ for $\rho \in \widehat{G/H_b}$ also do not change the boundary conditions. 
Thus the operators $M^0_g$ and $E^0_{\rho}$ for $g \in H_b$ and $\rho \in \widehat {G/H_b}$ should be viewed as partially localized (condensed) because each hom space between two boundary conditions contains them. 
These `local operators' form a Lagrangian algebra (see Remark \ref{rem:H_symmetric_Lagrangian_algebra})
\[
A(H_b) = \bigoplus_{\substack{g \in H_b \\ \rho \in \widehat{G/H_b}}} \mathcal{O}_{(g, \rho)} ,
\]
and the hom spaces between symmetry completely broken boundary conditions are $A(H_b)$-modules. Therefore, 
the categorical symmetry of symmetry partially broken boundary conditions is equivalent to the category $\FZ_1(\rep(G))_{A(H_b)}$. More explicitly,
% The structure of the enriched category $\bc[\FZ_1(\rep(G))_{A(H_b)}]{\rep(H_b)}$ of $H_b$-symmetric boundary conditions is determined by
%          \[
%          \Hom(\mathcal{S_{[\rho]}}, \mathcal{S_{[\sigma]}})= 
%          \bigoplus_{\substack{g \in H_b \\ \phi \in \widehat{G/H_b}}} \mathcal{O}_{(g, \phi \cdot \sigma \rho^{-1})} =  (\bigoplus_{\substack{g \in H_b \\ \phi \in \widehat{G/H_b}}} \mathcal{O}_{(g, \phi)}) \otimes (\bigoplus_{k\in G/H_b} \mathcal{O}_{(k, \sigma \rho^{-1})}) = A(H_b) \otimes \bigoplus_{k\in G/H_b} \mathcal{O}_{(k, \sigma \rho^{-1})}, 
%          \]
\begin{align*}
&\Hom(\mathcal{S_{[\rho]}}, \mathcal{S_{[\sigma]}}) = \bigoplus_{\substack{g \in G \\ \phi \in \widehat{G/H_b}}} \mathcal{O}_{(g, \phi \cdot \sigma \rho^{-1})} \\
=&\bigl( \bigoplus_{[k] \in G/H_b} \mathcal{O}_{(k, \sigma \rho^{-1})} \bigr) \otimes \bigl( \bigoplus_{\substack{g \in H_b \\ \phi \in \widehat{G/H_b}}} \mathcal{O}_{(g, \phi)} \bigr)  \\ 
=&\bigl( \bigoplus_{[k] \in G/H_b} \mathcal{O}_{(k, \sigma \rho^{-1})} \bigr) \otimes A(H_b)\\
\mapsto & \bigoplus_{[k] \in G/H_b} \mathcal M_{([k],[\sigma\rho^{-1}])} \in {}_{F_{H_b}} \rep(G)_{F_{H_b}}.
\end{align*}
    %   L(\bigoplus_{h\in {G/H_b}} \mathcal{O}_{(h, \sigma\rho^{-1})}) = \bigoplus_{[h]\in {G/H_b}} S_{[h], [\sigma\rho^{-1}]}
where $[\rho],[\sigma] \in \hat{H_b}$ and $[k] \in G/H_b$ (see Remark \ref{rem:groupoid_abelian}).
In other words, the 1+1D bulk topological sector of operators $\FZ_1(\rep(G))$ is condensed to be a 0+1D topological sector of operators $\FZ_1(\rep(G))_{A(H_b)} \simeq {}_{F_{H_b}} \rep(G)_{F_{H_b}}$.
% Moreover, there is a monoidal equivalence
% \begin{align*}
% {}_{F_H} \rep(G)_{F_H} & \simeq \FZ_1(\rep(G))_{A(H_b)} \\
% \mathcal M_{([g],[\rho])} & \mapsto \mathcal{O}_{(g,\rho)} \otimes A(H_b) ,
% \end{align*}
%% arXiv version

%% SB version
%where $\mathcal M_{([g],[\rho])}$ is the simple object in ${}_{F_H} \rep(G)_{F_H}$ corresponding to $[g] \in G/H_b$ and $[\rho] \in \hat H_b$ (see Remark B.8 in Supplementary material). 

And the action of categorical symmetry ${}_{F_{H_b}} \rep(G)_{F_{H_b}}$ on the underlying category $\rep(H_b)$ is given by 
\[
    \mathcal M_{([k],[\sigma\rho^{-1}])}\odot \mathcal{S}_{[\rho]} = \mathcal{S}_{[\sigma]}
\]

So the topological skeleton of this $G$-symmetry completely broken 0+1D boundary is enriched category 
$$\bc[{}_{F_{H_b}} \rep(G)_{F_{H_b}}]{\rep(H_b)}$$.

\begin{rem}
% By comparing with example \ref{expl:rep_H_enriched_in_bimodule}, we see that this enriched category $\bc[{}_{F_{H_b}} \rep(G)_{F_{H_b}}]{\rep(H_b)}$ is the one obtained from the canonical construction of the obvious action of ${}_{F_{H_b}} \rep(G)_{F_{H_b}}$ on $\rep(G)_{F_{H_b}} \simeq \rep(H_b)$.
The invaraint operators acting on the 0+1D boundary sectors of states are given by $\Hom(\mathcal{S}_{[1]}, \mathcal{S}_{[1]})$, which form a 1d condensable algebra in ${}_{F_{H_b}} \rep(G)_{F_{H_b}}$.
Its full center is just the ground state algebra $A(G)\coloneqq \Hom(\mathcal S_1,\mathcal S_1) \in \FZ_1(\rep(G))$ of the symmetry preserving 1+1D bulk phase.
    % \mynote{Merge it into 1d condensation?}
    % \mynote{See Chinese NOTE}
    % 可以注意到这三个例子，边界上其实都是condense了G分次的东西。这是由于bulk都是condense了M。所以在bdy上是平行的？
    This relation corresponds to open-closed duality discussed in \ref{sec:GS_algebra}.
\end{rem}

We can also give a picturesque explanation of the above boundary phases through holographic duality (see figure \ref{fig:boundary}). In figure \ref{fig:boundary} (a), the $0$d `corner' described by $\rep(H_b)$ is the invertible domain wall between two $1$d boundaries of the quantum double $\FZ_1(\rep(G))$, which can be obtained through 1d condensation $\rep(G)_{F_{H_b}}$.
$L$ is the functor of 2d condensation which maps the topological sector of operators in $\FZ_1(\rep(G))$ to its boundary, as introduced in \eqref{right_A_module}. 

After topological Wick rotation, this 2d condensation process corresponds to the "shrink" of operator space from $\FZ_1(\rep(G))$ to $\FZ_1(\rep(G))_{A(H_b)}$. 
And the invertible $0$d domain wall $\rep(H_b)$ becomes the underlying category of the enriched category $\bc[{}_{F_{H_b}} \rep(G)_{F_{H_b}}]{\rep(H_b)}$ of boundary conditions,
% and the boundary phase should form the enriched category $\bc[{}_{F_{H_b}} \rep(G)_{F_{H_b}}]{\rep(H_b)}$, 
as illustrated in figure \ref{fig:boundary} (b). 

\begin{figure*}
\centering
\begin{tikzpicture}
            \filldraw[fill=gray!20, draw=white] (-7.5,0) rectangle (-3.5,2);
            \draw[very thick](-7.5,0)--(-3.5,0);
            \draw[very thick] (-7.5,0)--(-7.5,2);
            \filldraw[fill=white] (-7.6,-0.1) rectangle (-7.4,0.1);
            % \draw[very thick, latex-] (-5.5,0)--(-3.5,0);

            \node at(-5.5,-0.5){$\rep(G)$};
            \node at(-7.5,-0.5){$\rep(H_b)$};
            \node[]at(-5.5, -1){(a)};
            \filldraw[fill=black, draw=black] (-6.5,1.5) circle (0.05) node[right]{$\mathcal{O}_{g,\rho}$};
            \filldraw[fill=black, draw=black] (-7.5,1.5) circle (0.05) node[left]{$L(\mathcal{O}_{g, \rho})$};
            \draw[very thick, -latex](-6.5,1.5)--(-7.4,1.5);
            % \draw[very thick, -latex] (-7.5,2)--(-7.5,1);
            \node at(-8.6,1){${}_{F_{H_b}} \rep(G)_{F_{H_b}}$};
            \node at(-5.5,0.8){$\FZ_1(\rep(G))$};
            % \node at(-9.7,0){\quad};

\begin{scope}[xshift=1.5cm]
            \draw[ -latex](-3, 1)--(-2.5,1);
            \draw[very thick](0.4, 0)--(4.4,0);

            \filldraw[fill=white, draw=black] (0.3,-0.1) rectangle (0.5,0.1);
            \node at(0.3, 0.4){$\bc[{}_{F_{H_b}} \rep(G)_{F_{H_b}}]{\rep(H_b)}$};
            \node at(2.9, -0.4){$\bc[\FZ_1(\rep(G))]{\rep(G)}$};
            \node[]at(2.4, -1){(b)};
            % \node at(-1,0){\quad};
\end{scope}
\end{tikzpicture}
\caption{
    The holographic explanation of the boundary phase described by $\bc[{}_{F_{H_b}} \rep(G)_{F_{H_b}}]{\rep(H_b)}$. (a) depicts a $2$d topological order described by $\FZ_1(\rep(G))$ together with two $1$d gapped boundaries $\rep(G)$ and ${}_{F_{H_b}} \rep(G)_{F_{H_b}}$ and a 0d gapped domain wall $\rep(H_b)$, while
    (b) illustrates 0+1D $H_b$-symmetric boundaries (with symmetry $G$ in the 1+1D bulk explicitly breaks to the subgroup $H_b \subseteq G$) of the trivial $1$d SPT bulk. The boundary phase (b) can be obtained from (a) by applying the topological Wick rotation.
}
\label{fig:boundary}
\end{figure*}

\begin{rem}
The bulk-to-wall map $L$ acts as the \emph{background changing functor} \cite{KYZZ21}, which changes the categorical symmetry (the background category) from $\FZ_1(\rep(G))$ to ${}_{F_{H_b}} \rep(G)_{F_{H_b}}$.
\end{rem}

\begin{rem}
The intuition of topological Wick rotation suggests that all possible enriched category of boundary conditions of the trivial 1d $G$-SPT order are determined by indecomposable $\rep(G)$-modules, which are classified by pairs $(H_b,[\omega])$ where $H_b \subseteq G$ is a subgroup and $[\omega] \in \mathrm{H}^2(H_b,U(1))$ is a 2-cohomology classes \cite{Ost03}. More precisely, the indecomposable module corresponding to $(H_b,[\omega])$ is the category $\rep(H_b,\omega)$ of finite-dimensional projective $H_b$-representations twisted by $\omega$. Hence, for a subgroup $H_b \subseteq G$ and a 2-cohomology class $[\omega]$, there are $H_b$-symmetric boundary conditions that are `twisted' by $[\omega]$. Such twisted boundary conditions should form an enriched category $\bc[{}_{F_{H_b}} \rep(G)_{F_{H_b}}]{\rep(H_b,\omega)}$. In a concrete lattice model, such twisted boundary conditions can be realized by attaching a projective representation to the boundary. We discuss the example $G = \Zb_2 \times \Zb_2$ in Section \ref{sec_Z2Z2_SPT}.
% In this work we only consider the `untwisted' sector, i.e., $[\omega]$ is trivial.
\end{rem}

%\begin{rem}
%    Another way to describe the `category of boundary conditions' $\CX$ is that, all $0$d topological phases living on the world line of $x$, together with instantons between them, form a category $\CX$. 
%In principle, we can just write down the canonical boundary Hamiltonian that preserves the symmetry we want, this would give us a distinguished element $x \in \CX$. Then we can act the non-local operators on it to get other boundary conditions.
%\end{rem}

\subsubsection{0+1D open boundaries of 1+1D symmetry breaking bulks}\label{general_bdy}

Now we consider the general case, where the $1+1$D bulk $\bc[\FZ_1(\rep(G))]{{}_{F_{H}} \rep(G)_{F_{H}}}$ is chosen to spontaneously break the symmetry $G$ to a subgroup $H$ (see section \ref{section:general_cases}), and we choose its boundary conditions explicitly breaking to a subgroup $H_b \subseteq G$. 
We can pick the canonical Hamiltonian that realizes $H_b$ symmetry on its boundary:
\be\label{general_bdy_Hamiltonian}
\mathcal{H} \coloneqq \sum_{i>0} (1 - X_{H}^i) + \sum_{i>0} (1 - Z_{H}^{i,i+1})  + (1-Z^{0}_{H_b}) +(1-X^{0}_{H_b}), 
\ee
and the ground state is 
\[
\frac{1}{\sqrt{\lvert H_b \rvert}} \sum_{h \in H_b} \lvert h \rangle_0 \otimes \lvert \psi_x\rangle_{i>0}
% (\bigotimes_{i>0} \frac{1}{\sqrt{\lvert H \rvert}} \sum_{g \in {x}} \lvert g \rangle_i),
\] 
$\forall x \in G/H$.
% The boundary observables in this lattice model can be checked following a similar procedure as the bulk SPT case. For $g \in H_b$, $U(g)M_g^0$ is a local operator since $H_b$-symmetry is preserved on the boundary. 
%     Note that the ground state of site $0$ is invariant under these local operators, what determines the sectors of states are the bulk states near the boundary. It is enough  to focus on the state $\lvert \psi_x \rangle$.
Just like the above case, on $H_b$-symmetric boundaries, the bulk operators $M^0_g$ and $E^0_{\rho}$ are partially localized (condensed) on the boundary up to $g \in H_b$ and $\rho \in \widehat {G/H_b}$, therefore the background category should be the category $\FZ_1(\rep(G))_{A(H_b)}$ of $A(H_b)$ modules, which is equivalent to ${}_{F_{H_b}} \rep(G)_{F_{H_b}}$.

%
%     \hspace*{\fill}
%%%%%%%%%%%%%%%%%%%%%%%%%%%%%%%%%%%%%%%%%%%%%
As the equivariantization technique introduced in section \ref{section:G_symmetry} has nothing to do with dimensions, we can view the underlying categories of these boundary phases as a $0$d topological orders imposed with symmetries. Since $G$ explicitly breaks to the subgroup $H_b$ on the boundary, the category of symmetric boundary conditions should be an $H_b$-equivariantization of the category of boundary conditions without symmetry. 

Following a similar procedure in section \ref{section:enriched_category}, we first view equation \eqref{general_bdy_Hamiltonian} as systems without symmetry. 
By ignoring the symmetry, the topological skeleton of the fine tuned bulk is $\Fun(\vect_{G/H},\vect_{G/H})$, as illustrated in section \ref{section:general_cases}. 
There are $n \coloneqq \lvert G \rvert / \lvert H \rvert$ topological sectors of states $\{\mathcal T'_{x}\}_{x \in G/H}$ in the boundary. Then the category of boundary conditions (without symmetry) of the fine tuned bulk is given by the category $\vect_{G/H}$. If we choose the boundary on the left side (the right side is similar) of $\Fun(\vect_{G/H}, \vect_{G/H})$, we have the following fusion rules, 
\begin{equation}\label{eq:1d_action_on_0d}
\mathcal T'_{x} \otimes \mathcal T_{y,z}  = \delta_{x,y} \mathcal T'_{z},  \qquad \forall x,y,z \in G/H.
\end{equation}
which endows the category $\vect_{G/H}$
with a structure of right $\Fun(\vect_{G/H}, \vect_{G/H})$-module.\footnote{Indeed, if we choose the boundary on the left side of the bulk, the category of boundary conditions should be the opposite category $\vect_{G/H}^\op$. As a category it is equivalent to $\vect_{G/H}$, so we use $\vect_{G/H}$ in order to simplify the notation.} See the bottom left figure.
% and the category of its $0$d boundary is $\vect_{G/H}$. 

\begin{figure}[H]
\centering
\begin{tikzpicture}[scale=0.8]
    \node[]at(-3.6, 1.3){\scriptsize 1d topological order};
    \node[]at(1.5, 1.3){\scriptsize Imposing symmetry};
    \draw[very thick](-5,0)--(-1,0);
    \node[scale=0.8]at(-2.5,0.3){$\mathrm{Fun}(\vect_{G/{H}}, \vect_{G/H})$};
    \filldraw[fill=white] (-5.1,-0.1) rectangle (-4.9,0.1);
    % \filldraw[fill=white] (-1.1,-0.1) rectangle (-0.9,0.1);
    \node[scale=0.8]at(-5, 0.3){$\vect_{G/{H}}$};
    % \node[]at(-1, 0.3){$\vect_{G/{H}}$};

    \draw[very thick](0,0)--(4,0) ;
    \filldraw[fill=white] (-0.1,-0.1) rectangle (0.1,0.1);
    % \filldraw[fill=white] (3.9,-0.1) rectangle (4.1,0.1);
    \node[scale=0.8]at(0.6, 0.3){$(\vect_{G/{H}})^{H_b}$};
    % \node[]at(3.7, 0.3){$(\vect_{G/{H}})^{\{e\}}$};
\end{tikzpicture}
\caption{The illustration of equivariantization process of the topological sector of states of a boundary phase. In which we have $(\vect_{G/H})^{H_b}$ equivalent to ${}_{F_{H_b}} \rep(G)_{F_{H}}$.}
\end{figure} 

Now we consider $H_b$-symmetric boundaries. 
The $H_b$-action on $\vect_{G/H}$ is induced from the $G$-action on $\vect_{G/H}$:
\[
U(h) \lvert \psi_{x} \rangle = \lvert \psi_{hx} \rangle , \qquad h \in H_b , \, x \in G/H .
\]
In other words, we have
\[
h\mathcal T'_{x} \coloneqq  \mathcal T'_{hx} .
\]
% Now we view \eqref{SPT_partially} as a system with $H_b$-symmetry. The $H_b$-action on  $\vect_{G/H}$ is induced by the $G$-action on states:
%% arXiv version
Then the category of $H_b$-symmetric boundaries is the equivariantization $(\vect_{G/H})^{H_b}$, which is equivalent to $\BMod_{F_{H_b}|F_{H}}(\rep(G)) \eqqcolon {}_{F_{H_b}} \rep(G)_{F_{H}}$. Readers may check remark \ref{rem:equivariantization_bdy} to see the proof of the following equivalence. 
%% SB version
%Then the category of $H_b$-symmetric boundaries is the equivariantization $(\vect_{G/H})^{H_b}$, which is equivalent to \\
%$\BMod_{F_{H_b}|F_{H}}(\rep(G)) \eqqcolon {}_{F_{H_b}} \rep(G)_{F_{H}}$. Readers may check remark B.13 in Supplementary material to see the proof of the following equivalence. 
% It is equivalent to the representation category of the action groupoid $\CG(G/H,H_b)$, where $H_b$ acts on $G/H$ by left translation. 

\begin{prop} \label{general_boundary}
Let $G$ be a finite group and $H, H_b \subseteq G$ be two subgroups of $G$. Then the category ${}_{F_{H_b}} \rep(G)_{F_{H}}$ is equivalent to the equivariantization $(\vect_{G/H})^{H_b}$.
\end{prop}

In order to see the action of background category ${}_{F_{H_b}} \rep(G)_{F_{H_b}}$ on the underlying category ${}_{F_{H_b}} \rep(G)_{F_{H}}$ (which is the black arrow drawn in figure \ref{fig:two_boundaries} (a)), we use three known actions (drawn in gray).

We label simple objects in ${}_{F_{H_b}} \rep(G)_{F_{H}}$ by $\mathcal P_{[g], [\rho]}$.
The action of $\Fun(\vect_{G/H}, \vect_{G/H})$ on $\vect_{G/H}$ (equation \eqref{eq:1d_action_on_0d}) induces action on their equivariantization, i.e. an action of ${}_{F_{H}} \rep(G)_{F_{H}}$ on ${}_{F_{H_b}} \rep(G)_{F_{H}}$.
By composing with the action of topological sectors of operators $\FZ_1(\rep(G))$ on ${}_{F_{H}} \rep(G)_{F_{H}}$ in the 1+1D bulk, we obtain the action from $\FZ_1(\rep(G))$ onto ${}_{F_{H_b}} \rep(G)_{F_{H}}$.
\[
\mathcal O_{(h, \sigma)} \odot \mathcal P_{[g], [\rho]} = \mathcal P_{[hg], [\sigma\rho]} ,
\]
Moreover, since the topological sector of operators ${}_{F_{H_b}} \rep(G)_{F_{H_b}} \simeq \FZ_1(\rep(G))_{A(H_b)}$ on the boundary is obtained from condensing the sectors of operators in the bulk, 
we can eventually obtain an categorical symmetry ${}_{F_{H_b}} \rep(G)_{F_{H_b}}$-action on the category of boundary conditions ${}_{F_{H_b}} \rep(G)_{F_{H}}$:
\[
\mathcal M_{[h], [\sigma]} \odot \mathcal P_{[g], [\rho]} = \mathcal P_{[hg], [\sigma\rho]} ,
\]
for all $g \in G$ and $\rho \in \hat G$, and $\mathcal M_{[h], [\sigma]}\in {}_{F_{H_b}} \rep(G)_{F_{H_b}}$. Thus the hom spaces between $H_b$-symmetric topological sectors of states are given by
\[
    \Hom(\mathcal P_{[g], [\rho]},\mathcal P_{[h], [\sigma]}) = \bigoplus_{\substack{k \in [hg^{-1}] \\ \lambda \in [\sigma \rho^{-1}]}} \mathcal{M}_{([k],[\lambda])},
\]
where $[hg^{-1}] $ is the equivalence class of $G/H_b$ and $[\sigma \rho^{-1}]$ is the equivalence class of $\hat H_b$.

It turns out that the observables of a $1$d bulk phase with symmetry $H \subseteq G$ with its boundary conditions that preserve a subgroup $H_b$ form the enriched category 
\be \label{general_bdy_enriched}
\bc[{}_{F_{H_b}} \rep(G)_{F_{H_b}}]{{}_{F_{H_b}} \rep(G)_{F_{H}}}.
\ee
Thus we have proved the following physical theorem:
\begin{pthm}
The $H_b$-symmetric boundary conditions of the $1$d bulk phase $\bc[\FZ_1(\rep(G))]{{}_{F_{H}} \rep(G)_{F_{H}}}$ forms an enriched category  $\bc[{}_{F_{H_b}} \rep(G)_{F_{H_b}}]{{}_{F_{H_b}} \rep(G)_{F_{H}}}$.
% Fix a subgroup $H \subseteq G$, all boundary conditions that preserves the symmetry $H$ form an enriched category.
\end{pthm}

% \begin{rem}\label{bdy-categorical symmetry}
%     Note that the topological sector of operators can also be described by equivariantization. In the boundary cases, it is just the equivariantization of the ``invisible bulk'', for example, in \eqref{general_bdy_enriched}, the categorical symmetry ${}_{F_{H_b}} \rep(G)_{F_{H_b}}$ is equivalent to the equivariantization $\Fun(\vect_{G/{H_b}},\vect_{G/{H_b}})^G$ . 
% \end{rem}

We depict a $H_b$-symmetric boundary condition of the $H$-partially broken $1$d bulk phase in figure \ref{fig:two_boundaries} (b), and figure \ref{fig:two_boundaries} (a) is the corresponding $2$d topological order $\FZ_1(\rep(G))$ with $1$d boundaries before topological Wick rotation. 
%  The invertible domain walls are just the underlying categories of the boundaries  

\begin{rem}
    % \mynote{1d condensation: a picture of dimensional reduction/fusion}
    The $0$d invertible domain wall between ${}_{F_{H_b}} \rep(G)_{F_{H_b}}$ and ${}_{F_{H}} \rep(G)_{F_{H}}$ is given by $\rep(G)_{F_{H_b}}\boxtimes_{\rep(G)} \rep(G)_{F_{H}} \simeq {}_{F_{H_b}} \rep(G)_{F_{H}}$\footnote{The Deligne's tensor product $\boxtimes$ physically corresponds to stacking two phases together.} \cite{KZ18}. 
\end{rem}

\begin{figure*}
\centering
\begin{tikzpicture}
                        \filldraw[fill=gray!20, draw=white] (-7.4,0) rectangle (-3.4,2);
                        \node[]at(-5.4, -0.9){(a)};

                        \draw[color=gray!80][thick, -latex](-6.45,1)--(-7.2,1);
                        \draw[color=gray!80][thick, -latex](-6.45,0.2)--(-7.2,0.2);

                        \draw[color=gray!80][thick, -latex](-6.4,0.9)--(-6.4,0.25);
                        \draw[thick, -latex](-7.2,0.9)--(-7.2,0.25);

                        \draw[very thick](-3.4,0)--(-7.4,0) ;
                        \draw[very thick](-7.4,0)--(-7.4,2) ;
                        % \draw[very thick](-3.4,0)--(-3.4,2) ;
                        \draw[very thick](-7.4,0)--(-5.4,0);
                        \node[]at(-5.4, -0.3){\text{\scriptsize ${}_{F_{H}} \rep(G)_{F_{H}}$}};
                        \draw[very thick](-7.4,2)--(-7.4,1) node[left]{\text{\scriptsize${}_{F_{H_b}} \rep(G)_{F_{H_b}}$}};
                        % \draw[very thick,-stealth](-3.4,2)--(-3.4,1) node[right]{\text{\scriptsize${}_{F_{H_c}} \rep(G)_{F_{H_c}}$}};
                        \node[]at(-5.4,1){$\FZ_1(\rep(G))$};
                        % \filldraw[fill=white] (-3.45,-0.05) rectangle (-3.35,0.05);
                        % \node[]at(-3.4, -0.3){\text{\scriptsize ${}_{F_{H}} \rep(G)_{F_{H_c}}$}};
                        \filldraw[fill=white] (-7.35,-0.05) rectangle (-7.45,0.05);
                        \node[]at(-7.4, -0.3){\text{\scriptsize ${}_{F_{H_b}} \rep(G)_{F_{H}}$}};
                        % \node[]at(-9.6,0){\quad};
                        % \node[]at(-1,0){\quad};

\begin{scope}[xshift=1.5cm]
                        \draw[ -latex](-3, 1)--(-2.5,1);
                        \draw[very thick](0.5,0)--(4,0) ;
                        \draw[very thick](0.5,0)--(2.5,0)node[below]{\scriptsize $\bc[\FZ_1(\rep(G))]{{}_{F_{H}} \rep(G)_{F_{H}}}$};
                        \filldraw[fill=white] (0.45,-0.05) rectangle (0.55,0.05);
                        \node[]at(0.5, 0.4){$\bc[{}_{F_{H_b}} \rep(G)_{F_{H_b}}]{{}_{F_{H_b}} \rep(G)_{F_{H}}}$};
                        % \filldraw[fill=white] (4.45,-0.05) rectangle (4.55,0.05);
                        % \node[]at(4.5,0.4){$\bc[{}_{F_{H_c}} \rep(G)_{F_{H_c}}]{{}_{F_{H}} \rep(G)_{F_{H_c}}}$};
                        \node[]at(2.5, -0.9){(b)};
                        \node[]at(-1.7,0){\quad};
                        % \node[]at(6,0){\quad};
\end{scope}
\end{tikzpicture}
\caption{\label{fig:two_boundaries}  
            The holographic explanation of the boundary phase.
            (a) depicts the $2$d quantum double model $\FZ_1(\rep(G))$ with two $1$d boundaries, and the bulk excitations have a natural action on the 0d domain wall ${}_{F_{H_b}} \rep(G)_{F_{H_b}}$ through two routes (depicted as the four arrows). 
            % which is just the the equivariantization $\Fun(\vect_{G/{H_b}},\vect_{G/{H}})^G$. 
            After topological Wick rotation, we get the categorical description of the $1$d bulk phase $\bc[\FZ_1(\rep(G))]{{}_{F_{H}} \rep(G)_{F_{H}}}$
            % $\bc[\FZ_1(\rep(G))]{{}_{F_{H}} \rep(G)_{F_{H}}}$ 
            with a boundary that breaks to $H_b \in G$, as depicted in (b).}
\end{figure*}

\begin{rem}
We can check that these $0$d boundary categorical descriptions are compatible with the 1d bulk descriptions from the boundary-bulk relation:
\[
\FZ_0 \bigl( \bc[{}_{F_{H_b}} \rep(G)_{F_{H_b}}]{{}_{F_{H_b}} \rep(G)_{F_{H}}} \bigr) \simeq \bc[\FZ_1(\rep(G))]{{}_{F_{H}} \rep(G)_{F_{H}}} .
\]
See \cite[section 4.3]{KYZZ21} for the definition of the $E_0$-center of an enriched category and the proof of the above equivalence.
\end{rem}

\subsection{Domain walls}

% \begin{enumerate}
%    \item \textbf{Invertible domain wall}: 
One can also consider a domain wall connecting two 1+1D gapped phases that break to subgroups $H_1$ (described by $\bc[\FZ_1(\rep(G))]{{}_{F_{H_1}}\rep(G)_{F_{H_1}}}$) and $H_2$ (described by $\bc[\FZ_1(\rep(G))]{{}_{F_{H_2}} \rep(G)_{F_{H_2}}}$) respectively.

We first consider an invertible domain wall between these two bulks, for which we mean there is no explicit symmetry breaking happens on site 0. We choose a canonical Hamiltonian constructed as 
\begin{align} \label{domain wall}
\mathcal{H}=\sum_{i<0}(1-X^i_{H_1}) + \sum_{i<0}(1-Z^{i,i-1}_{H_1}) \nonumber \\
+ \sum_{i\geq 0}(1-X^i_{H_2}) + \sum_{i\geq 0}(1-Z^{i,i+1}_{H_2}),
\end{align}
where $Z^{i,i-1}_{H_1} \coloneqq \frac{\lvert H_1 \rvert}{\lvert G \rvert} \sum_{\rho \in \widehat{G/H_1}} E_\rho^i (E_\rho^{i-1}) ^\dagger$.
The ground state is
    \[
        \lvert\psi_{{x_1},{x_2}}\rangle \coloneqq (\bigotimes_{i< 0} \frac{1}{\sqrt{\lvert H_1 \rvert}} \sum_{g \in {x_1}} \lvert g \rangle_i)\otimes (\bigotimes_{i\geq 0} \frac{1}{\sqrt{\lvert H_2 \rvert}} \sum_{g \in {x_2}} \lvert g \rangle_i) , 
    \]
$\forall x_1 \in G/H_1 , x_2 \in G/H_2$.

Since the $1$d Hilbert space extends to infinity on both sides.
$E^0_{\rho}$ and $M^0_g$ are not localized (or condensed) on this domain wall anymore. 
There are just some interchange of topological sectors of operators between left and right sides. 
Or to say, the trivial action of $A(H_1)=\bigoplus_{\substack{g \in H_1 \\ \rho \in \widehat{G/H_1}}} \mathcal{O}_{(g, \rho)}$ on the left ground state is replaced by the trivial action $A(H_2)=\bigoplus_{\substack{g \in H_2 \\ \rho \in \widehat{G/H_2}}} \mathcal{O}_{(g, \rho)}$ on the right ground state.
So the categorical symmetry is still $\FZ_1(\rep(G))$ as in the bulk cases. 

% We can find the explicit hom spaces here
% \[
%         \Hom(\mathcal P_{[g], [\rho]},\mathcal P_{[h], [\sigma]}) = \bigoplus_{\substack{k \in [hg^{-1}] \\ \lambda \in [\sigma \rho^{-1}]}} \mathcal{O}_{([k],[\lambda])} .
%     \]

    %   Since all boundaries of $\FZ_1(\rep(G))$ are Morita equivalent, the analysis follows from the above discussions directly: 
In order to find the underlying category, it is also straightforward to use the equivariantization technique. For equation \eqref{domain wall}, the bulk excitations without symmetry on each side is $\Fun(\vect_{G/{H_1}}, \vect_{G/{H_1}})$ and $\Fun(\vect_{G/{H_2}}, \vect_{G/{H_2}})$. 
And $\Fun(\vect_{G/{H_1}}, \vect_{G/{H_2}})$ is the $0$d invertible domain wall between them. There are $n_1 \times n_2 = \lvert G \rvert / \lvert H_1 \rvert \times \lvert G \rvert / \lvert H_2 \rvert$ simple topological sectors of state $\{\mathcal R_{x,y}\}_{x \in G/H_1, y \in G/H_2}$ generated by the ground state $\lvert\psi_{{x_1},{x_2}}\rangle$. Moreover, we have the following fusion rules:
\begin{gather*}
\mathcal R_{x,y} \otimes \mathcal T_{z,w} = \delta_{y,z} \mathcal R_{x,w}, \quad \forall x\in G/H_1, \, y,z,w \in G/H_2 , \\
\tilde{\mathcal T}_{x,y} \otimes \mathcal R_{z,w} = \delta_{y,z} \mathcal R_{x,w}, \quad \forall x,y,z\in G/H_1, \, w \in G/H_2 .
\end{gather*}
These fusion rules endow the category $\Fun(\vect_{G/{H_1}}, \vect_{G/{H_2}})$ with a structure of $\Fun(\vect_{G/{H_1}}, \vect_{G/{H_1}})$-$\Fun(\vect_{G/{H_2}}, \vect_{G/{H_2}})$-bimodule.

Note that Hamiltonian  \eqref{domain wall} is spontaneously breaking. So we can view the domain wall as a 0+1D topological order with $G$-symmetry.
The $G$ action on $\Fun(\vect_{G/{H_1}}, \vect_{G/{H_2}})$ is induced by the $G$ action on states:
    \[
    U(g)\lvert\psi_{{x_1},{x_2}}\rangle =\lvert\psi_{{gx_1},{gx_2}}\rangle ,\quad x_1\in G/H_1, x_2 \in G/H_2.
    \] 
    Or in other words:
    \[
    g(\mathcal R_{x,y})=\mathcal R_{gx,gy}.
    \]     
Thus, the $G$-symmetric topological sectors of states of the domain wall is equivalent to the equivariantization $\Fun(\vect_{G/{H_1}}, \vect_{G/{H_2}})^G$, which is equivalent to ${}_{F_{H_1}} \rep(G)_{F_{H_2}}$,
    % the underlying category of the $G$-symmetric enriched fusion category should be given by $(\vect_{G/H_a})^{H_b} \simeq \vect_{G/H_b}^{H_a} {}_{F_{H_b}} \rep(G)_{F_{H_a}}$
as illustrated below.
% in figure \ref{domain wall equivariantization}

    %   the left bulk now really exists and can not be ``folded up'' onto the time direction anymore.

% \begin{figure}[H]
% \centering
% \includegraphics{figures/domain_wall_equivariantization.pdf}
% \end{figure}
\begin{figure}[H]
\centering
    \begin{tikzpicture}[scale=0.9]
        \small
        \draw[very thick](-4, 0)--(4,0);
        \node at(-3, -0.4){\scriptsize $\Fun(\vect_{G/H_1},\vect_{G/H_1})^G$};
        \node at(3, -0.4){\scriptsize $\Fun(\vect_{G/H_2},\vect_{G/H_2} )^G$};
        \node at(0, 0.4){\scriptsize $\Fun(\vect_{G/H_1},\vect_{G/H_2})^G\simeq {}_{F_{H_1}} \rep(G)_{F_{H_2}}$};
        \filldraw[fill=white, draw=black] (-0.1,-0.1) rectangle (0.1,0.1); 
\end{tikzpicture}
%    \caption{\label{domain wall equivariantization} Topological sector of states of an invertible domain wall.}
\end{figure}

One can check that the invertible domain walls between two $1$d gapped phases 
$\bc[\FZ_1(\rep(G))]{{}_{F_{H_1}} \rep(G)_{F_{H_1}}}$ and $\bc[\FZ_1(\rep(G))]{{}_{F_{H_2}} \rep(G)_{F_{H_2}}}$ form an enriched category 
\[
    \bc[\FZ_1(\rep(G))]{{}_{F_{H_1}} \rep(G)_{F_{H_2}}} .
\] 
Mathematically, ${}_{F_{H_1}} \rep(G)_{F_{H_2}}$ is an invertible $({}_{F_{H_1}} \rep(G)_{F_{H_1}},{}_{F_{H_2}} \rep(G)_{F_{H_2}})$-bimodule \cite{EO04}. We also depict the topological Wick rotation process of this invertible domain wall in the following figure.
        %figure \ref{fig:inv_domain_wall}. 
        
% \begin{figure}[H]
%     \subfigure{
%         \begin{minipage}[t]{0.45\linewidth}
%                     \includegraphics{figures/inv_domain_wall_a.pdf}
%                 \end{minipage}
%             }
%             \subfigure{
%                 \begin{minipage}[t]{0.45\linewidth}
%                     \includegraphics{figures/inv_domain_wall_b.pdf}
%                 \end{minipage}
%             }
%         %  \caption{\label{fig:inv_domain_wall} Invertible Domain Wall}
%     \end{figure}

\begin{figure}[H]
\centering
    \subfigure{
        \begin{minipage}[t]{0.45\linewidth}
                    \adjustbox{scale=0.7}{
                \begin{tikzpicture}
                    \scriptsize
                    \filldraw[fill=gray!20, draw=white] (-8,0) rectangle (-4,2);
                    \draw[very thick](-8,0)--(-4,0);
                    \filldraw[fill=white] (-6.1,-0.1) rectangle (-5.9,0.1);
                    % \draw[very thick, -latex] (-8,0)--(-7,0);
                    % \draw[very thick, -latex] (-5.8,0)--(-5,0);
                    \node at(-8,-0.5){${}_{F_{H_1}} \rep(G)_{F_{H_1}}$};
                    \node at(-5.9,-0.5){${}_{F_{H_1}} \rep(G)_{F_{H_2}}$};
                    \node at(-3.6,-0.5){${}_{F_{H_2}} \rep(G)_{F_{H_2}} \qquad$};
                    \node at(-6,1){$\FZ_1(\rep(G))$};
                    \node at(-2,0){\quad};
                \end{tikzpicture}}
                \end{minipage}
            }
            \subfigure{
                \begin{minipage}[t]{0.45\linewidth}
                    \adjustbox{scale=0.7}{
                \begin{tikzpicture}
                    \scriptsize
                    \draw[ -latex](-1.5, 0.5)--(-1,0.5);
                    \filldraw[fill=gray!0, draw=white] (0,0) rectangle (4,2);
                    \draw[very thick](0, 0)--(4,0);
                    \filldraw[fill=white, draw=black] (2.1,-0.1) rectangle (1.9,0.1) node[above]{$\bc[\FZ_1(\rep(G))]{{}_{F_{H_1}} \rep(G)_{F_{H_2}}}$};
                    % \draw[very thick,-latex](0,0)--(1, 0); 
                    % \node at(-0.1,-0.4){$\bc[\FZ_1(\rep(G))]{{}_{F_{H_1}} \rep(G)_{F_{H_1}}}$};
                    % \draw[very thick,-latex](2.2,0)--(3, 0); 
                    % \node at(4, -0.4){$\bc[\FZ_1(\rep(G))]{{}_{F_{H_2}} \rep(G)_{F_{H_2}}}$};
                    \node at(2,-0.66){\quad};
                \end{tikzpicture}}
                \end{minipage}
            }
        %  \caption{\label{fig:inv_domain_wall} Invertible Domain Wall}
\end{figure}
    If two phases across the domain wall are the same, i.e. when $H=H_1=H_2$, the invertible domain wall $\bc[\FZ_1(\rep(G))]{{}_{F_H} \rep(G)_{F_H}}$ is just the same category as the
    1d bulk, namely, it becomes a trivial domain wall.       
% \end{enumerate}

% \mynote{Is all non-invertible domain walls "explicitly" symmetry broken?}

%     \begin{rem}
% %            One can check that the boundary-bulk relations holds for these domain walls (invertible or non-invertible) by the ``folding trick'', where for the invertible one, we have
%         One can check that the boundary-bulk relations holds by the `folding trick' \cite{KZ22a}:
%         \begin{gather*}
%         \FZ_0(\bc[\FZ_1(\rep(G))]{{}_{F_{H_1}} \rep(G)_{F_{H_2}}}) \simeq\\
%         \bc[\FZ_1(\rep(G))\boxtimes \overline{\FZ_1(\rep(G))}]{\bigl({}_{F_{H_1}} \rep(G)_{F_{H_1}}\boxtimes ({}_{F_{H_2}} \rep(G)_{F_{H_2}})^{\rev}\bigr)}.
%         \end{gather*}
%         ~
%     \end{rem}

% \begin{enumerate}
% \setcounter{enumi}{1}       
%    \item \textbf{non-invertible domain wall}: 
\vspace{2em}
We now consider the non-invertible domain walls between two 1+1D bulks, which we mean there is some explicit symmetry breaking happening on site 0. 

For 'purely' explicitly symmetry breaking case, we choose a canonical Hamiltonian constructed as 
{\small
\begin{align*}\label{pure_explicit_bdy_broken}
\mathcal{H} \coloneqq \sum_{i<0} (1 - X_{H_1}^i) + \sum_{i<0} (1 - Z_{H_1}^{i,i+1})  + (1-Z^{0}_{H_b}) +(1-X^{0}_{H_b})\\
+\sum_{i>1} (1 - X_{H_2}^i) + \sum_{i>1} (1 - Z_{H_2}^{i,i+1})  + (1-Z^{1}_{H_b'}) +(1-X^{1}_{H_b'}),
\end{align*}}
and the ground state is 
\begin{align*}
    \lvert \psi_x\rangle_{i<0} \otimes \frac{1}{\sqrt{\lvert H_b \rvert}} \sum_{h \in H_b} \lvert h \rangle_0 
    \otimes \frac{1}{\sqrt{\lvert H_b' \rvert}} \sum_{h \in H_b'} \lvert h' \rangle_1 \otimes \lvert \psi_y \rangle_{i>1},
\end{align*}
$\forall x \in G/H_1$ and $\forall y\in G/H_2$,
which can be viewed as piecing together two of the symmetric explicit breaking 0+1D boundaries, see bottom figure:
\begin{figure}[H]
\centering
\begin{tikzpicture}
        % \draw[ -latex](-7, -1)--(-6.5,-1);
        \draw[very thick](-4, -2)--(-0.3,-2);
        \draw[very thick](4, -2)--(0.3,-2);

        \filldraw[fill=white, draw=black] (-0.4,-2.1) rectangle (-0.2,-1.9);
        \node at(-1.8, -1.3){ $\bc[{}_{F_{H_b}} \rep(G)_{F_{H_b}}]{{}_{F_{H_1}} \rep(G)_{F_{H_b}}}$};
        \filldraw[fill=white, draw=black] (0.2,-2.1) rectangle (0.4,-1.9);
        \node at(0,-1.3){$\boxtimes$};
        \node at(1.8,-1.3){$\bc[{}_{F_{H_b'}} \rep(G)_{F_{H_b'}}]{{}_{F_{H_b'}} \rep(G)_{F_{H_2}}}$};
        \draw[dashed](-0.8,-2.3) rectangle (0.8, -1.7);
        %  \node at(0,-3.5){Domain wall: $\bc[{}_{F_{H_a}} \rep(G)_{F_{H_a}} \boxtimes \rep(G)_{F_{H_l}|F_{H_l}}]  {\rep(G)_{F_{H_c}|F_{H_a}}} \boxtimes {\rep(G)_{F_{H_l}|F_{H_m}}}$};
        %\draw[very thick](-3,-2)--(-1.5, -2); \node at(-5,-2.4){$\bc[\FZ_1(\rep(G))]{{}_{F_{H_1}} \rep(G)_{F_{H_1}}}$};
        %\draw[very thick](0.5,-2)--(0.75, -2); \node at(5, -2.4){$\bc[\FZ_1(\rep(G))]{{}_{F_{H_2}} \rep(G)_{F_{H_2}}}$};
        % \node at(0, -3.5){\large $(b)$};
\end{tikzpicture}
\caption{It is obvious to see that this domain wall is just the stacking of left and right gapped open boundaries analyzed in section \ref{general_bdy}.}
\end{figure}

Except for this invertible domain wall.
there are other kinds of domain walls controlled by $K \subseteq G$.
Or to say, there are some categorical entanglement \cite{XY24} induced by the topological sectors of operators.
The 2d topological order before topological Wick rotation is depicted in the follow figure (see the dashed frame part). In which the 'interconnected' part can be obtained through a 2d condensation from $\FZ_1(\rep(G))$ to $\FZ_1(\rep(K))$.

% \begin{figure}[H]
% \centering
% \includegraphics{figures/domain_wall_GKG.pdf}
% % \label{fig:domain_wall}
%     % should be written as $\bc[{}_{F_{H_a}} \rep(G)_{F_{H_a}} \boxtimes \rep(G)_{F_{H_l}|F_{H_l}}]  {\rep(G)_{F_{H_c}|F_{H_a}}} \boxtimes {\rep(G)_{F_{H_l}|F_{H_m}}}$, which }
% \end{figure}
\begin{figure}[H]
\centering

\adjustbox{scale=0.7}{\begin{tikzpicture}
            \scriptsize
            \filldraw[fill=gray!20, draw=white] (-5,-2) rectangle (5,1);
            \filldraw[fill=gray!10, draw=white] (-0.9,-2) rectangle (0.9,1);
            \draw[very thick](-5, -2)--(-1,-2);
            \draw[very thick](5, -2)--(1,-2);
            \draw[](-0.8, -2)--(0.8,-2);

            \filldraw[fill=white, draw=black] (-1,-2.1) rectangle (-0.8,-1.9);
            % \node at(-1.8, -1.3){$\bc[{}_{F_{H_a}} \rep(G)_{F_{H_a}}]{\rep(G)_{F_{H_c}|F_{H_a}}}$};
            \filldraw[fill=white, draw=black] (0.8,-2.1) rectangle (1,-1.9);
            % \node at(1.8,-1.3){$\bc[\rep(G)_{F_{H_l}|F_{H_l}}]{\rep(G)_{F_{H_l}|F_{H_m}}}$};
            \draw[dashed](-1.4,-2.3) rectangle (1.4, -1.7);
            %    \node at(0,-2.7){non invertible wall};
            \draw[](-0.9,-1.9)--(-0.9, 1); 
            \draw[](0.9,-1.9)--(0.9, 1);
            \node at(0,-0.5){\scriptsize $\FZ_1(\rep(K))$};
            \node at(-3,-0.5){\scriptsize $\FZ_1(\rep(G))$};
            \node at(3,-0.5){\scriptsize $\FZ_1(\rep(G))$};
            % \draw[](7,0)--(7,0.5);
            % \node at(0, -2.8){\large $(a)$};
            
            % \draw[very thick](0.5,-2)--(0.75, -2); \node at(5, -2.4){$\bc[\FZ_1(\rep(G))]{\rep(G)_{F_{H_m}|F_{H_m}}}$};
        \end{tikzpicture}
        }
% \label{fig:domain_wall}
    % should be written as $\bc[{}_{F_{H_a}} \rep(G)_{F_{H_a}} \boxtimes \rep(G)_{F_{H_l}|F_{H_l}}]  {\rep(G)_{F_{H_c}|F_{H_a}}} \boxtimes {\rep(G)_{F_{H_l}|F_{H_m}}}$, which }
\end{figure}

% \mynote{categorical entanglement}
%    \begin{figure}[H]
%        \begin{center}
%            \begin{tikzpicture}[scale=0.6]
%                \scriptsize
%                \filldraw[fill=gray!20, draw=white] (-5,-2) rectangle (5,1);
%                \draw[very thick](-5, -2)--(-1,-2);
%                \draw[very thick](5, -2)--(1,-2);
%                \draw[](-0.8, -2)--(0.8,-2);

%                \filldraw[fill=white, draw=black] (-1,-2.1) rectangle (-0.8,-1.9);
%                % \node at(-1.8, -1.3){$\bc[{}_{F_{H_a}} \rep(G)_{F_{H_a}}]{\rep(G)_{F_{H_c}|F_{H_a}}}$};
%                \filldraw[fill=white, draw=black] (0.8,-2.1) rectangle (1,-1.9);
%                % \node at(1.8,-1.3){$\bc[\rep(G)_{F_{H_l}|F_{H_l}}]{\rep(G)_{F_{H_l}|F_{H_m}}}$};
%                \draw[dashed](-1.4,-2.3) rectangle (1.4, -1.7);
%             %    \node at(0,-2.7){non invertible wall};
%                \draw[](-0.9,-1.9)--(-0.9, 1); 
%                \draw[](0.9,-1.9)--(0.9, 1);
%                \node at(0,-0.5){\scriptsize $\FZ_1(\rep(H))$};
%                \node at(-3,-0.5){\scriptsize $\FZ_1(\rep(G))$};
%                \node at(3,-0.5){\scriptsize $\FZ_1(\rep(G))$};
%                % \draw[very thick](0.5,-2)--(0.75, -2); \node at(5, -2.4){$\bc[\FZ_1(\rep(G))]{\rep(G)_{F_{H_m}|F_{H_m}}}$};
%            \end{tikzpicture}
%        \end{center}
%    \end{figure}

% The dashed frame part after topological wik rotation could give more domain walls. 
For example, if we choose $\FZ_1(\rep(K)) = \vect$. This domain wall after topological Wick rotation is just the above 'purely explicit breaking one. If we choose $\FZ_1(\rep(K)) = \FZ_1(\rep(G))$, this would become an invertible domain wall.

% \input{conclusion}
% !TeX root = main.tex
% !TeX program = pdfLaTeX

\section{Physical examples}\label{section:example}

\subsection{Transverse field Ising chain}\label{section:Ising}

We first revisit the 1+1D Ising model, in which Kong, Wen and Zheng directly find the observables to show that the $\Zb_2$ SPT phase can be describe by ${}^{\FZ_1(\rep(\mathbb{Z}_2))}\rep(\mathbb{Z}_2)$ and the spontaneous symmetry-breaking phase can be described by ${}^{\FZ_1(\rep(\mathbb{Z}_2))}\vect_{\mathbb{Z}_2}$ in \cite{KWZ22}. We will first review their methods, then show the another way in finding the topological sectors of states through the equivariantization method invented in section \ref{section:equivariantization}.

So consider a 1d Ising chain: $\CH_{\text{tot}}=\otimes_{i\in \mathbb{Z}}\mathbb{C}^2_i$,
% $ \quad
%     \vcenter{\hbox{
%             \adjustbox{scale=0.7}{
%                 \begin{tikzpicture}
%                     \draw[thick] (-4,0)--(4,0);
%                     \draw[](-3,-0.2)node[below]{$-3$}--(-3,0.2) ;
%                     \draw[](-2,-0.2)node[below]{$-2$}--(-2,0.2) ;
%                     \draw[](-1,-0.2)node[below]{$-1$}--(-1,0.2) ;
%                     \draw[](0,-0.2)node[below]{$0$}--(0,0.2) ;
%                     \draw[](3,-0.2)node[below]{$3$}--(3,0.2) ;
%                     \draw[](2,-0.2)node[below]{$2$}--(2,0.2) ;
%                     \draw[](1,-0.2)node[below]{$1$}--(1,0.2) ;
%                 \end{tikzpicture}}}}
% $
with the Hamiltonian
\be
\mathcal{H}^{Ising} = -\sum_i gX^i - \sum_i J Z^i Z^{i+1},
\ee
where $X^i$ and $Z^i$ are Pauli matrices, i.e. $Z^i\lvert\uparrow \rangle_i =\lvert\uparrow \rangle_i ,  Z^i\lvert\downarrow \rangle_i =-\lvert\downarrow \rangle_i$ and $X^i\lvert\pm\rangle_i = \pm \lvert\pm \rangle_i $ for
$\lvert\pm\rangle_i = \frac{1}{\sqrt{2}}(\lvert\uparrow \rangle_i \pm \lvert\downarrow \rangle_i)$.
% and $|-\rangle_i = \frac{1}{\sqrt{2}}(\lvert\uparrow \rangle_i -\lvert\downarrow \rangle_i).$
This model is equivalent to taking $G= \Zb_2:=\{e,a\}$ for the model \eqref{eq:Hamiltonian} we constructed in section \ref{section:lattice_model}.
% \eqref{eq:Hamiltonian}.
It has a global $\mathbb{Z}_2$ onsite symmetry:
$$U=\bigotimes_{i\in \mathbb{Z}}X^i$$

A symmetric operator should commute with $U=\otimes_{i\in \mathbb{Z}}X^i$, (i.e. $[P, U]=0$). 
We can explicitly find four $U$-symmetric topological sectors of operators listed as follows:
\begin{enumerate}
\item $\bfone:=\mathcal{O}_{e,1}$ consists of $U$-symmetric local operators;
\item $\bfm:=\mathcal{O}_{a,1}$ is generated by $M^i = \prod_{k \leq i}X^k$;
\item $\bfe:=\mathcal{O}_{e,-1}$ is generated by $E^{i} = \prod_{k \geq i}(Z_kZ_{k+1})$;
\item $\bff:=\mathcal{O}_{a,-1}$ is generated by $M^i E^{j}$.
\end{enumerate}
% Note that $\one$ and $Z^iZ^{i+1}$ are $U$-symmetric local operators, and although $Z^i$ breaks the $U$-symmetry as a local operator, it can be viewed as a $U$-symmetric non-local operator because $Z^i=\otimes_{k\geq i}(Z_k Z_{k+1})$. In other words, we can say $Z^i$ catches all the corrected local properties of $Z_{i, \infty}$ near the site $i$.
% Why use $\one, e, m, f$ to label the topological sector of operators? Because they actually reproduce the category of 2d toric code model $\FZ_1(\rm Rep(\mathbb{Z}_2))$.
$M^i$ and $E^i$ here are just $M_g^i$ and $E_{\rho}^i$ in section \ref{sec:topo_operator} by taking $G = \Zb_2$.

Notice that these topological sectors of $U$-symmetric
operators automatically satisfy the same fusion rule as the excitations of the 2d toric code model, i.e. 
$$ \bfe \otimes \bfe = \bfone, \, \bfm \ot \bfm = \bfone, \, \bfe \ot \bfm = \bff = \bfm \ot \bfe, \, \bff \ot \bff = \bfone $$

Moreover, if we first create a pair of "$\bfm$ particles" at site $i$ and $j$ for $i < j$ 
% (by applying $M^i M^j$ to $\lvert\Omega\rangle$), 
then apply $E_k$ for $i < k < j$, then annihilate two $\bfm$-particles to obtain $M_i M_j E_k M_i M_j = -E_k$, we can recover the double braiding between $\bfe$ and $\bfm$ in the 2d toric code model, i.e.
\[
\bfm \otimes \bfe \xrightarrow{-1} \bfm \otimes \bfe .
\]
% which is precisely the double braiding between $e$ and $m$ in .
As a consequence, these topological sectors of operators $\bfone, \bfe, \bfm ,\bff$ provide a physical
realization of the category of 2d toric code model $\FZ_1(\rm Rep(\mathbb{Z}_2)) =: \CT\CC$, 
% which is what we need for the background category $\CB$ in ${}^{\CB} \CS$,
as we expected.

% Moreover, for $i<k<j$ we have
% \[
% M^{i} M^{j} E^k M^{j} M^{i} = -E^k .
% \]
% This means that the double braiding of two topological sectors $\bfm$ and $\bfe$ is
% \[
% \bfm \otimes \bfe \xrightarrow{-1} \bfm \otimes \bfe .
% \]

Now we consider the two cases that can be realized by tuning parameters $J$ and $B$.

\subsubsection{Ising chain realizing \texorpdfstring{$\mathbb{Z}_2$}{Z2} SPT order}

First consider $J=0$ and $g\approx 1$. In this case, the Hamiltonian is
\[
\mathcal{H}=-\sum_i gX^i.
\]

% The system is gapped, and 
% The ground state is
% $$\lvert\Omega\rangle = \lvert\cdots ++++\cdots\rangle.$$
The ground state is the product state 
% $$\lvert \Omega \rangle \coloneqq \medotimes_i \lvert \leftarrow \rangle_i,$$ 
\[\lvert\Omega\rangle = \lvert\cdots ++++\cdots\rangle,\] which corresponding to the trivial SPT phase discussed in section \ref{eq:Hamiltonian_symmetry_preserving}. 

% If we do not impose any symmetry, the only topological sector of states is the vacuum
% sector, denoted by $\one$. And the only topological sector of operators is the trivial one.
% % \begin{rem}
% %     Without imposing any symmetry, all non-local operators can be confined by adding proper small perturbations. For example,$ m_i=\otimes_{k\leq i}X^k$ is confined by adding the perturbation term $\sum_i KZ^i$ to the Hamiltonian.
% % \end{rem}
% As a consequence, the phase is the trivial $1d$ topological order and can be mathematically described by $\rm Vec$ (or $\rm {}^{Vec}Vec$).

% The ground state $\lvert\Omega\rangle $ preserves the $U$-symmetry. 

Now we act the four topological sectors of operators on the ground state $\lvert\Omega\rangle$, we see the total Hilbert space splits into two irreducible topological sectors of states (note that $M^i := \otimes_{k\leq i}X^k$ only has a trivial action on $\lvert\Omega \rangle$):
% 这里也有直和问题（我们在setup theory 的时候也需要artificially 地把直和加上去，不加上去internal hom 没法定义，为什么不能简单看成有两种可能）
\begin{enumerate}
    \item The trivial sector $\one$ is generated by the vacuum state $\lvert\Omega \rangle$;
            % and is viewed as a trivial particle or a $\one$-particle.
    \item A non-trivial sector $E$: the lowest energy states are given by
    $$E^i\lvert\Omega \rangle = \lvert\cdots ++-_i +++ \cdots\rangle, \quad \forall i \in \mathbb{Z},$$
            %   each of which represents an $e$-particle located at site $i$.
\end{enumerate}
For two $e$ topological sectors of operators acting on site $i$ and $j$, we have
% (the distance between $i$ and $j$ is smaller than the correlation length),
$E^{i} E^{j}\lvert\Omega\rangle = Z^i Z^j \lvert\Omega\rangle \in \one$ since $Z^i Z^j$ is a local operator.
% $= \lvert\cdots ++-_i +++-_j++ \cdots\rangle \in \one$
This implies the following fusion rules: $\one \otimes E=E\otimes\one=E, E\otimes E=\one$, which coincide with those in the fusion category $\rep(\mathbb{Z}_2)$.

% That is the underlying category $\CS$ in ${}^{\CB} \CS$.
%  of $\mathbb{Z}_2$-representations.

% Let us denote the space of $U$-symmetric operators that map from $x$ to $y$ by ${\Hom}^{SPT}_{bulk}(x,y)$, then for $x, y=\one, E$,

% we immediately obtain
It is also straightforward to see that the space of $U$-symmetric operators which map between these two sectors of sates $\one$ and $E$ are given by:
\begin{gather*}
{\Hom}^{SPT}_{Ising}(\one,\one)=\bfone \oplus \bfm\\
{\Hom}^{SPT}_{Ising}(\one,E)={\Hom}^{SPT}_{Ising}(E,\one)=\bfe\oplus \bff\\
{\Hom}^{SPT}_{Ising}(E,E)=\bfone \oplus \bfm.
\end{gather*}

This result is just equation \eqref{SPT_hom} for $\Zb_2$ case, which exactly matches the enriched fusion category $\rm {}^{\FZ_1(Rep(\mathbb{Z}_2))}Rep(\mathbb{Z}_2)$ obtained from the canonical
construction. This convinces theorem \ref{thm_SPT} in case of $G=\Zb_2$.
% Thus we recover the following theorem \cite{KWZ22}:
\begin{thm}[\cite{KWZ22}]
    The $1d$ $\mathbb{Z}_2$ SPT order of Ising chain can be mathematically described by the enriched fusion category $\rm {}^{\FZ_1(Rep(\mathbb{Z}_2))}Rep(\mathbb{Z}_2)$.
\end{thm}

\begin{rem}\label{Is_SPT_eqvar}
    We can also check the topological sectors of states form fusion category $\rep(\Zb_2)$ using the equivariantization technique.
    
    % Consider the symmetric phase of the 1d Ising chain. The Hamiltonian is
    % \[
    % H = - \sum_i X^i .
    % \]
    % The ground state is the product state $\lvert \Omega \rangle \coloneqq \medotimes_i \lvert \leftarrow \rangle_i$. Thus it realizes the trivial topological order without symmetry, whose topological excitations form a fusion category equivalent to $\vect$.
    
    Note that without symmetry, $\lvert\Omega\rangle$ realizes a trivial 1d topological order, whose topological excitations form a fusion category equivalent to $\vect$.
    If we act a local operator $Z^j$, $Z^j \lvert \Omega \rangle = \lvert \cdots ++-_i+++ \cdots \rangle$ generates the same topological sector of state as $\lvert \Omega \rangle$ (i.e. the tensor unit $\one$ in $\vect$).
    
    Now we view this system as a topological order with the $\Zb_2$-symmetry realized by $U \coloneqq \medotimes_i X^i$. 
    We can obtain $Z^j \lvert \Omega \rangle$ by adding a symmetric local trap $2 X_j$ at a site $j$, i.e. $Z^j \lvert \Omega \rangle$ is the ground state of  $
    \mathcal{H}' = X_j - \sum_{i \neq j} X^i
    $. But since $X_j$ anti-commutes with $Z^j$, we have
    $$
    U Z^j \lvert \Omega \rangle = -Z^j U \lvert \Omega \rangle = -Z^j \lvert \Omega \rangle .
    $$
    In other words, the topological excitation generated by $Z^j \lvert \Omega \rangle$ should be described not only by $\one \in \vect$ but also by a morphism $u = -1 \colon \one \to \one$, which measures the nontrivial $\Zb_2$-charge of $\Z^j \lvert \Omega \rangle$. 
    % Recall $T$
    As a consequence, there are two simple topological excitations $\{\one, u=1\}$  and $\{\one, u=-1\}$ in this topological order with the $\Zb_2$-symmetry, and they form a fusion category equivalent to $\vect^{\Zb_2} \simeq \rep(\Zb_2)$. 
    % More generally, for any group $G$, the equivariantization $\vect^G$ of $\vect$ equipped with the trivial $G$-action is equivalent to $\rep(G)$.
\end{rem}

\begin{rem}
From the holographic point of view, $\rep(\mathbb{Z}_2)$ is just the smooth boundary of toric code model $\CT \CC$. The trivial action of $M^i$ on $\lvert\Omega \rangle$ can be interpreted as a condensation of the "$\bfm$-particles" (or equivalently, the ground state algebra $A_{\bfm}=\bfone\oplus \bfm = {\Hom}^{SPT}_{Ising}(\one,\one)$) in the categorical symmetry $\CT\CC$. In other words, the ground state algebra is $A_{\bfm}$.
In this way we get the underlying category $\rep(\mathbb{Z}_2) \simeq \FZ_1(\rep(\mathbb{Z}_2))_{A_{\bfm}}$. See the following figure.
\end{rem}

\begin{figure}[H]
\centering
\begin{tikzpicture}[scale=0.8]
        \filldraw[fill=gray!20, draw=white] (-8,0) rectangle (-4,2);
        \draw[very thick](-8, 0)--(-4,0);
        % \filldraw[fill=white, draw=black] (-6.1,-0.1) rectangle (-5.9,0.1);
        \node[scale=0.9] at(-6,0.8){$\CT\CC := \FZ_1(\rep(\mathbb{Z}_2))$};
        % \node at(-4.3,-0.4){$\vect_{\mathbb{Z}_2}$};
        \node[scale=0.8] at(-6, -0.2){Smooth bdy $\rep(\mathbb{Z}_2)$};

        \draw[very thick](-1.5,0)--(2.5,0);
        % \draw[draw=none](-8,0)--(-7,0);
        % \filldraw[fill=white] (1.9,-0.1) rectangle (2.1,0.1);
        \node at(0.3,0.3){${}^{\CT\CC}\rep(\mathbb{Z}_2)$};
        % \node at(4,0.5){${}^{\FZ_1(\rep(\mathbb{Z}_2))}\vect_{\mathbb{Z}_2}$};

        \node[scale=0.9] at(-6.8,1.7){\scriptsize 2d Toric code};
        \node[scale=0.9] at(-0.3,1.7){\scriptsize 1+1D Ising SPT};
        \draw[-latex](-3,1)--(-2.4,1);
\end{tikzpicture}
\caption{By condensing $\bfone \oplus \bfm$ in 2d toric code model $\CT\CC$, we can have a 1d smooth gapped boundary $\rep(\Zb_2)\simeq \FZ_1(\rep(\mathbb{Z}_2))_{\bfone\oplus \bfm}$. 
    After topological Wick rotation, the excitations $\bfone, \bfe, \bfm, \bff \in \CT\CC$ become the topological sectors of operators, and the excitations $\one, E \in \rep(\Zb_2)$ becomes the topological sectors of states of 1+1D Ising SPT. Their mapping behaviors exactly meet with enriched fusion  category $\rm {}^{\FZ_1(Rep(\mathbb{Z}_2))}Rep(\mathbb{Z}_2)$, leading to a macroscopic description of the Ising SPT phase.
%  The trivial action of $M^i$ on $\lvert\Omega \rangle$ can be interpreted as a condensation of the "m-particles" (or equivalently, the Lagrangian algebra $A_m=1\oplus m$) in the categorical symmetry provided by the vacuum sector of states.
    }
\end{figure}

\subsubsection{Ising chain realizing \texorpdfstring{$\Zb_2$}{Z2} symmetry breaking order}

    On the other hand, we can consider the case when $g = 0$ and $J \thickapprox 1$, so the Hamiltonian is
\[\mathcal{H}=- \sum_i J Z^i Z^{i+1}\]
        which realizes the symmetry breaking case.
The global symmetry is still $U = \otimes_i X^i$, and
\[
        \lvert \Omega_{\uparrow} \rangle=\lvert\cdots \uparrow \uparrow \uparrow \cdots \rangle \quad \text{and} \quad \lvert \Omega_{\downarrow} \rangle=\quad \lvert\cdots \downarrow\downarrow\downarrow \cdots \rangle
\]
    are two-fold degenerate ground states representing $U$-symmetry broken phases.
    
    If we do not impose symmetry and ignore perturbations,
    %  and ignore perturbations, 
    the total Hilbert space splits into four sectors $\CH_{ab}$ for $a, b = \uparrow, \downarrow$, where $\CH_{ab}$ is spanned by states $(\otimes_{k<i}|a\rangle_k)(\otimes_{k \geqslant i}|b\rangle_k)$ for $i \in \mathbb{Z}$. We denote the topological sector associated to $\CH_{ab}$ by $s_{ab}$, $s_{ab}$ has fusion rule
    \begin{equation*}
        s_{ab}\otimes s_{cd} = \delta_{bc} s_{ad}.
    \end{equation*}
    % Note that when the ground states is viewed as a 1d topological order without symmetry, 
    The topological sectors of states form a multi-fusion 1-category
    $$\mathrm{Mat}_2(\vect) = \rm Fun(Vec_{\mathbb{Z}_2} ,Vec_{\mathbb{Z}_2} ).$$
    % that consists of four simple objects $s_{\uparrow \uparrow}, s_{\uparrow \downarrow}, s_{\downarrow \uparrow}, s_{\downarrow \downarrow}$.

    % Once we impose the $U$-symmetry, The topological excitation $\one$ generated by the ground state subspace is not simple, which should be
    % \begin{equation*}
    %     \one:=s_{\uparrow \uparrow}\oplus s_{\downarrow \downarrow}
    % \end{equation*}

    Now we view this system as a topological order with the $\Zb_2$-symmetry realized by $U \coloneqq \medotimes_i X^i$. 
    In this case,
    each $s_{ab}$ can not be a topological excitation because 
    symmetry $U$ would flip each ground state and the domain wall in between, i.e. $U(s_{\uparrow \uparrow}) = s_{\downarrow \downarrow}, U(s_{\uparrow \downarrow}) = s_{\downarrow \uparrow}$ and vice versa. However, $$\one:=s_{\uparrow \uparrow}\oplus s_{\downarrow \downarrow}$$ is a topological excitation. 
    
    After acting topological sectors of operators $\bfone, \bfe, \bfm, \bff$ on vacuum $\one$, there emerges a non-trivial sector $$M:= s_{\uparrow \downarrow}\oplus s_{\downarrow \uparrow}.$$
    Note that here $\bfe$ only has a trivial action on the vacuum, so the morphism $u_{\one}:U(\one) := \one \to \one$ is identity.
    Similarly, $M$ has identity morphism $u_{M}:U(M) = M \to M$. 
    Thus $\{\one, u_{\one}=1\}$  and $\{M, u_{M}=1\}$ exhaust all symmetric topological excitations.
    The fusion rules are given by $\one \otimes M = M \otimes \one = M, M \otimes M = \one$, which coincides with those in $\rm Vec_{\mathbb{Z}_2}$.
    
    The above analysis is just equivariantization technique introduced in section \ref{section:equivariantization}.  As a consequence, the topological excitations $\{\one, u_{\one}=1\}$  and $\{M, u_M=1\}$ form a fusion category equivalent to the equivariantization $\Fun(\vect_{\Zb_2},\vect_{\Zb_2})^{\Zb_2} \simeq \vect_{\Zb_2}$.

    % Using the same analysis as in the SPT case, we obtain
    The space of $U$-symmetric operators  map between these two sectors of sates $\one$ and $M$ are clearly given by:
\begin{gather*}
            {\Hom}^{SB}_{Ising}(\one, \one)=\bfone\oplus \bfe\\
            {\Hom}^{SB}_{Ising}(\one, M)={\Hom}^{SB}_{Ising}(M, \one) = \bfm\oplus \bff \\
            {\Hom}^{SB}_{Ising}(M, M)=\bfone\oplus\bfe
\end{gather*}
Again, this result is just equation \eqref{SCB_hom} for $\Zb_2$ case, which exactly matches the enriched fusion category $\rm {}^{\FZ_1(Rep(\mathbb{Z}_2))}\vect_{\Zb_2}$ obtained from the canonical
construction. This convinces theorem \ref{thm_cmplbreak} in case of $G=\Zb_2$.
        \begin{thm}[\cite{KWZ22}]
            The $\Zb_2$ symmetry breaking phase of Ising chain can be described mathematically by the enriched fusion category $\rm {}^{\FZ_1(Rep(\mathbb{Z}_2))}Vec_{\mathbb{Z}_2}.$
        \end{thm}

        \begin{rem}
        From the holographic point of view, $\vect_{\Zb_2}$ is just the smooth boundary of toric code model $\CT \CC$. The trivial action of $E^{i}$ on $\lvert\Omega \rangle$ can be interpreted as a condensation of the "$\bfe$-particles" (or equivalently, the ground state algebra $A_{\bfe}=\bfone\oplus \bfe = {\Hom}^{SB}_{Ising}(\one, \one)$) in the categorical symmetry. In other words, the Lagrangian algebra is $A_{\bfe}$.
In this way we get the underlying category $\rep(\mathbb{Z}_2) \simeq \FZ_1(\rep(\mathbb{Z}_2))_{A_{\bfe}}$. See the following figure.
\end{rem}

\begin{figure}[H]
\centering
\begin{tikzpicture}[scale=0.8]
                \filldraw[fill=gray!20, draw=white] (-8,0) rectangle (-4,2);
                \draw[very thick, dashed](-8, 0)--(-4,0);
                % \filldraw[fill=white, draw=black] (-6.1,-0.1) rectangle (-5.9,0.1);
                \node[scale=0.9] at(-6,0.8){$\CT\CC := \FZ_1(\rep(\mathbb{Z}_2))$};
                % \node at(-4.3,-0.4){$\vect_{\mathbb{Z}_2}$};
                \node[scale=0.8] at(-6, -0.3){Rough bdy $\vect_{\mathbb{Z}_2}$};
        
                \draw[very thick](-1.5,0)--(2.5,0);
                % \draw[draw=none](-8,0)--(-7,0);
                % \filldraw[fill=white] (1.9,-0.1) rectangle (2.1,0.1);
                \node at(0.3,0.3){${}^{\CT\CC}\vect_{\mathbb{Z}_2}$};
                % \node at(4,0.5){${}^{\FZ_1(\rep(\mathbb{Z}_2))}\vect_{\mathbb{Z}_2}$};
        
                \node[scale=0.9] at(-6.8,1.7){\scriptsize \text{2d Toric code}};
                \node[scale=0.9] at(-0.3,1.7){\scriptsize \text{1+1D Ising chain}};
                \draw[-latex](-3,1)--(-2.4,1);
\end{tikzpicture}

\caption{By condensing $\bfone \oplus \bfe$ in 2d toric code model $\CT\CC$, we can have a 1d rough boundary $\vect_{\Zb_2}\simeq \FZ_1(\rep(\mathbb{Z}_2))_{\bfone\oplus \bfe}$. 
            After topological Wick rotation, excitations $\bfone, \bfe, \bfm, \bff \in \CT\CC$ become the topological sectors of operators, and the excitations $\one, M \in \vect_{\Zb_2}$ becomes the topological sectors of states of the symmetry breaking phase of 1+1D Ising chain. Their mapping behaviors exactly meet with enriched fusion  category $\rm {}^{\FZ_1(Rep(\mathbb{Z}_2))}\vect_{\Zb_2}$, leading to a macroscopic description of the Ising symmetry breaking phase.}
\end{figure}

    \begin{rem}
        Using method discussed in section \ref{section:boundaries}, we can interpret the $0+1$D gapped boundaries of this phases. Detailed analysis can be found in \cite{KWZ22}, here we show an alternative way of finding the topological sectors of states using equivariantization:
        \begin{itemize}
            \item If we view the Ising SPT system as 1d topological order without symmetry, which is $\vect$, its 0d open boundaries should also be $\vect$. But if we impose $\Zb_2$ symmetry on one of its boundaries, we got the sectors of states from $\vect^{\Zb_2} \simeq \rep(\Zb_2)$, similar to remark \ref{Is_SPT_eqvar}. For symmetry breaks on the boundary the topological sectors of states still form ${\vect}^{\{e\}}\simeq \vect$ 
            % \begin{figure}[H]
            %     \centering
            %     \includegraphics{figures/Ising_boundary_a.pdf}
            % \end{figure}  
            \begin{figure}[H]
                \centering
                \begin{tikzpicture}[scale=0.8]
                    \tiny
                    \node[]at(-3.6, 1.3){\scriptsize 1d topological order};
                    \node[]at(1.5, 1.3){\scriptsize Imposing symmetry};
                    \draw[very thick](-5,0)--(-1,0) ;
                    \filldraw[fill=white] (-5.1,-0.1) rectangle (-4.9,0.1);
                    \filldraw[fill=white] (-1.1,-0.1) rectangle (-0.9,0.1);
                    \node[]at(-5, 0.3){$\vect$};
                    \node[]at(-3, 0.3){$\vect$};
                    \node[]at(-1, 0.3){$\vect$};
        
                    \draw[very thick](0,0)--(4,0) ;
                    \filldraw[fill=white] (-0.1,-0.1) rectangle (0.1,0.1);
                    \filldraw[fill=white] (4.1,-0.1) rectangle (3.9,0.1);
                    \node[]at(0.7, 0.3){${\vect}^{\Zb_2} \simeq \rep(\Zb_2)$};
                    \node[]at(4, 0.3){${\vect}^{\{e\}}$};
        
                \end{tikzpicture}
            \end{figure}  
    \item Similarly, if we view the Ising symmetry breaking order as 1d topological order without symmetry, which is $\rm Fun(Vec_{\mathbb{Z}_2}, Vec_{\mathbb{Z}_2})$, its 0d open boundaries can be described by $\vect_{\Zb_2}$ with objects $s'a$ satisfying $s'_a \otimes s_{bc} = \delta_{ab},\forall a,b,c=0,1$. If we impose $\Zb_2$ symmetry on one of its boundaries, after equivariantization, we are left with only one object $(\one \oplus M, u=1)$, which is $(\vect_{\Zb_2})^{\Zb_2} \simeq \vect$.
    For symmetry breaks on the boundary the topological sectors of states still form ${\vect_{\Zb_2}}^{\{e\}}\simeq \vect_{\Zb_2}$ .
    
    % \begin{figure}[H]
    %     \centering
    %     \includegraphics{figures/Ising_boundary_b.pdf}
    % \end{figure} 
    \begin{figure}[H]
        \centering
        \begin{tikzpicture}[scale=0.8]
        \tiny
        \node[]at(-3.6, 1.3){\scriptsize 1d topological order};
        \node[]at(1.5, 1.3){\scriptsize Imposing symmetry};
        \draw[very thick](-5,0)--(-1,0);
        \node[]at(-3,0.3){$\rm Fun(Vec_{\mathbb{Z}_2}, Vec_{\mathbb{Z}_2})$};
        \filldraw[fill=white] (-5.1,-0.1) rectangle (-4.9,0.1);
        \filldraw[fill=white] (-1.1,-0.1) rectangle (-0.9,0.1);
        \node[]at(-5, 0.3){$\rm Vec_{\mathbb{Z}_2}$};
        \node[]at(-1, 0.3){$\rm Vec_{\mathbb{Z}_2}$};
    
        \draw[very thick](0,0)--(4,0) ;
        \filldraw[fill=white] (-0.1,-0.1) rectangle (0.1,0.1);
        \filldraw[fill=white] (3.9,-0.1) rectangle (4.1,0.1);
        \node[]at(0.7, 0.3){$(\rm Vec_{\mathbb{Z}_2})^{\Zb_2}\simeq \vect$};
        \node[]at(3.7, 0.3){$(\rm Vec_{\mathbb{Z}_2})^{\{e\}}$};
        \end{tikzpicture}
    \end{figure} 
         This is an example to show that equivariantization technique introduced in section \ref{section:equivariantization} is also valid in 0d. 
        \end{itemize}
    \end{rem}

\subsection{Kramers-Wannier dual of Ising chain}\label{section:KW}

One can also start from a dual model and find the exact same topological skeleton for "dual phases". For abelian cases, these dualities are related to an automorphism (or a non-trivial invertible domain wall) within the categorical symmetry. Here we show the simplest example of 1+1D transverse field Ising model.
% , general discussions are performed in outlooks \ref{section:conclusion} .

A nonlocal mapping of Pauli matrices known as the Kramers-Wannier duality transformation can be done as follows:
\begin{gather*}
    \mathbf{X}^i = Z^i Z^{i+1}\\
\mathbf{Z}^i \mathbf{Z}^{i+1} = X^{i+1}
\end{gather*}

Or we can also write $\mathbf{Z}^i = \prod_{j\leq i} X^i$. This transformation is also known as the $\Zb_2$ gauging.
The newly defined Pauli matrices obey the same algebraic relations as the original Pauli matrices, in which we denote $\lvert \uparrow_{\text{\tiny KW}}\rangle $ and $\lvert \downarrow_{\text{\tiny KW}}\rangle$ as the eigenstates of $\mathbf{Z}$, and
$\lvert +_{\text{\tiny KW}}\rangle$ and $\lvert -_{\text{\tiny KW}}\rangle$ as the eigenstates of $\mathbf{X}$.
The Hamiltonian is simply:
\be
\mathcal{H}^{KW} = -\sum_i g \mathbf{Z}^i \mathbf{Z}^{i+1}  - \sum_i J\mathbf{X}^{i+1},
\ee
in which the coupling parameter \( g \) is dual to parameter \(J\), and the critical Ising point, in which $g=J$ is a self dual point. See figure \ref{DW}.
% Their spectrum are the same. 
% There is a duality between the SPT phase and the symmetry spontaneously breaking phase. 
% As a result of these, 
The degeneracy and $\mathbb{Z}_2$ symmetry properties of the spontaneously breaking and SPT phases are changed under the Kramers-Wannier duality.

% For example, the
% critical point for the 1d Ising chain is not only $\mathbb{Z}_2$ symmetric, but also
% has an additional $\mathbb{Z}_2$ dual symmetry, denoted by $\widetilde{\mathbb{Z}_2}$, see fig.\ref{DW}.
%  [12, 36,39]'

% \begin{figure}[H]\label{DW}
%     \includegraphics[scale=0.5]{DW_phase.jpg}
% \end{figure}

The only difference is that the $\Zb_2$ onsite symmetry of Ising chain becomes
$$U=\bigotimes_{i}\mathbf{X}^i,$$
in KW model. In some literature, people understood it as $\rep (\Zb_2)$ symmetry \cite{SS24,CAW24,LL24}.

\begin{figure}[H]
\centering
\begin{tikzpicture}
        \draw[very thick](-4,0)--(4,0);
        \draw[fill=gray!70, draw=gray!70](-4,0)rectangle(0,-0.2);
        \draw[fill=gray!30, draw=gray!30](4,0)rectangle(0,0.2);
        \draw[fill=black](0,0)circle(0.1);
        \node[]at(-4.1, 0.4){$\mathcal{H}^{Ising}$};
        \node[]at(-2, 0.4){$\mathbb{Z}_2$ \text{SSB}};
        \node[]at(2, 0.4){$\mathbb{Z}_2$ \text{symmetric}};
        \node[]at(-4.1, -0.5){$\mathcal{H}^{KW}$};
        \node[]at(-2, -0.5){$\rep(\Zb_2)$ \text{symmetric}};
        \node[]at(2, -0.5){$\rep(\Zb_2)$ \text{SSB}};
        % \node[]at(4.1, -0.5){$B/J$};
\end{tikzpicture}
\caption{The $\mathbb{Z}_2$ symmetry is explicit in the $\mathcal{H}^{Ising}$ description, while dual symmetry $\rep(\Zb_2)$ is explicit in the $\mathcal{H}^{KW}$ description. These two models actually has both the $\mathbb{Z}_2$ symmetry and $\rep(\Zb_2)$ symmetry, namely, we can start from either one to find the topological sectors of operators form $\FZ_1 (\rep(\Zb_2))$. 
    % The ground state usually spontaneously breaks one of the symmetries, except at the critical point, where categorical symmetry
    % $\FZ_1 (\rep(\Zb_2))$ is still in presence.
    % both the symmetry and the dual symmetry (the full 
    % $\mathbb{Z}_2\vee \widetilde{\mathbb{Z}}_2$ 
    % symmetry) are not spontaneously broken. 
    }
\label{DW}
\end{figure}

It is clear that the topological sectors of operators are the same as Ising.
In which we have:
\begin{enumerate}
    \item $\bfone$ consists of $U$-symmetric local operators;
    \item $\bfe$=$\bfm_{\text{\tiny KW}}$ is generated by $\bfM^i = \prod_{k \leq i}\mathbf{X}^k = E^i$;
    \item $\bfm$=$\bfe_{\text{\tiny KW}}$ is generated by $\bfE^{i} = \prod_{k \geq i}(\mathbf{Z}_k \mathbf{Z}_{k+1}) = M^i$;
    \item $\bff:=\mathcal{O}_{a,-1}$ is generated by $\bfM^i \bfE^{j}$.
\end{enumerate}

Note that $\bfm_{\text{\tiny KW}}$ ($\bfe_{\text{\tiny KW}}$) in KW model corresponds to the $\bfe$ ($\bfm$) in Ising model.
This is related to the braided-equivalence ($\bfe-\bfm$ exchange) of toric code model before topological Wick rotation.

\begin{itemize}
    \item 
    In case of $J =1, g =0$, $\mathcal{H} = -\sum_i \mathbf{X}^{i+1}$, we have the trivial sector of states generated by the symmetry preserving ground state:
    $$\lvert\Omega\rangle = \lvert\cdots +_{\text{\tiny KW}}+_{\text{\tiny KW}}+_{\text{\tiny KW}} \cdots\rangle \in \one$$
    Since $\mathbf{X^i} = {Z}^i{Z}^{i+1}$, we have $\lvert \uparrow^i\uparrow^{i+1}\rangle= \lvert +_{\text{\tiny KW}}\rangle$ and $\lvert \downarrow^i\downarrow^{i+1}\rangle=\lvert +_{\text{\tiny KW}}\rangle$.
    And the non-trivial topological sector of states is generated by applying an $\bfm$ particle on site $i$:
     $$\bfE^i \lvert\Omega\rangle = \lvert\cdots +_{\text{\tiny KW}}+_{\text{\tiny KW}} -_{i \text{\tiny KW}} +_{\text{\tiny KW}}+_{\text{\tiny KW}} \cdots\rangle \in M$$
    
     In this KW SPT case the topological sectors of states form fusion category $\vect_{\Zb_2}$, and it is obvious that the enriched category should be ${}^{\FZ_1(\rep(\Zb_2))}\vect_{\Zb_2} $.

    \item For $J =0, g =1$, we have $\mathcal{H} = -\sum_i \mathbf{Z}^i \mathbf{Z}^{i+1}$, and the ground states are two fold degenerate
     \begin{align*}
         \lvert\Omega_{\uparrow_{\text{\tiny KW}}}\rangle = \lvert\cdots \uparrow_{\text{\tiny KW}}\uparrow_{\text{\tiny KW}}\uparrow_{\text{\tiny KW}} \cdots\rangle \\   \lvert \Omega_{\downarrow_{\text{\tiny KW}}} \rangle= \lvert\cdots \downarrow_{\text{\tiny KW}}\downarrow_{\text{\tiny KW}}\downarrow_{\text{\tiny KW}} \cdots \rangle
     \end{align*}
     % Here, i.e. $\mathbf{Z}\lvert \uparrow_{\text{\tiny KW}}\rangle =\lvert \uparrow_{\text{\tiny KW}}\rangle$ and $\mathbf{Z}\lvert \downarrow_{\text{\tiny KW}}\rangle =-\lvert \downarrow_{\text{\tiny KW}}\rangle $.
     Consider $\mathbf{Z}^i\mathbf{Z}^{i+1}\lvert \uparrow_{\text{\tiny KW}}^{i}\uparrow_{\text{\tiny KW}}^{i+1}\rangle =\lvert \uparrow_{\text{\tiny KW}}^{i}\uparrow_{\text{\tiny KW}}^{i+1}\rangle$, since $X^i=\mathbf{Z}^i\mathbf{Z}^{i+1}$, thus the state $\lvert \uparrow_{\text{\tiny KW}}^{i}\uparrow_{\text{\tiny KW}}^{i+1}\rangle$ should be the eigenstate of $X^i$, i.e. $\lvert \uparrow_{\text{\tiny KW}}^{i}\uparrow_{\text{\tiny KW}}^{i+1}\rangle=\lvert +_i\rangle$.
     Similarly, we should have $\lvert \downarrow_{\text{\tiny KW}}^{i}\downarrow_{\text{\tiny KW}}^{i+1}\rangle=\lvert +_i\rangle$, which concides with the ground state of the Ising SPT phase.

    Again according to the analysis in previous subsection, we can find two topological sector of states $\one$ and $E$, which form the fusion category $\rep(\Zb_2)$.
    It is also obvious that the topological skeleton of this SSB phase should be ${}^{\FZ_1(\rep(\Zb_2))}\rep(\Zb_2)$.
\end{itemize}

\begin{rem}
    These topological sectors of states' categorical descriptions should also be checkable using equivariantization. For example, the symmetry preserving case that has trivial product state before applying symmetry is just $\vect$, and we have $\vect ^{\rep (\Zb_2)} \simeq \vect_{\Zb_2}$ after imposing $\rep(\Zb_2)$ symmetry ($\rep(G)$ equivariantization is equivalent to $G$ de-equivariantization). Also for the symmetry breaking case, we have the topological sectors of states as $\fun(\rep(\Zb_2), \rep(\Zb_2)) ^{\rep (\Zb_2)} \simeq \rep (\Zb_2)$.
\end{rem}

\begin{table}[H]
    \centering
  Categorical symmetry \,  $\CT \CC := \FZ_1(\rep(\mathbb{Z}_2))  \qquad \qquad \qquad \qquad$
    \begin{tabular}{|c|c|c|c}
        \cline{1-3}
        \diagbox[width=8em, height=3em]{topo. \\ states}{lattice \\sym.} & $\vect_{\mathbb{Z}_2}$ & $\rep(\mathbb{Z}_2)$ & ~\\
        \cline{1-3}
        $\vect_{\mathbb{Z}_2}$ \rule{0pt}{12pt}& \cellcolor{gray!20} Ising SSB & \cellcolor{gray!20} KW SPT &  \cellcolor{gray!20} ${}^{\FZ_1(\rep(\mathbb{Z}_2))}\vect_{\mathbb{Z}_2}$\\[3pt]
        % \multicolumn{2}{|c|}{Ising SSB  wws} & \arraybackslash test \\
        \cline{1-3}
        \hline
        $\rep(\mathbb{Z}_2)$ \rule{0pt}{12pt} &\cellcolor{gray!20} Ising SPT &\cellcolor{gray!20} KW SSB & \cellcolor{gray!20} ${}^{\FZ_1(\rep(\mathbb{Z}_2))}\rep(\mathbb{Z}_2)$ \\[3pt]
        \cline{1-3}

    \end{tabular}
    \caption{
        The $\Zb_2$-action on transverse field Ising model can aso be viewed as $\vect_{\Zb_2}$-action.
            Under KW duality, the $\Zb_2$ onsite symmetry becomes a $\rep(\Zb_2)$ symmetry.
            The categorical symmetry is always $\FZ_1(\rep(\Zb_2))$, and spectrum in these two cases would be the same.
    An enriched category description not only describes the SPT (SSB) phase of Ising, but also the KW dual SSB (SPT) phase under duality.
            % According to \cite{JW20}, the categorical symmetry of a G-symmetric system is a combination of the symmetry and the dual symmetry.
            }
\end{table}

See section \ref{section:conclusion} for general discussion on duality and enriched categories.

\subsection{\texorpdfstring{$\Zb_n$}{Zn} clock model}
The next example is the quantum clock model, which is a generalization of the transverse-field Ising chain. It is defined on a lattice with $ n $ states on each site. The Hamiltonian of this model is typically known as \cite{ESR80,Rad18}
    % \[
    %     \mathcal{H} = -J \left(\sum_{\langle i,j \rangle} (Z^\dagger_i Z^j + Z^i Z^\dagger_j) + g \sum_j (X_j + X_j^\dagger)\right)
    % \]
\begin{equation}\label{eq:Z_n_clock}
    \mathcal{H}_{c} = - g \sum_i \left({\tilde{X}}^i+ (\tilde{X}^i)^{\dagger} \right) - J \sum_i \left({\tilde{Z}}^{i} (\tilde{Z}^{i+1})^{\dagger} + (\tilde{Z}^{i})^{\dagger} {\tilde{Z}}^{i+1} \right).
\end{equation}

The clock operators ${\tilde{X}}^i$ and ${\tilde{Z}}^{i}$ are $ n \times n $ generalizations of the Pauli matrices $X$ and $Z$ satisfying 
\begin{align*}
    \tilde{Z}^j \tilde{X}^k = \mathrm{e}^{\frac{2\pi \mathrm{i}}{n}\delta_{j,k}} \tilde{X}^k \tilde{Z}^j \quad (\tilde{X}^i)^n = (\tilde{Z}^j)^n = 1, 
\end{align*}
where $\delta_{j,k}$ is 1 if $ j $ and $ k $ are the same site and zero otherwise. 
% $ J $ is a prefactor with dimensions of energy, and $ g $ is another coupling coefficient that determines the relative strength of the external field compared to the nearest neighbor interaction.
The model obeys a global $ \mathbb{Z}_n $ symmetry, which is generated by the unitary operator $ U_X = \prod_i \tilde{X}^i $, where the product is over every site of the lattice. 
% In other words, $ U_X $ commutes with the Hamiltonian.

Now we focus on one site $i$ and omits the site index. 
We choose a basis $\lvert 0\rangle, \lvert 1\rangle,\dots,\lvert n-1\rangle$ for $\CH^i$, such that 
\begin{align*}
    \tilde{X}\lvert k\rangle=\lvert k+1\rangle \quad \tilde{Z}\lvert k\rangle =(\mathrm{e}^\frac{2\pi\mathrm{i}}{n})^k\lvert k\rangle.
\end{align*} 
It is clear that these operators satisfy the relations of generalized Pauli matrices, i.e
$\tilde{Z}\tilde{X}\lvert k\rangle =\tilde{Z} \lvert k+1\rangle=(\mathrm{e}^{\frac{2\pi\mathrm{i}}{n}})^{k+1}\lvert k+1\rangle=\mathrm{e}^{\frac{2\pi\mathrm{i}}{n}}\tilde{X}((\mathrm{e}^{\frac{2\pi\mathrm{i}}{n}})^k\lvert k\rangle)=\mathrm{e}^{\frac{2\pi\mathrm{i}}{n}}\tilde{X}\tilde{Z}\lvert k\rangle$.

The clock model 
has two gapped phases parameterized by $g$ and $J$. When $g \approx 1, J\approx 0$, it realizes the disordered phase. When $g \approx 0, J\approx 1$, it realizes the ordered phase.

To figure out the relation between the clock model and our constructed model \eqref{eq:Hamiltonian},
let $a$ be the generator of $\mathbb{Z}_n$, then basis in our model are $\lvert e\rangle,\lvert a\rangle, \lvert a^2\rangle,\dots,\lvert a^{n-1}\rangle$.
Recall that $L_{a^k}\lvert a^l\rangle=\lvert a^{k+l}\rangle$.
Under the isomorphism $\lvert k\rangle \mapsto\lvert a^{k}\rangle$, we see $\tilde{X}$ is indeed $L_a$ in our model. 
Also $\tilde{X}^{\dagger}=L_{a^{n-1}}$, since $\tilde{X}\tilde{X}^{\dagger}=\id$.
Note that the representation $\rho_a\in\rep(\mathbb{Z}_n)$ such that $\rho_a(a)=\mathrm{e}^{\frac{2\pi\mathrm{i}}{n}}$ is the generator of $\rep(\Zb_n)$.
It is not hard to see $Z_{\rho_a}\lvert a^k\rangle= \rho_a(a)^k\lvert a^k\rangle$, and hence $\tilde{Z}=Z_{\rho_a}$.
Therefore, $\tilde{Z}^{\dagger}=(Z_{\rho_a})^{n-1}$.

Note that these two models have the same global symmetry, since $\tilde{X}=L_a$.
As a consequence, the topological sectors of operators of $\Zb_n$ clock model should also be $\FZ_1(\rep(\Zb_n))$, because the $\Zb_n$ clock model has the same generator of operators as our construction.

% coincides with ours when $n=2$ and $n=3$.
When $ n = 2 $, the quantum clock model is identical to the transverse-field Ising model analyzed in the last subsection. 

\begin{expl}[clock model with $\Zb_3$ symmetry]

When $n=3$, i.e. $G=\mathbb{Z}_3:=\{e,a,a^2\}$, the clock model is also known as the three-states Potts model \cite{Wu1982, AOS98}. Hamiltonian \eqref{eq:Z_n_clock} also coincides with our construction precisely:
\begin{itemize}
    \item 
    % when we take
    % $J\gg g$ in Hamiltonian \eqref{eq:Z_n_clock}, we obtain  $\mathcal{H}_c=-\sum\limits_i \left(\tilde{X}^i+(\tilde{X}^i)^{\dagger}\right)$. 
    % On the other hand, 
    % when we consider  i.e.
    If we take $H=\Zb_3$ in Hamiltonian \eqref{eq:Hamiltonian}, we have $\mathcal{H}=\sum\limits_i \left(1-\sum\limits_{g\in \Zb_3}L_g^i\right)$.
    Note that $L_e=1$, so we have $\mathcal{H}^i=-L_{a}^i-L_{a^2}^i=-\tilde{X}^i-(\tilde{X}^i)^{\dagger}=\mathcal{H}^i_c$ on each site. This reveals $g\gg J$ in Hamiltonian \eqref{eq:Z_n_clock}
    This shows that the trivial $\Zb_3$ SPT phase realized in \eqref{eq:Hamiltonian} is just the disordered phase of quantum clock model.

    \item 
    % when we take $J\ll g$ in Hamiltonian \eqref{eq:Z_n_clock}, we obtain $\mathcal{H}_c=-\sum\limits_i\left({\tilde{Z}}^{i} (\tilde{Z}^{i+1})^{\dagger} + (\tilde{Z}^i)^{\dagger} {\tilde{Z}}^{i+1} \right)$. On the other hand, 
    If we take $H=\{e\}$ in Hamiltonian \eqref{eq:Hamiltonian}, we have $\mathcal{H}= \sum\limits_i (1 - \sum\limits_{\rho\in\hat{\Zb_3}}Z_{\rho}^i(Z_{\rho}^{i+1})^{\dagger}) = \mathcal{H}_c$ for $g\ll J$.   
    % These two terms coincide.
    This show that the symmetry breaking phase realized in \eqref{eq:Hamiltonian} is just the ordered phase of quantum clock model.
 
\end{itemize}
    
% Moreover when $ n = 3 $, the quantum clock model is equivalent to the quantum three-state Potts model, we would illustrate why we have this situation from categorical perspective in the next subsection.

The process of finding the macroscopic observables for $n=3$ has no difference from section \ref{section:lattice_model} for $G=\Zb_3$,
%  we still explicitly write down these sectors and states
% the topological sectors of operators can be written down directly:
% \begin{itemize}
%     \item the trivial sector $\bfone:=\mathcal{O}_{e,1}$ consists of $U$-symmetric local operators,
%     \item sectors generated by $M^i:=\prod_{k\leq i} \tilde{X}^k$, $(M^i)^2=\prod_{k\leq i} \tilde{X}^k^{\dagger}$,
%     \item sectors generated by $E^i:=\prod_{k\geq i} \tilde{Z}_k\tilde{Z}_{k+1}^{\dagger}$, $(E^i)^2:=\prod_{k \geq i} \tilde{Z}^{\dagger}_k\tilde{Z}_{k+1}$,
%     \item and sectors generated by $E^i$ and $M^i$: $E^i M^i$, $E^i (M^i)^2$, $(E^i)^2 M^i$, $(E^i)^2 (M^i)^2$.
% \end{itemize}
In which we have the enriched fusion category ${}^{\FZ_1(\rep(\Zb_3))}\rep(\Zb_3)$ to describe the disordered phase and 
${}^{\FZ_1(\rep(\Zb_3))}\vect_{\Zb_3}$ to describe the ordered phase.

\end{expl}

\begin{expl}[clock model with $\Zb_p$ symmetry]
    When $n=p$ for $p$ a prime number, even though Hamiltonian \eqref{eq:Z_n_clock} lack some operators comparing to our construction, they still realize the same phases (which are described by the same topological skeletons ${}^{\FZ_1(\rep(\Zb_p))}\rep(\Zb_p)$ and ${}^{\FZ_1(\rep(\Zb_p))}\vect_{\Zb_p}$.
    \begin{itemize}
        \item For $g\gg J$ in Hamiltonian \eqref{eq:Z_n_clock}, $\mathcal{H}^i_c = -\tilde{X}^i-(\tilde{X}^i)^{\dagger}\neq \mathcal{H}^i=-\sum\limits_{k=1}^{p-1}L_{a^k}^i$ on each site.
        However, the ground state is only controlled by generator of operators $\tilde{X}=L_a$ for prime cases. 
        % other operators are redundant.
        Thus, even though $\mathcal{H}_c$ has less operators compare to $\mathcal{H}$, these two Hamiltonians again has the same ground state $\lvert \Omega_{\Zb_p}\rangle$.
        And we can obtain the same topological sector of states by applying the topological sectors of operators in $\FZ_1(\rep(\Zb_p))$: trivial sector $\one$ generated by $\lvert \Omega_{\Zb_p}\rangle$ and sector $E^k$ generated by states $E^k\lvert\Omega_{\Zb_p} \rangle$, $\forall k \in \mathbb{Z}_p$, whose non-trivial sites are given by
            \begin{align*}
                \frac{1}{\sqrt{p}}\left(\sum\limits_{k=0}^{p-1}\mathrm{e}^{\frac{2\pi\mathrm{i}k}{n}}\lvert a^k\rangle_i\right), 
            \end{align*}
            respectively.
                    
The sectors of states form fusion category $\rep(\Zb_p)$. The topological skeleton of this phase is ${}^{\FZ(\rep(\Zb_p))} \rep(\Zb_p)$, in which the trivial $\Zb_p$ SPT phase realized in \eqref{eq:Hamiltonian} is just the disordered phase of p-states quantum clock model.

        \item when we take $g\ll J$ in Hamiltonian \eqref{eq:Z_n_clock}, 
        % again we obtain $\mathcal{H}= \sum\limits_i \left(1 - \sum\limits_{\rho\in\hat{\Zb_4}}Z_{\rho}^i(Z_{\rho}^{i+1})^{\dagger}\right) \neq c_c$.
        % These two terms coincide.
        The p-fold degenerate ground states $\{\lvert \psi_{a^k}\rangle \mid a^k\in\Zb_p\}$ of $\mathcal{H}_c$ (recall that $\lvert \psi_g\rangle:=\bigotimes_i \lvert g\rangle_i$) are the same as those of $\mathcal{H}_c$. According to a similar argument as the above case, the symmetry breaking phase realized in \eqref{eq:Hamiltonian} is just the ordered phase of quantum clock model. And the topological skeleton of this phase is ${}^{\FZ(\rep(\Zb_p))} \vect_{\Zb_p}$, where the sectors of states form fusion category $\vect_{\Zb_p}$. See the following picture.
        
        % We can check this using equivariantization:
        % before imposing $\Zb_4$ symmetry, the topological sector of states are generated by $\lvert \psi_{g,h,i}\rangle $ for all $g,h\in\Zb_4$, which form the category $\Fun(\vect_{\Zb_4},\vect_{\Zb_4})$.
        % The $\Zb_4$-equivariantization of this category is indeed $\vect_{\Zb_4}$.
    \end{itemize}
\end{expl}
\begin{figure}[H]
    \centering
    \begin{tikzpicture}[scale=0.7]
        \tiny
        \filldraw[fill=gray!20, draw=white] (-8,0) rectangle (-4,2);
        \draw[very thick](-8, 0)--(-4,0);
        \filldraw[fill=white, draw=black] (-6.1,-0.1) rectangle (-5.9,0.1);
        \node[] at(-6,0.8){$\FZ_1(\rep(\Zb_p))$};
        \node at(-4.3,-0.4){$\vect_{\Zb_p}$};
        \node at(-7.5, -0.4){$\rep(\Zb_p)$};

        \draw[very thick](-1.2,0)--(3,0);
        \draw[draw=none](-9,0)--(-8,0);
        \filldraw[fill=white] (0.7,-0.1) rectangle (0.9,0.1);
        \node at(-1,0.5){${}^{\FZ_1(\rep(\Zb_p))}\rep(\Zb_p)$};
        \node at(2.3,0.5){${}^{\FZ_1(\rep(\Zb_p))}\vect_{\Zb_p}$};

        % \node at(-7,1.7){\text{2d Toric code}};
        \node at(0.4,1.7){\text{1+1D $\Zb_p$ clock model}};
        \draw[-latex](-3.5,1)--(-2.9,1);
        
    \end{tikzpicture}
\end{figure}

\begin{expl}[clock model with $\Zb_4$ symmetry]
  But the lacking of operators in Hamiltonian \eqref{eq:Z_n_clock} would result in the lacking of 
% But things are different for $G = \Zb_4$ and other non-prime cases, in which we have 
partially symmetry breaking phases comparing with our Hamiltonian \eqref{eq:Hamiltonian} for $G = \Zb_4$ and other non-prime cases.
% Again the Hamiltonian \eqref{eq:Z_n_clock} lack some operators comparing to our model, 
This is because for non-prime cases, not only the generator of operators $\tilde{X}=L_a$ determines the ground state, but also some generators of subgroup operators 
(e.g. $L_{a^2}$ can generate $\Zb_2$ in $\Zb_4$ case).
Indeed, in $\Zb_n$-clock model, the clock model minimally add operators (i.e. Hermitian conjugation of $\tilde{X}$ and $\tilde{Z}$) in Hamiltonian such that it is well-defined.
Our model admit all possible operators that are related to the symmetry.

We illustrate the phases in case of $\Zb_4$:

\begin{itemize}
    \item When we take $g\gg J$ ($g\ll J$) in Hamiltonian \eqref{eq:Z_n_clock}, i.e. the disordered (ordered) phase of $\Zb_4$ clock model, 
    the ground state(s) is again determined by the generator of operators $\tilde{X}=L_a$ ($\tilde{Z}=Z_{\rho_a})$.
    So the analysis of $\Zb_p$ case still work, and we again obtain the enriched category descriptions as ${}^{\FZ(\rep(\Zb_4))} \rep(\Zb_4)$ (${}^{\FZ(\rep(\Zb_4))} \vect_{\Zb_4}$).

    \item Note that the our model \eqref{eq:Hamiltonian} admits a $\Zb_2$ partially symmetry breaking phase if we set $H=\Zb_2 \subseteq \Zb_4$.
    The corresponding Hamiltonian is {\footnotesize
        \begin{align*} \mathcal{H}=\sum\limits_i \left(1-\frac{1}{2}\sum\limits_{g\in \Zb_2}L_g^i\right) -\sum\limits_i \left(1 - \frac{1}{2} \sum\limits_{\rho\in\hat{\Zb_2}}Z_{\rho}^i(Z_{\rho}^{i+1})^{\dagger}\right),
        \end{align*}
    }and the ground states are $\{\bigotimes_i(\lvert e \rangle_i + \lvert a^2 \rangle_i),\bigotimes_i (\lvert a\rangle_i +\lvert a^3\rangle_i)\}$, the topological sectors of states form ${}_{F_{\Zb_2}}\rep(\Zb_4)_{F_{\Zb_2}}\simeq \rep(\Zb_2)\boxtimes \vect_{\Zb_2}$. See the dashed line part in figure \ref{fig:Z_4}.
    However, $\mathcal{H}_c$ cannot be tuned to create such ground state, since $\mathcal{H}_c$ miss operators $L_{a^2}$ and $Z_{\rho_a^2}$.
    Hence the clock model \eqref{eq:Z_n_clock} cannot realize this $\Zb_2$ partially symmetry breaking phase. However this gapped phase is "hidden" inside the gapless phase transition point for $g=J$, see outlook \ref{section:conclusion} for some discussions.

    % Note that Lagrangian algebra just corresponds to the Hamiltonian, which is $\bfone \oplus {\bfe}^2 \oplus {\bfm}^2 \oplus {\bfe \bfm}^2$.
 
\end{itemize}

\begin{figure}[H]
\centering
\begin{tikzpicture}
        \filldraw[draw=none, fill=gray!20](-2,0) rectangle (2,2);
        \draw[very thick](-2,0)--(-0.7,0);
        \draw[very thick](0.7,0)--(2,0);
        \draw[](-0.7,0)--(0.7,0);
        \draw[dashed](-0.9,-0.2)rectangle(0.9,0.2);
        \node at(0, 1){$\FZ(\rep(\mathbb{Z}_4))$};
        \filldraw[fill=white](-0.6,-0.1) rectangle (-0.8,0.1);
        \filldraw[fill=white](0.6,-0.1) rectangle (0.8,0.1);
        \node at(-1.4, -0.3){$\rep(\mathbb{Z}_4)$};
        \node at(1.4, -0.3){$\vect_{\mathbb{Z}_4}$};
\end{tikzpicture}
\caption{The phase transition point between symmetry preserving case and symmetry completely broken phase can be understood in two ways: one is to completely break $G$ (in this case $\Zb_4$), other one is to first partially break to some subgroup $H \subseteq G$ (here $\Zb_2$), then break $H$ again.}
\label{fig:Z_4}
\end{figure}

% Indeed, for $n\geq 4$, the $\Zb_n$-clock model also has only two phases: disordered and ordered phase, which correspond to trivial SPT and SSB phase in our model.
% The disappearance of partially broken phases in $\Zb_n$-clock model is due to the lacking of $G$-symmetric operators in Hamiltonian \eqref{eq:Z_n_clock}.

% \begin{rem}
%     However, the $\Zb_n$-clock model can be tuned to realize the phase transition between the ordered phase and disordered phase.
%     To realize phase transition in our model, we need to add some parameters in Hamiltonian that is related to the subgroup $H\subseteq G$.
% \end{rem}

% When $ n = 4 $, the model is again equivalent to the Ising model. When $ n> 4 $, strong evidence has been found that the phase transitions exhibited in these models should be certain generalizations of Kosterlitz-Thouless transition, whose physical nature is still largely unknown.

\end{expl}

\subsection{\texorpdfstring{$\Zb_2 \times \Zb_2$}{Z2Z2} SPT with non-trivial two-cocycle} \label{sec_Z2Z2_SPT}

\subsubsection{Cluster state}

When $G=\Zb_2 \times \Zb_2 \coloneqq \{e,a,b,ab\}$, it has five subgroups $\{e\}$, $\Zb_2 \times \{e\}$, $\{e\} \times \Zb_2$, $\Zb_2^f \coloneqq \{e,ab\}$ and $\Zb_2 \times \Zb_2$. They correspond to the symmetry completely broken phase, three $\Zb_2$ partially symmetry breaking phases and two SPT phases respectively. The description of four symmetry breaking phases and trivial SPT phase can be analyzed directly by the method introduced in section \ref{section:general_cases}, so we omit here.

Now we focus on the non-trivial 1+1D SPT order, 
% which is also the simplest non-trivial SPT phase. 
% we can directly use analysis in section \ref{section:general_cases} to find the enriched categorical descriptions of these phases.
which can be understood as the non-trivial $\Zb_2\times \Zb_2$-SPT phase of the Haldane chain, or the cluster model \cite{PBTO12,CLV14}. 
% also known as the Haldane chain realizing the non-trivial $\Zb_2 \times \Zb_2$ SPT order.

\medskip
First we consider the cluster model. The Hilbert space of the cluster model is $\CH=\bigotimes_i\CV_i$ where $\CV_i \coloneqq \mathbb{C}^2$ and the Hamiltonian is given by
\begin{align}\label{eq:Hamil_cluster}
        \mathcal{H}=-\sum_i Z^{i-1} X^i Z^{i+1},
\end{align}
here $X^i$ and $Z^i$ are Pauli matrices acting on each site. And the onsite $\Zb_2\times \Zb_2$ symmetry is given by $U_e=\bigotimes_i X^{2i}$ and $U_o=\bigotimes_i X^{2i+1}$, where the subscripts $_e$ and $_o$ denote the even and odd sites, respectively.

We define $\CH_i \coloneqq \CV_{2i-1} \otimes \CV_{2i}$, then there is an isomorphism of $\Zb_2 \times \Zb_2$-representations
\begin{gather*}
    \Cb[\Zb_2 \times \Zb_2] = \CH_i \simeq \CV_{2i-1} \otimes \CV_{2i} \\
    \lvert e \rangle \mapsto \lvert \uparrow \rangle \otimes \lvert \uparrow \rangle \quad
    \lvert a \rangle \mapsto \lvert \downarrow \rangle \otimes \lvert \uparrow \rangle\\
    \lvert b \rangle \mapsto \lvert \uparrow \rangle \otimes \lvert \downarrow \rangle \quad 
    \lvert ab \rangle \mapsto \lvert \downarrow \rangle \otimes \lvert \downarrow \rangle
\end{gather*}
where the $\Zb_2 \times \Zb_2$-action on $\CH_i$ is given by
\[
a \mapsto X_{2i-1} , \quad b \mapsto X_{2i} .
\]
Therefore, the Hilbert space and the $G$-action of the cluster model are the same as those in Section \ref{section:lattice_model}. So the categorical symmetry, i.e., the category of topological sectors of operators, is still $\FZ_1(\rep(\Zb_2 \times \Zb_2))$. Under the isomorphism $\CH_i \simeq \CV_{2i-1} \otimes \CV_{2i}$, the operators $L_a$ and $L_b$ are identified with $X_{2i-1}$ and $X_{2i}$, respectively. The dual group of $\Zb_2 \times \Zb_2$ is also isomorphic to $\Zb_2 \times \Zb_2$, and we denote the two generators by $\phi_a$ and $\phi_b$, where $\phi_b$ maps $a$ to $-1$ and $b$ to $1$, and $\phi_a$ maps $a$ to $1$ and $b$ to $-1$. Then the operators $Z_{\phi_b}^i$ and $Z_{\phi_a}^i$ are identified with $Z^{2k-1}$ and $Z^{2k}$. Hence, the topological sectors of operators are generated by the following symmetric non-local operators:
\begin{itemize}
\item $\bfe_1$: generated by $E_{\phi_b}^i \coloneqq \prod_{k\geq i}Z^{2k-1} Z^{2k+1} = Z^{2i-1}$,
\item $\bfe_2$ generated by $E_{\phi_a}^i \coloneqq \prod_{k\geq i}Z^{2k} Z^{2k+2} = Z^{2i}$;
\item $\bfm_1$ generated by $M^i_{a} \coloneqq \prod_{k<i} L_a^k = \prod_{k<i} X^{2k-1}$,
\item $\bfm_2$ generated by  $M^i_{b} \coloneqq \prod_{k<i} L_b^k = \prod_{k<i} X^{2k}$;
\item The simple topological sectors of operators are generated by the composition of above four kinds of symmetric non-local operators, e.g. $\bfe_1\bfm_1 \eqqcolon \bff_1, \bfe_2\bfm_2 \eqqcolon \bff_2 , \bfe_1 \bfe_2 \bfm_2$.
\end{itemize}

As we have seen in Section \ref{section:lattice_model}, the topological sectors of operators that act on the (topological sector of the) ground state invariably form a Lagrangian algebra in the categorical symmetry. In the cluster model, the ground state $\lvert \Omega \rangle$ is the common eigenstate of operators $Z^{i-1} X^i Z^{i+1}$ for all $i$ with eigenvalue $1$. Therefore, an operator acts on $\lvert \Omega \rangle$ invariably if and only if it commutes with $Z^{i-1} X^i Z^{i+1}$ for all $i$. For example, the operator $Z^{2k}$ does not commute with $Z^{2i-1} X^{2i} Z^{2i+1}$, thus $\bfe_2$ does not act on the ground state invariably. It is not hard to see that the only simple topological sectors of operators that act on the ground state invariably are $\bfone,\bfe_1 \bfm_2,\bfe_2 \bfm_1,\bff_1 \bff_2$. Therefore, the ground state algebra is the Lagrangian algebra $\one \oplus \bfe_1 \bfm_2 \oplus \bfe_2 \bfm_1 \oplus \bff_1 \bff_2$. The Lagrangian algebras in $\FZ_1(\rep(\Zb_2 \times \Zb_2))$ are listed in Example \ref{expl_Lagrangian_algebra_Z2Z2}, and we see that the ground state algebra corresponds to the nontrivial cohomology class $[\omega] \in H^2(\Zb_2 \times \Zb_2;U(1)) \simeq \Zb_2$. In other words, the ground state algebra is $A(\Zb_2 \times \Zb_2,\omega)$. This suggests that the cluster model realizes the nontrivial $\Zb_2 \times \Zb_2$ SPT phase. Thus we obtain the following result.

\begin{pthm}
The non-trivial $\Zb_2 \times \Zb_2$ symmetry protected topological order can be described by the enriched fusion category $\bc[\FZ_1(\rep(\Zb_2 \times \Zb_2))]{\FZ_1(\rep(\Zb_2 \times \Zb_2))_{A(\Zb_2 \times \Zb_2,\omega)}}$.
\end{pthm}

\subsubsection{Boundaries of Haldane chain}

To study the 0+1D boundary theories of this phase, it is simpler to consider the Haldane chain. First we note that there is a unique projective representation $(W,\rho)$ associated to $[\omega] \in H^2(\Zb_2 \times \Zb_2;U(1))$, where $W = \Cb^2$ with the basis $\{\lvert \uparrow \rangle,\lvert \downarrow \rangle\}$ and $\rho \colon \Zb_2 \times \Zb_2\to \mathrm{GL}(W)$ is given by
\begin{gather*}
    \rho(e) =\begin{pmatrix}
        1 &0\\
        0 &1
    \end{pmatrix} \quad 
    \rho(a)=\begin{pmatrix}
        0 &1\\
        1 &0
    \end{pmatrix}\\
    \rho(b)=\begin{pmatrix}
        1 &0\\
        0 &-1
    \end{pmatrix} \quad 
    \rho(ab)=\begin{pmatrix}
        0 &1\\
        -1 &0
    \end{pmatrix}
\end{gather*}
The local Hilbert space of the Haldane chain is $\CH_i \coloneqq \Cb[\Zb_2 \times \Zb_2]$. We also have an isomorphism of $\Zb_2 \times \Zb_2$-representations
\begin{gather*}
    \Cb[\Zb_2 \times \Zb_2] = \CH_i \simeq  W_{2i-1} \otimes W_{2i} \\
    \lvert e \rangle \mapsto \lvert \uparrow \rangle \otimes \lvert \uparrow \rangle + \lvert \uparrow \rangle \otimes \lvert \downarrow \rangle \quad
    \lvert a \rangle \mapsto \lvert \downarrow \rangle \otimes \lvert \downarrow \rangle + \lvert \downarrow \rangle \otimes \lvert \uparrow \rangle \\
    \lvert b \rangle \mapsto \lvert \uparrow \rangle \otimes \lvert \uparrow \rangle - \lvert \uparrow \rangle \otimes \lvert \downarrow \rangle \quad 
    \lvert ab \rangle \mapsto \lvert \downarrow \rangle \otimes \lvert \downarrow \rangle - \lvert \downarrow \rangle \otimes \lvert \uparrow \rangle
\end{gather*}
In other words, under this isomorphism, the operator $L_g^i$ is identified with $\rho(g)_{2i-1} \rho(g)_{2i}$ for all $g \in \Zb_2 \times \Zb_2$. Also, it is not hard to see that the operator $Z_{\phi_b}^i$ and $Z_{\phi_a}^i$ are identified with $\rho(b)_{2i-1}$ and $\rho(a)_{2i}$, respectively. 
% acting on $\Cb[\Zb_2 \times \Zb_2]$ is defined by
%\[
%\lvert e \rangle \mapsto \lvert e \rangle , \quad \lvert a \rangle \mapsto - \lvert a \rangle , \lvert b \rangle \mapsto \lvert b \rangle , \quad \lvert ab \rangle \mapsto - \lvert ab \rangle .
%\]
%under the isomorphism $\Cb[\Zb_2 \times \Zb_2] \simeq W \otimes W$, it acts on $W \otimes W$ by
%\begin{align*}
%\lvert \uparrow \rangle \otimes \lvert \uparrow \rangle & \mapsto \lvert \uparrow \rangle \otimes \lvert \uparrow \rangle \\
%\lvert \uparrow \rangle \otimes \lvert \downarrow \rangle & \mapsto \lvert \uparrow \rangle \otimes \lvert \downarrow \rangle \\
%\lvert \downarrow \rangle \otimes \lvert \uparrow \rangle & \mapsto - \lvert \downarrow \rangle \otimes \lvert \uparrow \rangle \\
%\lvert \downarrow \rangle \otimes \lvert \downarrow \rangle & \mapsto - \lvert \downarrow \rangle \otimes \lvert \downarrow \rangle
%\end{align*}
Therefore, the topological sectors of operators in the Haldane chain are generated by the following symmetric non-local operators:
\begin{itemize}
\item $\bfe_1$: generated by $E_{\phi_b}^i \coloneqq \prod_{k \geq i} Z_{\phi_b}^k Z_{\phi_b}^{k+1} = Z_{\phi_b}^i = \rho(b)_{2i-1} $,
\item $\bfe_2$ generated by $E_{\phi_a}^i \coloneqq \prod_{k \geq i} Z_{\phi_a}^k Z_{\phi_a}^{k+1} = Z_{\phi_a}^i = \rho(a)_{2i}$;
\item $\bfm_1$ generated by $M^i_{a} \coloneqq \prod_{k<i} L_a^k = \prod_{k<i} \rho(a)_{2k-1} \rho(a)_{2k}$,
\item $\bfm_2$ generated by  $M^i_{b} \coloneqq \prod_{k<i} L_b^k = \prod_{k<i} \rho(b)_{2k-1} \rho(b)_{2k}$.
\end{itemize}
Note that $\rho(b)_{2i}$ is also contained in $\bfe_1$ because $\rho(b)_{2i-1} \rho(b)_{2i} = L_b^i$ is a symmetric local operator. Similarly, $\rho(a)_{2i-1}$ is contained in $\bfe_2$.

Define an operator $P_{i,i+1}$ acting on $\CH_i \otimes \CH_{i+1}$ by
\begin{multline*}
P_{i,i+1} = \bigl( \CH_i \otimes \CH_{i+1} = W \otimes W \otimes W \otimes W \\
\xrightarrow{1 \otimes X_G \otimes 1} W \otimes W \otimes W \otimes W = \CH_i \otimes \CH_{i+1} \bigr) ,
\end{multline*}
where $X_G = (1 + L_a + L_b + L_{ab})/4$ is the projector to the state $\lvert e \rangle + \lvert a \rangle + \lvert b \rangle + \lvert ab \rangle$. Under the isomorphism $\Cb[\Zb_2 \times \Zb_2] \simeq W \otimes W$, it is the projector to $\lvert \uparrow \rangle \otimes \lvert \uparrow \rangle + \lvert \downarrow \rangle \otimes \lvert \downarrow \rangle$. The Hamiltonian is defined by
\[
\mathcal H = -\sum_i P_{i,i+1} .
\]

Similar to the cluster model, the ground state of the Haldane chain is the common eigenstate of $P_{i,i+1}$ for all $i$ with eigenvalue $1$. If we take the total Hilbert space to be $\bigotimes_i W_{2i} \otimes W_{2i+1}$, the ground state is the `tensor product state' of $\lvert \uparrow \rangle \otimes \lvert \uparrow \rangle + \lvert \downarrow \rangle \otimes \lvert \downarrow \rangle$ on $W_{2i} \otimes W_{2i+1}$. So the only simple topological sectors of operators that act on the ground state invariably (i.e., commute with $P_{i,i+1}$) are $\bfone,\bfe_1 \bfm_2,\bfe_2 \bfm_1,\bff_1 \bff_2$. Thus we obtain the ground state algebra $A(\Zb_2 \times \Zb_2,\omega) = \bfone \oplus \bfe_1 \bfm_2 \oplus \bfe_2 \bfm_1 \oplus \bff_1 \bff_2$ again, which suggests that the Haldane chain realizes the nontrial $\Zb_2 \times \Zb_2$ SPT phase.

\medskip
Now let us study the boundary theories of the Haldane chain. We only focus on the symmetry preseving boundary conditions.

The most obvious symmetry preserving boundary is given by the Hilbert space $\bigotimes_{i \geq 1} \CH_i$ and the Hamiltonian $H = - \sum_{i \geq 1} P_{i,i+1}$ (see the following figure, where the big circle denotes the interaction $P_{i,i+1}$). 
% \begin{figure}[H]
% \centering
% \includegraphics{figures/Haldane_boundary_a.pdf}
% \end{figure}
\begin{figure}[H]
    \centering
    \begin{tikzpicture}[scale=0.9]

        % Ellipses and circles
        % \draw (0,0) ellipse (1cm and 0.6cm);
        % \filldraw[fill=gray!20] (-0.6,0) circle (0.29cm);
        \filldraw[fill=gray!20] (0.6,0) circle (0.29cm);
        
        \draw (2.4,0) ellipse (1cm and 0.6cm);
        \filldraw[fill=gray!20] (1.8,0) circle (0.29cm);
        \filldraw[fill=gray!20] (3.0,0) circle (0.29cm);
        
        \draw (4.8,0) ellipse (1cm and 0.6cm);
        \filldraw[fill=gray!20] (4.2,0) circle (0.29cm);
        \filldraw[fill=gray!20] (5.4,0) circle (0.29cm);

        \node at(6.7, 0){$\cdots$};
        
        % Labels
        % \node at (-0.6,0) {\tiny 2$i$-2};
        \node at (0.6,0) {\tiny 1};
        \node at (1.8,0) {\tiny 2};
        \node at (3.0,0) {\tiny 3};
        \node at (4.2,0) {\tiny 4};
        \node at (5.4,0) {\tiny 5};

        \draw[-] (1,0) -- (1.4,0);
        \draw[-] (3.4,0) -- (3.8,0);
        % \draw[-] (-1,0) --(-1.4,0);
        \draw[-] (5.8,0)--(6.2,0);

        % \draw[-] (-0.2,0) --(0.2,0);
        \draw[-] (2.2,0) --(2.6,0);
        \draw[-] (4.6,0) --(5,0);

    \end{tikzpicture}
\end{figure}
The ground state is two-fold degenerate, and if we identify the Hilbert space with $\bigotimes_{i \geq 1} W_{2i-1} \otimes W_{2i}$, the ground state subspace can be identified with $W_1$ because the projectors $P_{i,i+1}$ only act on the other local Hilbert spaces. This suggests that the category of symmetry preserving boundary conditions is equivalent to $\rep(\Zb_2 \times \Zb_2,\omega)$ and $W_1$ corresponds the unique simple objects. This category can also be obtained from equivariantization $\vect^{\Zb_2 \times \Zb_2} \simeq \rep(\Zb_2 \times \Zb_2,\omega)$, where the $\Zb_2 \times \Zb_2$-action on $\vect$ is twisted by $\omega$ because the boundary Hilbert space $W_1$ is a projective representation.

Also, on this boundary, the operators $M_g^i = \prod_{k \leq i} L_g^k$ for all $g \in \Zb_2 \times \Zb_2$ is the product of finitely many symmetric local operators. So they are local on this boundary. These operators form the topological sectors of operators $\bfone \oplus \bfm_1 \oplus \bfm_2 \oplus \bfm_1 \bfm_2$, which is the Lagrangian algebra $A(\Zb_2 \times \Zb_2) \in \FZ_1(\rep(\Zb_2 \times \Zb_2))$. Therefore, the categorical symmetry on the symmetry preserving boundary is given by
\[
\FZ_1(\rep(\Zb_2 \times \Zb_2))_{A(\Zb_2 \times \Zb_2)} \simeq \rep(\Zb_2 \times \Zb_2) .
\]
Thus, the topological skeleton of this boundary should be described by $\bc[\rep(\Zb_2\times \Zb_2)]{\rep(\Zb_2\times \Zb_2,\omega)}$.

There is also an unusual symmetry preserving boundary, called the $\omega$-twisted symmetry preserving boundary. The Hilbert space is given by $W_0 \otimes \bigotimes_{i \geq 1} \CH_i = \bigotimes_{i \geq 0} W_i$. The $\Zb_2 \times \Zb_2$ action is given by $U(g) \coloneqq \rho(g)_0 \prod_{i \geq 1} L_g^i$. The Hamiltonian is given by $H = - \sum_{i \geq 0} P_{i,i+1}$, where $P_{0,1}$ is the projector $X_G$ acting on $W_0 \otimes W_1 \simeq \Cb[\Zb_2 \times \Zb_2]$. The following figure depicts this $\omega$-twisted symmetry preserving boundary.
% \begin{figure}[H]
% \centering
% \includegraphics{figures/Haldane_boundary_b.pdf}
% \end{figure}
\begin{figure}[H]
    \centering
\begin{tikzpicture}[scale=0.9]

    % Ellipses and circles
    \draw (0,0) ellipse (1cm and 0.6cm);
    \filldraw[fill=gray!20] (-0.6,0) circle (0.29cm);
    \filldraw[fill=gray!20] (0.6,0) circle (0.29cm);

    \draw (2.4,0) ellipse (1cm and 0.6cm);
    \filldraw[fill=gray!20] (1.8,0) circle (0.29cm);
    \filldraw[fill=gray!20] (3.0,0) circle (0.29cm);

    \draw (4.8,0) ellipse (1cm and 0.6cm);
    \filldraw[fill=gray!20] (4.2,0) circle (0.29cm);
    \filldraw[fill=gray!20] (5.4,0) circle (0.29cm);

    \node at(6.7, 0){$\cdots$};

    % Labels
    \node at (-0.6,0) {\tiny 0};
    \node at (0.6,0) {\tiny 1};
    \node at (1.8,0) {\tiny 2};
    \node at (3.0,0) {\tiny 3};
    \node at (4.2,0) {\tiny 4};
    \node at (5.4,0) {\tiny 5};

    \draw[-] (1,0) -- (1.4,0);
    \draw[-] (3.4,0) -- (3.8,0);
    % \draw[-] (-1,0) --(-1.4,0);
    \draw[-] (5.8,0)--(6.2,0);

    \draw[-] (-0.2,0) --(0.2,0);
    \draw[-] (2.2,0) --(2.6,0);
    \draw[-] (4.6,0) --(5,0);

    \end{tikzpicture}
\end{figure}
Note that this lattice model is the same as the symmetric boundary of the trivial SPT phase studied in Section \ref{sec_boundary_trivial_SPT} (but the symmetry is different). Thus the category of the topological sectors of states is $\rep(\Zb_2 \times \Zb_2)$. However, the categorical symmetry of this boundary is different. For example, the operator $M_a^i = \prod_{k \leq i} L_a^k$ is not symmetric because $L_a^0 = \rho(a)_0$ does not commute with $U(b)$, but the operator $E_{\phi_a}^0 M_a^i = \prod_{k=1}^{2i} L_a^k$ is a symmetric local operator. Then we find the topological sector of operators that become local on this unusual boundary is $\bfone \oplus \bfe_1 \bfm_2 \oplus \bfe_2 \bfm_1 \oplus \bff_1 \bff_2$, which is the Lagrangian algebra $A(\Zb_2 \times \Zb_2,\omega)$. Hence, the categorical symmetry on the $\omega$-twisted symmetry preserving boundary is given by
\[
\FZ_1(\rep(\Zb_2 \times \Zb_2))_{A(\Zb_2 \times \Zb_2,\omega)} .
\]
Thus, the topological skeleton of this boundary should be described by $\bc[\FZ_1(\rep(\Zb_2 \times \Zb_2))_{A(\Zb_2 \times \Zb_2,\omega)}]\FZ_1(\rep(\Zb_2 \times \Zb_2))_{A(\Zb_2 \times \Zb_2,\omega)}$.

\section{Ground state algebra as the ground state subspace on a circle}\label{sec:GS_algebra}

In this section we discuss the notion of the ground state algebra and show that it is indeed the ground state subspace on a circle $S^1$ (with all possible twisted periodic boundary conditions). This is also the reason why we call it the ground state algebra.

Recall the ground state algebra $A$ of a 1+1D quantum liquid phase $\bc[\CB]{\CS}$ is defined in Section \ref{sec_phase_condenstion} as the operators that acts on the topological sector of the ground state invariably. The holographical duality (or the intuition of topological Wick rotation) and the anyon condensation theory implies that $A$ is a Lagrangian algebra in the categorical symmetry $\CB$ and the fusion category $\CS$ is equivalent to the module category $\CB_A$.

In the anyon condensation theory, it is known that the Lagrangian algebra $A$ can be obtained by shrinking a hole in the topological order $\CB$ with boundary $\CS$ \cite{Kon14,AKZ17,KZZZ24} (see the following figure).

\begin{figure}[H]
\centering
\[
\begin{array}{c}
\begin{tikzpicture}
\fill[gray!20] (0.5,0.5) rectangle (2.5,2.5) node[black,below left] {$\CB$} ;
\fill (1.5,1.5) circle (0.08) node[above] {$A$} ;
\end{tikzpicture}
\end{array}
=
\begin{array}{c}
\begin{tikzpicture}
\fill[gray!20] (0.5,0.5) rectangle (2.5,2.5) node[black,below left] {$\CB$} ;
\filldraw[very thick,fill=white] (1.5,1.5) circle (0.4) ;
\node at (1.8,0.9) {$\CS$} ;
\end{tikzpicture}
\end{array}
\]
\caption{The Lagrangian algebra $A \in \CB$ can be obtained by shrinking a hole with boundary $\CS$.}
\label{fig_Lagrangian_algebra_hole}
\end{figure}

After the topological Wick rotation, this hole becomes a circle $S^1$. So we obtain a 1+1D phase $\bc[\CB]{\CS}$ on a circle (see the following figure). Then the shrinking process becomes shrinking the 1+1D phase on the circle. The result is a 0+1D phase, that is, a quantum mechanic system, which is just the ground state subspace of the phase on the circle.

\begin{figure}[H]
\centering
\[
\begin{array}{c}
\begin{tikzpicture}
\fill[gray!20] (0.5,0.5) rectangle (2.5,2.5) node[black,below left] {$\CB$} ;
\filldraw[very thick,fill=white] (1.5,1.5) circle (0.4) ;
\node at (1.8,0.9) {$\CS$} ;
\end{tikzpicture}
\end{array}
\rightsquigarrow
\begin{array}{c}
\begin{tikzpicture}[scale=0.8]
\fill[gray!20,opacity=0.5] (0,0) .. controls (0,0.3) and (1,0.3) .. (1,0)--(1,2) .. controls (1,2.3) and (0,2.3) .. (0,2)--cycle ;
\draw[very thick,dashed] (0,0) .. controls (0,0.3) and (1,0.3) .. (1,0) ;
\fill[gray!20,opacity=0.5] (0,0) .. controls (0,-0.3) and (1,-0.3) .. (1,0)--(1,2) .. controls (1,1.7) and (0,1.7) .. (0,2)--cycle ;
\draw[very thick] (0,0) .. controls (0,-0.3) and (1,-0.3) .. (1,0) node[right] {$\CS$} ;
\node[right] at (1,1) {$\CB$} ;
\end{tikzpicture}
\end{array}
\]
\caption{The topological Wick rotation suggests that the ground state algebra $A$ can be viewed as the ground state subspace on the circle $S^1$.}
\label{fig_Lagrangian_algebra_TWR_cylinder}
\end{figure}

We consider the 1+1D gapped phases with onsite $G$ symmetries and thus $\CB = \FZ_1(\rep(G))$. To put such a phase on a global manifold $M$, we also need to specify a $G$-bundle or a $G$-connection on $M$. Physically the $G$-connection is the same as a $G$-gauge field, and we need to couple the system with $G$-symmetry to the fixed gauge field to put it on $M$. For $M = S^1$, the $G$-bundles on $S^1$ are classified by the conjugacy classes of $G$, i.e., $G$-fluxes. Indeed, the $G$-gauge fields on $S^1$ can be distinguished by their holonomies, which are elements in $G$. However, the holonomy depends on the reference point, and if we change the reference point the holonomy may changed by a conjugation. Equivalently, if we do a gauge transformation of $G$-gauge fields, the holonomy may be changed by a conjugation. So the different $G$-bundles can be viewed as the insertions of $G$-fluxes and are usually called the twisted periodic boundary conditions.

Here the ground state subspace on $S^1$, which is still denoted by $A$, is the ground state subspace with all possible $G$-gauge fields or twisted periodic boundary conditions. In other words, there is a $G$-grading
\[
A = \bigoplus_{g \in G} A_g ,
\]
where $A_g$ is the ground state subspace on $S^1$ with a $G$-gauge field whose holonomy is $g$. As we discussed above, $A_h$ and $A_{ghg^{-1}}$ are isomorphic.

Moreover, the global $G$-symmetry operators $U(g)$ can also act on the ground state subspace. However, if there is a $G$-gauge field with holonomy $h$, after a global action of $U(g)$ we obatin a gauge field with holonomy $ghg^{-1}$. In other words, $U(g)$ changes the twisted periodic boundary condition from $h$ to $ghg^{-1}$. Therefore, $U(g)$ maps the space $A_h$ to $A_{ghg^{-1}}$.

Mathematically, an object in $\FZ_1(\vect_G) \simeq \FZ_1(\rep(G))$ is a vector space $V$ equipped with a $G$-grading $V = \bigoplus_{g \in G} V_g$ and a $G$-representation $\rho(g) \in \mathrm{GL}(V)$ such that $\rho(g)(V_h) \subseteq V_{ghg^{-1}}$ for all $g,h \in G$. Hence, we see that the ground state subspace $A$ of a 1+1D gapped phase with $G$-symmetry is naturally an object in $\FZ_1(\rep(G)) = \CB$, where the $G$-grading is given by the fluxes or the twisted periodic boundary conditions, and the $G$-action is given by the global $G$-action.

\begin{expl}
Let us consider the Ising chain. For the $\Zb_2$-symmetric phase, the ground state is $\lvert \cdots +++ \cdots \rangle$. This ground state can be put on $S^1$ with arbitrary fluxes. Indeed, the anti-periodic boundary condition does not affect the Hamiltonian $\mathcal H = -\sum_i X^i$. Therefore, the ground state subspace $A$ on $S^1$ is given by $A_e = \Cb$ and $A_a = \Cb$. The global $\Zb_2$-symmetry acts on the ground state invariably, so $A$ is equipped with the trivial $\Zb_2$-action. This gives $A = \bfone \oplus \bfm \in \FZ_1(\rep(\Zb_2))$.

For the symmetry breaking phase, the ground state is two-fold degenerate. In this case we can not put the ground state on $S^1$ with the anti-periodic boundary condition, because the anti-periodic boundary condition is equivalent to add a domain wall $M$, which can not live alone on a closed manifold. So in this case $A_a = 0$ and $A_e = \Cb_2$ with the $\Zb_2$-action permuting two basis. This gives $A = \bfone \oplus \bfe \in \FZ_1(\rep(\Zb_2))$.

So we see that in this two examples the ground state subspaces on $S^1$ (with all twisted periodic boundary conditions) are exactly the ground state algebra discussed in the previous sections.
\end{expl}

\begin{expl}
Let us consider the $\Zb_2 \times \Zb_2$ SPT orders. For the trivial SPT, similar to the Ising chain, the ground state is invaraint under the symmetry action and can be put on the circle with arbitrary twisted periodic boundary conditions. Thus we obtain the ground state space $\bfone \oplus \bfm_1 \oplus \bfm_2 \oplus \bfm_1 \bfm_2$, which is the same as $A(\Zb_2 \times \Zb_2)$.

For the nontrivial SPT order, we first consider the Haldane chain on a circle with an $a$-twisted periodic boundary condition. Such a twisted periodic boundary condition is equivalent to inserting an $a$-flux or adding an $\bfm_1$ defect, which can be realized by the string operator $M_a^i = \prod_{k \leq i} L_a^k$. This string operator commutes with the usual Hamiltonian of the Haldane chain, except on the end of the string. Indeed, recall that the Hamiltonian is $\mathcal H = -\sum_i P_{i,i+1}$ with
\[
P_{i,i+1} = 1 \otimes X_G \otimes 1 = 1 \otimes (1+L_a+L_b+L_{ab})/4 \otimes 1
\]
acting on $\CH_i \otimes \CH_{i+1} \simeq W \otimes \Cb[G] \otimes W$. For the projective representation $W = (W,\rho)$, we have
\[
\rho(a) \rho(g) = \begin{cases}
\rho(g) \rho(a) , & g = e,a; \\
-\rho(g) \rho(a) , & g = b,ab .
\end{cases}
\]
Therefore, we have $L_g^i P_{i,i+1} = P_{i,i+1}^{\phi_b} L_g^i$, where $P_{i,i+1}^{\phi_b} = 1 \otimes X_G^{\phi_b} \otimes 1$ with
\[
X_G^{\phi_b} = (1 + L_a - L_b - L_{ab})/4 = \sum_{g \in \Zb_2 \times \Zb_2} L_g / 4.
\]
It is easy to see that $X_G^{\phi_b}$ is a projector to the state $\lvert e \rangle + \lvert a \rangle - \lvert b \rangle - \lvert ab \rangle$. Under the isomorphism $\Cb[\Zb \times \Zb_2] \simeq W \times W$, it is the projector to $\lvert \uparrow \rangle \otimes \lvert \downarrow \rangle + \lvert \downarrow \rangle \otimes \lvert \uparrow \rangle$. The original ground state $\lvert \Omega \rangle$ of the Haldane chain is the common eigenstate of $P_{j,j+1}$ for all $j$ with eigenvalue $1$. Then we see that $M_a^i \lvert \Omega \rangle$ is not an eigenstate of $P_{i,i+1}$, but the eigenstate of $P_{i,i+1}^{\phi_b}$. So the $a$-twisted periodic boundary condition in the Haldane chain can be realized by replacing $P_{i,i+1}$ in the Hamiltonian by $P_{i,i+1}^{\phi_b}$. If we take the total Hilbert space to be $\bigotimes_j W_{2j} \otimes W_{2j+1}$, the $a$-twisted ground state $\lvert \Omega_a \rangle$ is also a `tensor product state' with $\lvert \uparrow \rangle \otimes \lvert \downarrow \rangle + \lvert \downarrow \rangle \otimes \lvert \uparrow \rangle$ on $W_{2i} \otimes W_{2i+1}$ and $\lvert \uparrow \rangle \otimes \lvert \uparrow \rangle + \lvert \downarrow \rangle \otimes \lvert \downarrow \rangle$ on $W_{2j} \otimes W_{2j+1}$ for $j \neq i$. It is easy to see that
\[
U(g) \lvert \Omega_a \rangle = \begin{cases}
\lvert \Omega_a \rangle , & g = e,a ; \\
- \lvert \Omega_a \rangle , & g = b,ab .
\end{cases}
\]
Hence, the $a$-twisted ground state spans a $\Zb_2 \times \Zb_2$-representation $\phi_b$. In other words, the $a$-twisted ground state carries the $\Zb_2 \times \Zb_2$-charge $\bfe_2$.

Similarly, we can check that the $b$-twisted ground state is unique and carries the $\Zb_2 \times \Zb_2$-charge $\bfe_1$, and the $ab$-twisted ground state is also unique and carries the $\Zb_2 \times \Zb_2$-charge $\bfe_1 \bfe_2$. So we obtain the ground state space of the nontrivial $\Zb_2 \times \Zb_2$ SPT on the circle is $\bfone \oplus \bfe_1 \bfm_2 \oplus \bfe_2 \bfm_1 \oplus \bff_1 \bff_2$. This is exactly the Lagrangian algebra $A(\Zb_2 \times \Zb_2,\omega)$.
\end{expl}

\begin{rem}
The dimension of a vector space $V \in \FZ_1(\rep(G))$ is equal to its Frobenius-Perron dimension \cite{ENO05}. Moreover, the Frobenius-Perron dimension of a Lagrangian algebra $A \in \FZ_1(\rep(G))$ is always equal to $\lvert G \rvert$. Thus we see that the ground state degeneracy of a 1+1D gapped quantum phase with symmetry $G$ on a circle (with all twisted periodic boundary conditions) is equal to $G$. 

Furthermore, when $G$ is abelian, by the construction of the Lagrangian algebra $A(H,\omega)$ (see Example \ref{expl:abelian_Lagrangian_center}), we have
\[
\dim A(H,\omega)_g = \begin{cases}
\lvert G \rvert / \lvert H \rvert , & g \in H , \\
0 , & \text{otherwise}.
\end{cases}
\]
This gives the ground state degeneracy on the circle with the $g$-twisted periodic boundary condition.
\end{rem}

\begin{rem}
We have not discussed the algebra structure of the ground state space on the circle. In the language of open-closed TQFT \cite{MS06}, the ground state space on $S^1$ is the closed TQFT (or closed string algebra), and its algebra structure is obviously given by the cobordism of pants. 
% In the Hamiltonian formultaion, it seems that this algebra structure is not very clear.

Moreover, given a 0+1D boundary, we can also consider the ground state space on a finite interval $[0,1]$. In the language of open-closed TQFT, the ground state space on $[0,1]$ is the open TQFT (or open string algebra). In a 2D open-closed TQFT or CFT, the open-closed duality \cite{MS06, RFFS07, KR09} states that the closed string algebra is the full center \cite{Dav10} of the open string algebra.

Futhermore, if we fix the categorical symmetry on the boundary (in our case this means that we need to specify that the symmetry on the boundary is explicitly broken to a subgroup $H$, possibly with a 2-cocycle twist), we can also consider the ground state spaces on $[0,1]$ with different boundary conditions. These ground state spaces are not only open string algebras, but also realize the whole enriched category of boundary conditions with the given categorical symmetry discussed in Section \ref{sec_boundary}.
\end{rem}

\begin{rem}
The finite group symmetry $G$ in the above discussion can also be generalized. For example, a 1+1D modular invariant CFT can be viewed as a quantum liquid phase with a gapless non-chiral symmetry, usually realized by a non-chiral algebra $V$, and the categorical symmetry is the UMTC $\CB \coloneqq \mathrm{Mod}_V$ of $V$-modules \cite{KZ20,KZ21}. In this case, the `ground state subspace on $S^1$' is usually known as the Hilbert space on $S^1$ of the CFT (or simply the closed CFT). It has been proved \cite{KR09} that the Hilbert space on $S^1$ of a modular invariant CFT is a Lagrangian algebra in $\mathrm{Mod}_V$.
\end{rem}

\section{Discussions and outlooks} \label{section:conclusion}

Category theory is becoming an indispensable tool in studying the quantum many-body systems. However,
it has been mostly used in the study of topological phases.
In this work, we have shown the capability of enriched fusion categories in describing $1$d gapped phases with symmetries, including symmetry breaking phases within traditional Landau's paradigm. 

For a 1+1D gapped phase
with abelian onsite symmetry $G$, 
its macroscopic properties can be summarized by the categorical symmetry $\FZ_1(\rep(G))$ and a Lagrangian algebra $A(H, \omega)$, in which $A(H, \omega)$ plays the role of invariant operators acting on the sector of ground state, such that
the enriched fusion category
$\bc[\FZ_1(\rep(G))]{\FZ_1(\rep(G))_{A_{(H, \omega)}}}$ can be read off from lattice model directly. Under this observation, we can distinguish different symmetry breaking phases as well as the trivial and non-trivial SPT orders.

We construct general lattice models with abelian onsite symmetry to show that the topological sectors of operators
form a braided fusion category $\FZ_1(\rep(G)) \simeq \FZ_1(\vect_G))$, and we invent equivariantization technique in section \ref{section:G_symmetry} to find different topological sectors of state, which turns out to be $\FZ_1(\rep(G))_{A_{(H)}}$ (depending on which subgroup $H \subset G$ the symmetry breaks to).
% In other words, we explicitly write down the mathematical structures
% of the SPT and symmetry breaking phases.
We also demonstrate that the 0d boundaries of these 1d phases form enriched
categories which are compatible with the 1d bulk descriptions. 
Well-known physical examples are also been performed.
These results meet with the predictions of holographic duality. But it is just a tip of the iceberg. For example, the idea of holographic duality can be expanded to other kinds of symmetries such as non-invertible ones.
Here we briefly explain how lattice duality should be unified under the same enriched categorical description.

\subsection{lattice duality under holographic duality}

\begin{table*}
    \centering
  Categorical symmetry \, \, $\FZ_1(\rep(\mathbb{Z}_2 \times \mathbb{Z}_2)) \qquad \qquad \qquad \qquad \qquad$
  \setlength{\fboxrule}{1.5pt}
  \setlength{\fboxsep}{3.5pt}
    \begin{tabular}{|c|c|c|c|c|c|c|c}
        \cline{1-7}
        \diagbox[width=14em, height=3.5em]{topo. states \\ with ground state algebra}{~\\lattice sym.} & \makecell{$\mathcal{H}^{\Zb_2 \times \Zb_2}$\\$ A_1 (A_6^{KT})$} & $A_2$ &$A_3$&$A_4$&\makecell{$\mathcal{H}^{\rep(\Zb_2 \times \Zb_2)}$\\$A_5$}&\makecell{$\mathcal{H}^{\Zb_2 \times \Zb_2,\omega}$\\$A_6$}& ~\\
        \cline{1-7}
        $A_1 = \bfone \oplus \bfe_1 \oplus \bfe_2 \oplus \bfe_1 \bfe_2$ \rule{0pt}{12pt}& \cellcolor{gray!20} SCB & \cellcolor{gray!20} $\mathbb{Z}_2$ &  \cellcolor{gray!20} $\mathbb{Z}_2$& \cellcolor{gray!20} $\mathbb{Z}_2$&  \cellcolor{gray!20} $\rm SPT_0$& \cellcolor{gray!20} SPT& \cellcolor{gray!20} ${}^{\FZ_1(\rep(\mathbb{Z}_2 \times \mathbb{Z}_2))}\vect_{\mathbb{Z}_2 \times \mathbb{Z}_2}$\\[3pt]
        \hline
        $A_2 = \bfone \oplus \bfe_1 \oplus \bfm_2 \oplus \bfe_1\bfm_2$ \rule{0pt}{12pt}& \cellcolor{gray!20} $\mathbb{Z}_{2}$ & \cellcolor{gray!20} SCB &  \cellcolor{gray!20} SPT& \cellcolor{gray!20} SPT&  \cellcolor{gray!20}  ${\mathbb{Z}_{2}}$& \cellcolor{gray!20}  $\mathbb{Z}_{2}$& \cellcolor{gray!20} ${}^{\FZ_1(\rep(\mathbb{Z}_2 \times \mathbb{Z}_2))}\FZ_1(\rep(\mathbb{Z}_2 \times \mathbb{Z}_2))_{A_2}$\\[3pt]
        \hline
        $A_3 = \bfone \oplus \bfm_1 \oplus \bfe_2 \oplus \bfm_1\bfe_2$ \rule{0pt}{12pt}& \cellcolor{gray!20} $\mathbb{Z}_{2}$ & \cellcolor{gray!20} SPT &  \cellcolor{gray!20} SCB& \cellcolor{gray!20} SPT&  \cellcolor{gray!20}  $\mathbb{Z}_{2}$& \cellcolor{gray!20}  $\mathbb{Z}_{2}$& \cellcolor{gray!20} ${}^{\FZ_1(\rep(\mathbb{Z}_2 \times \mathbb{Z}_2))}\FZ_1(\rep(\mathbb{Z}_2 \times \mathbb{Z}_2))_{A_3}$\\[3pt]
        \hline
        $A_4 = \bfone \oplus \bfe_1\bfe_2 \oplus \bfm_1\bfm_2 \oplus \bff_1\bff_2$ \rule{0pt}{12pt}& \cellcolor{gray!20} $\mathbb{Z}_{2}$ & \cellcolor{gray!20} SPT &  \cellcolor{gray!20} SPT& \cellcolor{gray!20} SCB&  \cellcolor{gray!20}  $\mathbb{Z}_{2}$& \cellcolor{gray!20}  $\mathbb{Z}_{2}$& \cellcolor{gray!20} ${}^{\FZ_1(\rep(\mathbb{Z}_2 \times \mathbb{Z}_2))}\FZ_1(\rep(\mathbb{Z}_2 \times \mathbb{Z}_2))_{A_4}$\\[3pt]
        \hline
        $A_5=\bfone \oplus \bfm_1 \oplus \bfm_2 \oplus \bfm_1\bfm_2$ \rule{0pt}{12pt}& \cellcolor{gray!20} $\rm SPT_0$ & \cellcolor{gray!20} $\mathbb{Z}_2$ &  \cellcolor{gray!20} $\mathbb{Z}_2$& \cellcolor{gray!20} $\mathbb{Z}_2$&  \cellcolor{gray!20}  SCB& \cellcolor{gray!20}  SPT& \cellcolor{gray!20} ${}^{\FZ_1(\rep(\mathbb{Z}_2 \times \mathbb{Z}_2))}\rep(\mathbb{Z}_2 \times \mathbb{Z}_2)$\\[3pt]
        \hline
        $A_6 = \bfone\oplus\bfe_1\bfm_2\oplus\bfe_2\bfm_1\oplus\bff_1\bff_2$ \rule{0pt}{12pt}& \cellcolor{gray!20} \fbox{\;$\rm SPT_1 $\;} & \cellcolor{gray!20} $\mathbb{Z}_2$ &  \cellcolor{gray!20} $\mathbb{Z}_2$& \cellcolor{gray!20} $\mathbb{Z}_2$&  \cellcolor{gray!20}  $\rm SPT_1$& \cellcolor{gray!20}  \fbox{\;$\rm  SCB $\;}& \cellcolor{gray!20} ${}^{\FZ_1(\rep(\mathbb{Z}_2 \times \mathbb{Z}_2))}\rep(\mathbb{Z}_2 \times \mathbb{Z}_2)_{A(G, \omega)}$\\[3pt]
        \hline
    \end{tabular}
    \caption{
    The modular tensor category $\FZ_1(\rep(\Zb_2\times \Zb_2))$ has six Lagrangian algebra, which we label by $A_1$ through $A_6$ \cite{CW22a}.
Through lattice models can have different symmetries up to Morita equivalence, phases in LWLL are classified by ground state algebras. For example,
the non-trival SPT order and the symmetry completely brokon phase
connected by Kennedy-Tasaki transformation can be both described by 
${}^{\FZ_1(\rep(\mathbb{Z}_2 \times \mathbb{Z}_2))}\rep(\mathbb{Z}_2 \times \mathbb{Z}_2)_{A(G, \omega)}$
}
\label{KT_transformation}
\end{table*}

As the enriched category description only cares about observables in the LWLL (or to say, it really captures what a (gapped) phase is in the macroscopic level), for a given lattice model with onsite symmetry $G$, we should also have the same categorical descriptions appear in the phases of its dual lattices, in which their energy specturm are the same. 
Dual models 
% are characterized by the same fusion category $\CD$,  but different choices of indecomposable module category $\CM$.
% Dual models are characterized by the same fusion category $\CD$, with the same recouping theory, but different choices of module category $\CM$. Duality is an isomorphism of the algebra of local symmetric operators.
have equivalent but distinct realizations of 
% (MPO) 
symmetries, 
characterized by $\CS_{\CM}^*$ \cite{LDOV21} with different choices of indecomposable module category $\CM$.
On the other hand, the indecomposable module categories of $\CS$ one to one correspond to the gapped boundaries of $\FZ_1(\CS)$ \cite{DMNO13} with different choices of condensable algebras $A_i$.
So just like the topological sectors of states,
the global lattice symmetries form Morita equivalent fusion categories that can be obtained by condensing Lagrangian algebras.

\begin{figure}[H]\label{general_dual}
\centering
\begin{tikzpicture}
        \draw[very thick](0,0) node[left]{$\mathcal{H}^{A_4}$}--(6,0);
        \draw[very thick, fill=white](1.4,-0.1) rectangle (1.6,0.1);
        \draw[very thick, fill=white](2.9,-0.1) rectangle (3.1,0.1);
        \draw[very thick, fill=white](4.4,-0.1) rectangle (4.6,0.1);
        \node[]at(0.7,0.3){${\FZ_1(\CS)}_{A_4}$};
        \node[]at(2.2,0.3){${\FZ_1(\CS)}_{A_1}$};
        \node[]at(3.7,0.3){${\FZ_1(\CS)}_{A_2}$};
        \node[]at(5.2,0.3){${\FZ_1(\CS)}_{A_3}$};

        \draw[very thick](0,1) node[left]{$\mathcal{H}^{A_2}$}--(6,1);
        \draw[very thick, fill=white](1.4,0.9) rectangle (1.6,1.1);
        \draw[very thick, fill=white](2.9,0.9) rectangle (3.1,1.1);
        \draw[very thick, fill=white](4.4,0.9) rectangle (4.6,1.1);
        \node[]at(0.7,1.3){${\FZ_1(\CS)}_{A_3}$};
        \node[]at(2.2,1.3){${\FZ_1(\CS)}_{A_4}$};
        \node[]at(3.7,1.3){${\FZ_1(\CS)}_{A_1}$};
        \node[]at(5.2,1.3){${\FZ_1(\CS)}_{A_2}$};

        \draw[very thick](0,2) node[left]{$\mathcal{H}^{A_2}$}--(6,2);
        \draw[very thick, fill=white](1.4,1.9) rectangle (1.6,2.1);
        \draw[very thick, fill=white](2.9,1.9) rectangle (3.1,2.1);
        \draw[very thick, fill=white](4.4,1.9) rectangle (4.6,2.1);
        \node[]at(0.7,2.3){${\FZ_1(\CS)}_{A_2}$};
        \node[]at(2.2,2.3){${\FZ_1(\CS)}_{A_3}$};
        \node[]at(3.7,2.3){${\FZ_1(\CS)}_{A_4}$};
        \node[]at(5.2,2.3){${\FZ_1(\CS)}_{A_1}$};

        \draw[very thick](0,3) node[left]{$\mathcal{H}^{A_1}$}--(6,3);
        \draw[very thick, fill=white](1.4,2.9) rectangle (1.6,3.1);
        \draw[very thick, fill=white](2.9,2.9) rectangle (3.1,3.1);
        \draw[very thick, fill=white](4.4,2.9) rectangle (4.6,3.1);
        \node[]at(0.7,3.3){${\FZ_1(\CS)}_{A_1}$};
        \node[]at(2.2,3.3){${\FZ_1(\CS)}_{A_2}$};
        \node[]at(3.7,3.3){${\FZ_1(\CS)}_{A_3}$};
        \node[]at(5.2,3.3){${\FZ_1(\CS)}_{A_4}$};
        % \node[]at(1.5,3.6){\text{\scriptsize \textcolor{blue}{gapless} }};
\end{tikzpicture}
\caption{
        Actually, the topological skeletons we obtain here describes a set of phases that can be achieved by dual models.
    Suppose there are 4 Lagrangian algebras $A_1$ to $A_4$ of the categorical symmetry $\FZ_1(\CS)$, then we would have 4 dual models with lattice symmetries $A_i$. Then the permutation of topological sectors of states $\FZ_1(\CS)_{A_i}$ defines the symmetry preseving / symmetry breaking properties of each model, leading to seemingly 16 gapped phases, but each four of them can be unified under one enriched fusion categories.
    }
\end{figure} 

Indeed, everytime we talk about the (onsite) symmetry of a concrete lattice model, it is that we fix a specific boundary (or Lagrangian algebra equivalently) from the categorical symmetry which forms a modular tensor category. In this paper, we implicitly fix the global onsite symmetry $G$ which can be regarded as fusion category $\vect_G \simeq \FZ_1(\rep(G))_{A(\{e\})}$, where $A(\{e\})$ is the Lagrangian algebra in $\FZ_1(\rep(G))$ contains all chargeons.

% and the classification of these dual models, just like the topological sectors of states, are also characterized by the boundaries of the holographic bulk.
% This is due to $G$-symmetry (or $\vect_G$ symmetry in categorical language) and $\rep(G)$ share the same bulk $\FZ_1(\rep(G))$.

If we choose other Lagrangian algebras in  $\FZ_1(\rep(G))$, we can have different lattice symmetries. Note that the lattice symmetries may be non-invertible and non-onsite (e.g. $\rep(G)$ for non-abelian cases). Concrete lattice constructions with these kind of symmetries can be found in literatures such as \cite{LDOV21,SS24,CAW24}.
A lattice model $\mathcal{H}$'s dual models and $\mathcal{H}$'s gapped phases are actually 'dual concepts' which are both aspects of the total categorical symmetry, and every enriched category describes a bunch of representations in dual lattices realizing equivalent phases.
% and should be organized under one principle.

% Dual models and one model's gapped phases are both aspects of the total categorical symmetry, which should be organized under one principle. A lattice models's dual models and the models' gapped phases are actually 'dual concepts'.

% This transformation can be generalized to the case
% of abelian groups $G \simeq H \time H$ of the form, where H is
% some group.

% These five phases correspond to anyon condensation in $\FZ_1(\rep(\Zb_2\times \Zb_2))$ via five different Lagrangian algebras: $\bfone \oplus \bfe_1 \oplus \bfe_2 \oplus \bfe_1\bfe_2$, $\bfone \oplus \bfe_1 \oplus \bfm_2 \oplus \bfe_1\bfm_2$, $\bfone \oplus \bfm_1 \oplus \bfe_2 \oplus \bfm_1\bfe_2$, $\bfone \oplus \bfe_1\bfe_2 \oplus \bfm_1\bfm_2 \oplus \bff_1\bff_2$, $\bfone \oplus \bfm_1 \oplus \bfm_2 \oplus \bfm_1\bfm_2$ (see \cite{CW22a} and ref therein).

In the example of Kramers-Wannier duality of transverse field Ising model \ref{section:KW}, we give a taste of this.
We can also consider the Kennedy-Tasaki duality \cite{KT92}.
KT transformation provides a map between a model with symmetry
protected topological order in the ground state, such as
the Haldane phase, to
a model without symmetry protected topological order.
And the four degenerate edge modes becomes four bulk degenerate ground states in the symmetry breaking phase afterwards.
See the two frames in table \ref{KT_transformation}, this two phases are both controlled by the ground state algebra $A(G, \omega)$.  
We also list other kinds of probabilities in $\Zb_2 \times \Zb_2$'s dual models.
0+1D boundaries can also be understood through a similar procedure, which may lead to non-trivial conclusions.

Other than explaining interesting phenomena emerge in dualities,
the immediate follow-up for Theorem$^{\text{ph}}$ is to check that the $1$d gapped phases with onsite non-abelian symmetries can also be unified through enriched fusion categories with ground state algebras.
 %\ref{pthm:topological_skeleton}. 
In fact, as equivariantization does not depend on symmetry $G$ abelian or not, the only difficulty remains is to check the topological sector of operators, namely, to see the non-local operators in a system of a general onsite symmetry $G$ also form $\FZ_1(\rep(G))$. 

Then one can try to replace the holographic bulk phase of figure \ref{fig:enriched_category} by the Levin-Wen model \cite{LW05}, which should give interesting fusion category symmetries). Actually, there are already constructions of the general 1d lattice models with arbitrary fusion category symmetries \cite{Ina22}. 
The generalization of our construction using quantum current is given recently in \cite{LZ24}.
% Boundaries with non-trivial 2-cocycles can also be readlily solved after that.
% In the non-abelian case, we cannot use the metric group any more. One possible way is to think of the equivariantization description for braided fusion categories. 
% An obvious question left in this paper is that:
% can we also use equivariantization to find the categorical symmetry $\CB$? We believe it is possible, since the principle of equivariantization is to find the categorical description of a topological order after imposing symmetry.
% Actually it is already emerged in the boundary cases, see remark \ref{bdy-categorical symmetry}. 
% In $0$d boundary cases, the categorical symmetry is equivalent to the equivariantization of the invisible bulk, which is in accordance with our intuition.
% Moreover, since $\Fun(\vect_{G/H}, \vect_{G/H})$ is anomaly-free, 
% So to recover $\FZ_1(\rep(G))$ might be to equivariantize the $2$d braided fusion category $\vect$ in a non-trivial way. 

One can also think of the fermionic cases, the starting point might be to think of the fermion condensation\cite{WW17} in $2$d topological order $\FZ_1(\rep(G, z)) = \FZ_1(\rep(G))$ ($z \in G$ is the fermion parity). Different from the bosonic cases, an interchange between two topological excitations should have a non-trivial phase factor $-1$.

% we would like to find the enriched category that describes a specific phase at a time in these models.

% One can also start another way around, to first pick a special lattice model like the Haldane chain \cite{Hal83} or AKLT model \cite{AKLT87}, then try to find the categorical descriptions of their phases. 

%The topological Wick rotation approach is also valid in higher dimensions and may be used in describing higher dimensional lattice models, .
Moreover, the enriched category description is valid in describing phase transition point \cite{KZ20, CJKYZ20}. It would be interesting to describe the phase transition points between 1d gapped quantum phases.
In the clock model example, we roughly discuss the 'step-like' phase transition behavior (see Figure \ref{fig:Z_4}).
The topological skeleton of phase transition points in partially symmetry breaking cases may be 'fused' in an interesting way to become the phase transition point of larger symmetry breaking subgroup.

More generally,
% Except for giving the categorical descriptions of theses lattice models, there are other mysterious things to be explored. 
% For example, 
% why the onsite symmetry can be viewed as a local quantum symmetry? 
% how to understand the `boundary-bulk relation' in spacetime between the topological sectors of operators and the states after the topological Wick rotation? And how to interpret the `braidings' of operators in spacetime...? 
since our work is an example of 
% the gauge Symmetries of the bulk theory that are mapped to global Symmetries of the boundary theory
the $n$+1D gauge theory with gapped boundaries that maps to nD gapped quantum liquid with global symmetries,
we want to see more works that fulfill the territory of holographic duality, especially in the realm of high energy physics. We should even look for a theory unifying both AdS/CFT dualities and the holographic dualities in condensed matter.
% Behind all these things, the deepest question might be, why these physical phases can be performed in such neat and beautiful structures.
For the adventurers, there is still a continent to be discovered.

\begin{acknowledgments}
We would like to thank Holiverse Yang and Liang Kong for helpful discussions. We also thank Jian Li and Wei Zhu for helpful insights. RX acknowledges Shenzhen Institute of Quantum Science and Engineering for hospitality during the visits. ZHZ is supported by NSFC under Grant No.~11971219 and Guangdong Basic and Applied Basic Research Foundation under Grant No.~2020B1515120100 and Guangdong Provincial Key Laboratory (Grant No.~2019B1212e03002).
\end{acknowledgments}

\begin{appendix}

% \input{appendix_pointed}
% !TeX root = supp.tex
% !TeX program = pdfLaTeX

\section{Pointed braided fusion categories} \label{appendix}

\subsection{Pointed braided fusion categories and pre-metric groups} \label{appendix:pointed_metric_group}

In this section we briefly review the relation between pointed braided fusion categories and pre-metric groups.

\begin{defn}
Let $\CC$ be a fusion category. An object $x \in \CC$ is called \emph{invertible} if it is invertible under the tensor product, i.e. there exists an object $y \in \CC$ such that $x \otimes y \simeq \one \simeq y \otimes x$. We say $\CC$ is \emph{pointed} if every simple object in $\CC$ is invertible.
\end{defn}

\begin{defn}
Let $G$ be an abelian group. A \emph{quadratic form} on $G$ is a map $q \colon G \to \mathbb C^\times$ such that $q(g) = q(g^{-1})$ and the symmetric function $b \colon G \times G \to \mathbb C^\times$ defined by
\[
    b(g,h) \coloneqq \frac{q(gh)}{q(g)q(h)}
\]
is a \emph{bicharacter}, i.e., $b(g_1 g_2,h) = b(g_1,h) b(g_2,h)$ for all $g_1,g_2,h \in G$. The symmetric bicharacter $b$ is called the \emph{associated bicharacter} of $q$. We say $q$ is \emph{nondegenerate} if $b$ is, i.e., $b(g,h) = 1$ for all $h \in G$ if and only if $g = e$ is the unit.

A \emph{pre-metric group} is a finite abelian group $G$ equipped with a quadratic form $q$. If $q$ is nondegenerate, then $(G,q)$ is called a \emph{metric group}.
\end{defn}

Let $\CC$ be a pointed braided fusion category. Then isomorphism classes of simple objects of $\CC$ form a finite abelian group under the tensor product, denoted by $G$. In other words, $\CC \simeq \vect_G$ as categories. For $g \in G$, define
\[
q(g) \coloneqq c_{x,x} \in \Aut_{\CC}(x \otimes x) \simeq \mathbb C^\times ,
\]
where $x$ is a simple object of $\CC$ whose isomorphism class is $g$. It follows from the pentagon and hexagon equations that $q$ is a quadratic form. The associated bicharacter of $q$ is equal to the double braiding:
\be \label{eq:double_braiding_metric_group}
b(g,h) = c_{y,x} \circ c_{x,y} \in \Aut_\CC(x \otimes y) \simeq \mathbb C^\times ,
\ee
where $x$ and $y$ are simple objects of $\CC$ whose isomorphism classes are $g$ and $h$, respectively. We call $(G,q)$ the \emph{associated pre-metric group} of $\CC$.

Conversely, there is a mathematical theorem \cite{JS93,DGNO10} states that two pointed braided fusion categories are equivalent if and only if their associated pre-metric groups are isomorphic. In other words, for any pre-metric group $(G,q)$ there exists a unique (up to equivalence) pointed braided fusion category, denoted by $\CC(G,q)$, such that its associated pre-metric group is isomorphic to $(G,q)$. Clearly the pointed braided fusion category $\CC(G,q)$ is nondegenerate if and only if the pre-metric group $(G,q)$ is nondegenerate, i.e., $(G,q)$ is a metric group.

Moreover, since $\CC(G,q)$ is pseudo-unitary, there is a unique spherical structure on $\CC(G,q)$ such that the quantum dimension of every simple object is $1$ \cite{ENO05}. Then $\CC(G,q)$ is a pre-modular category with this spherical structure, and it is modular if and only if $(G,q)$ is a metric group. Its $S$ matrix is given by
\[
S_{g,h} = b(g,h^{-1}) = b(g^{-1},h) ,
\]
and the $T$ matrix is given by
\[
T_g = q(g) .
\]

\begin{expl} \label{expl:abelian_center_metric}
Let $G$ be a finite abelian group. Then the Drinfeld center $\FZ_1(\rep(G)) \simeq \FZ_1(\vect_G)$ is a nondegenerate pointed braided fusion category and the associated metric group is $(G \times \hat G,q)$, where $\hat G$ is the dual group of $G$ and the quadratic form $q$ is defined by $q(g,\rho) \coloneqq \rho(g)$. We use $\{\mathcal O_{(g,\rho)}\}_{g \in G,\rho \in \hat G}$ to denote a representative of isomorphism classes of simple objects of $\FZ_1(\rep(G))$. Then the double braiding of two simple objects $\mathcal O_{(g,\rho)}$ and $\mathcal O_{(h,\sigma)}$, by \eqref{eq:double_braiding_metric_group}, is equal to $\rho(h) \sigma(g)$.
\end{expl}

\subsection{Condensable algebras in pointed braided fusion categories} \label{appendix:condensable_pointed}

In this section we briefly review the classification of condensable algebras (in particular, Lagrangian algebras) in pointed braided fusion categories.

\begin{defn}
Let $\CC$ be a braided fusion category and $A = (A,\mu,\eta)$ be an algebra in $\CC$. Then
\bnu[(a)]
\item $A$ is \emph{separable} if there exists an $(A,A)$-bimodule map $e \colon A \to A \otimes A$ such that $\mu \circ e = \id_A$;
\item $A$ is \emph{connected} if $\dim \Hom_\CC(\one,A) = 1$;
\item $A$ is \emph{commutative} if $\mu \circ c_{A,A} = \mu$.
\enu
A \emph{condensable algebra} is a commutative connected separable algebra. Moreover, a \emph{Lagrangian algebra} in a nondegenerate braided fusion category $\CC$ is a condensable algebra $A$ such that $\fpdim(A)^2 = \fpdim(\CC)$, or equivalently, the category $\CC_A^0$ of local right $A$-modules in $\CC$ is equivalent to $\vect$.
\end{defn}

\begin{defn}
Let $(G,q)$ be a pre-metric group. The associated bicharacter of $q$ is denoted by $b$. For any subgroup $H \subseteq G$, define the \emph{orthogonal complement} of $H$ as
\[
    H^\perp \coloneqq \{g \in G \mid b(g,h) = 1 , \, \forall h \in H\} .
\]
We say a subgroup $H \subseteq G$ is \emph{isotropic} if $q(h) = 1$ for all $h \in H$. It follows that an isotropic subgroup $H$ satisfies $H \subseteq H^\perp$. A \emph{Lagrangian subgroup} of a metric group is an isotropic subgroup $H$ such that $H = H^\perp$.
\end{defn}

\begin{thm}[\cite{DGNO10,DMNO13,FSV13,DS18}]
Let $(G,q)$ be a pre-metric group and $A$ be a condensable algebra in $\CC(G,q)$.
\bnu[(1)]
\item Suppose $\{\Cb_g\}_{g \in G}$ is a representative of isomorphism classes of simple objects of $\CC(G,q)$. Then $\dim \Hom_{\CC(G,q)}(\Cb_g,A)$ is either $0$ or $1$.
\item Define the \emph{support} of $A$ to be the set
\[
    \Supp A \coloneqq \{g \in G \mid \dim \Hom_{\CC(G,q)}(\Cb_g,A) = 1\} .
\]
Then the support $\Supp A$ is an isotropic subgroup of $(G,q)$.
\item For any isotropic subgroup $H$ of $(G,q)$, there exists a unique (up to isomorphism) condensable algebra $A_H$ in $\CC(G,q)$ such that $\Supp A_H = H$. In other words, there is a one-to-one correspondence between isomorphism classes of condensable algebras in $\CC(G,q)$ and isotropic subgroups of $(G,q)$. As an object $A_H$ is isomorphic to $\bigoplus_{h \in H} \Cb_h$, and the multiplication is induced by the group multiplication:
\begin{align*}
    A \otimes A \simeq \bigl( \bigoplus_{g \in H} \Cb_g \bigr) \otimes \bigl( \bigoplus_{h \in H} \Cb_h \bigr) =\\
        \bigoplus_{g,h \in H} \Cb_g \otimes \Cb_h \xrightarrow{\bigoplus \delta_{gh,k}} \bigoplus_{k \in H} \Cb_k = A .
\end{align*}
    
\item For any isotropic subgroup $H \subseteq G$, the braided fusion category $\CC(G,q)_{A_H}^0$ of local right $A_H$-modules is pointed, and the associated pre-metric group is the quotient group $H^\perp / H$ equipped with the quadratic form $\bar q([g]) \coloneqq q(g)$. In particular, there is a one-to-one correspondence between isomorphism classes of Lagrangian algebras in $\CC(G,q)$ and Lagrangian subgroups of $(G,q)$.
\enu
\end{thm}

\begin{expl} \label{expl:abelian_Lagrangian_center}
Let $G$ be a finite abelian group. Recall example \ref{expl:abelian_center_metric} that the metric group associated to $\FZ_1(\rep(G)) \simeq \FZ_1(\vect_G)$ is $(G \times \hat G,q)$ where $q(g,\rho) = \rho(g)$. For any subgroup $H \subseteq G$, the dual group $\widehat{G/H}$ naturally embeds into $\hat G$ and $H \times \widehat{G/H}$ is clearly a Lagrangian subgroup of $(G \times \hat G,q)$. We denote the corresponding Lagrangian algebra by $A(H) \in \FZ_1(\rep(G))$.

More generally, suppose $H \subseteq G$ is a subgroup and $\tau \colon H \times H \to \mathbb C^\times$ is an \emph{anti-symmetric bicharacter}, i.e. a bicharacter satisfying $\tau(h,h) = 1$ for all $h \in H$. Then
\[
    S(H,\tau) \coloneqq \{(h,\rho) \in G \times \hat G \mid h \in H , \, \rho(k) = \tau(h,k) , \, \forall k \in H\}
\]
is a Lagrangian subgroup of $(G \times \hat G,q)$. Conversely, every Lagrangian subgroup of $(G \times \hat G,q)$ is equal to $S(H,\tau)$ for some subgroup $H \subseteq G$ and anti-symmetric bicharacter $\tau$.

For any finite abelian group $H$ and 2-cocycle $\omega \in Z^2(H;\mathrm{U}(1))$, one can verify that
\begin{align*}
    H \times H & \to \mathbb C^\times                                \\
    (g,h)      & \mapsto \frac{\omega(g,h)}{\omega(h,g)}
\end{align*}
defines an anti-symmetric bicharacter on $H$. Also, this anti-symmetric bicharacter only depends on the cohomology class of $\omega$. Thus we have a group homomorphism from $\mathrm{H}^2(H,U(1))$ to the group of anti-symmetric bicharacters on $H$. It is well-known that this group homomorphism is indeed an isomorphism (because every symmetric 2-cocycle of a finite group is a coboundary). Thus for every subgroup $H$ and $\omega \in H^2(H;U(1))$ there is a Lagrangian algebra $A(H,\omega)$ defined by the Lagrangian subgroup $S(H,\tau)$.

As a conclusion, there is a one-to-one correspondence between isomorphism classes of Lagrangian algebras in $\FZ_1(\rep(G)) \simeq \FZ_1(\vect_G)$ and pairs $(H,[\omega])$, where $H \subseteq G$ is a subgroup and $[\omega] \in \mathrm{H}^2(H,U(1))$ is a 2-cohomology class. This coincides with the general classification result in \cite{Dav10a}.
\end{expl}

\begin{expl} \label{expl_Lagrangian_algebra_Z2Z2}
As an example, we list the Lagrangian algebras in $\FZ_1(\rep(G))$ when $G = \Zb_2 \times \Zb_2$. We denote the elements in $G$ by $\bfone,\bfm_1,\bfm_2,\bfm_1 \bfm_2$, and denote the elements in $\hat G$ by $\one,\bfe_1,\bfe_2,\bfe_1 \bfe_2$ with $(\bfm_i,\bfe_j) = (-1)^{\delta_{i,j}}$ for $i,j = 1,2$. Then there are 6 Lagrangian algebras in $\FZ_1(\rep(G))$ (i.e., 6 Lagrangian subgroups in $(G \times \hat G,q)$):
\bit
\item The Lagrangian algebra corresponding to the trivial subgroup $H = \{\bfone\}$ is $\bfone \oplus \bfe_1 \oplus \bfe_2 \oplus \bfe_1 \bfe_2$.
\item There are 3 $\Zb_2$ subgroups of $G$ generated by $\bfm_1,\bfm_2,\bfm_1 \bfm_2$, respectively. Note that $H^2(\Zb_2;U(1)) = 0$ is trivial. Then the Lagrangian algebra corresponding to these $\Zb_2$ subgroups are $\bfone \oplus \bfm_1 \oplus \bfe_2 \oplus \bfm_1\bfe_2$, $\bfone \oplus \bfm_2 \oplus \bfe_1 \oplus \bfm_2\bfe_1$ and $\bfone \oplus \bfm_1 \bfm_2 \oplus \bfe_1 \bfe_2 \oplus \bff_1\bff_2$, respectively, where $\bff_i \coloneqq \bfe_i \bfm_i$ for $i = 1,2$.
\item We have $H^2(\Zb_2 \times \Zb_2;U(1)) \simeq \Zb_2$, and the anti-symmetric bicharacter corresponding to the nontrivial 2-cohomology class is given by $\tau(\bfm_i,\bfm_j) = (-1)^{1-\delta_{i,j}}$. Then the Lagrangian algebra $A(\Zb_2 \times \Zb_2)$ corresponding to $H = G$ and the trivial 2-cohomology class is $\bfone \oplus \bfm_1 \oplus \bfm_2 \oplus \bfm_1 \bfm_2$, and the Lagrangian algebra $A(\Zb_2 \times \Zb_2,\omega)$ corresponding to $H = G$ and the nontrivial 2-cohomology class is given by $\one \oplus \bfm_1 \bfe_2 \oplus \bfm_2 \bfe_1 \oplus \bff_1 \bff_2$.
\eit
\end{expl}

\section{Equivariantization} \label{appendix:equivariantization}

%% arXiv version
In this appendix we briefly review the notion of equivariantization and prove Proposition \ref{prop:bimodule=equivariantization}.
%% SB version
%In this appendix we briefly review the notion of equivariantization and prove Proposition 3.9.

\subsection{Group actions on categories and equivariantization} \label{appedix:actions_equivariantization}

Let $\CC$ be a category. The autoequivalences of $\CC$ and natural isomorphisms between them form a monoidal category, denoted by $\Aut(\CC)$. An \emph{action} of a group $G$ on $\CC$ is defined to be a monoidal functor $T \colon G \to \Aut(\CC)$, where $G$ is viewed as a monoidal category with only identity morphisms.

More precisely, by the definition of a monoidal functor, a $G$-action on $\CC$ consists of a set of equivalences $\{T_g \colon \CC \to \CC\}_{g \in G}$ and a set of natural isomorphisms $\{\gamma_{g,h} \colon T_g \circ T_h \Rightarrow T_{gh}\}_{g,h \in G}$, such that the following diagram commutes:
\be \label{diag:G_action_appendix}
\begin{array}{c}
\xymatrix{
(T_g T_h) T_k \ar@{=}[rr] \ar@{=>}[d]_{\gamma_{g,h} 1} & & T_g (T_h T_k) \ar@{=>}[d]^{1 \gamma_{h,k}} \\
T_{gh} T_k \ar@{=>}[r]^{\gamma_{gh,k}} & T_{ghk} & T_g T_{hk} \ar@{=>}[l]_{\gamma_{g,hk}}
}
\end{array}
\ee

\begin{rem}
If $\CC$ is a monoidal category, a \emph{monoidal action} of a group $G$ on $\CC$ is a monoidal functor from $G$ to the monoidal category $\Aut_\otimes(\CC)$ of monoidal autoequivalences of $\CC$. If $\CC$ is a braided monoidal category, a \emph{braided action} of a group $G$ on $\CC$ is a monoidal functor from $G$ to the monoidal category $\Aut^\br(\CC)$ of braided monoidal autoequivalences of $\CC$ .
\end{rem}

\begin{defn}\label{defn:equivariantization}
Let $G$ be a group and $\CC$ be a category equipped with a $G$-action. The \emph{equivariantization} $\CC^G$ of $\CC$ is the category consisting of the following data:
\bit
\item The objects of $\CC^G$ are pairs $(x,\{u_g\}_{g \in G})$, where $x \in \CC$ and $u_g \colon T_g(x) \to x$ is an isomorphism such that the following diagram commutes:
\be \label{diag:equivariant_object_appendix}
\begin{array}{c}
\xymatrix{
T_g T_h (x) \ar[r]^{(\gamma_{g,h})_x} \ar[d]^{T_g(u_h)} & T_{gh}(x) \ar[d]^{u_{gh}} \\
T_g(x) \ar[r]^{u_g} & x
}
\end{array}
\ee
\item A morphism $f \colon (x,\{u_g\}_{g \in G}) \to (y,\{v_g\}_{g \in G})$ is a morphism $f \colon x \to y$ in $\CC$ such that the following diagram commutes:
\be \label{diag:equivariant_operator_appendix}
\begin{array}{c}
\xymatrix{
T_g (x) \ar[r]^{T_g(f)} \ar[d]^{u_g} & T_g(y) \ar[d]^{v_g} \\
x \ar[r]^{f} & y
}
\end{array}
\ee
\eit
\end{defn}

There is an obvious forgetful functor $\CC^G \to \CC$ defined by $(x,\{u_g\}_{g \in G}) \mapsto x$.

\begin{rem}
Let $G$ be a group. If $\CC$ is a (braided) monoidal category equipped with a (braided) monoidal $G$-action, then the equivariantization $\CC^G$ is also a (braided) monoidal category and the forgetful functor $\CC^G \to \CC$ is also a (braided) monoidal functor.
\end{rem}

\begin{expl} \label{expl:vect_equivariantization_rep_G}
Let $G$ be a group. Consider the trivial $G$-action on the category $\vect$ of finite-dimensional vector spaces. More precisely, all functors $T_g$ are identity functors and all natural transformations $\gamma_{g,h}$ are identity natural transformations. Then an object $(V,\{v_g\}_{g \in G})$ is a finite-dimensional vector space $V$ equipped with linear isomorphisms $v_g \colon V \to V$ such that $v_{gh} = v_g \circ v_h$. In other words, $(V,\{v_g\}_{g \in G})$ is a finite-dimensional $G$-representation. Similarly one can verify that morphisms between equivariant objects are homomorphisms between $G$-representations. Hence the equivariantization $\vect^G$ is equivalent to the symmetric monoidal category $\rep(G)$ of finite-dimensional $G$-representations, and the forgetful functor $\rep(G) \simeq \vect^G \to \vect$ is the usual forgetful functor.
\end{expl}

%% arXiv version
\subsection{The proof of Proposition \ref{prop:bimodule=equivariantization}} \label{appendix:equivariantization_proof}
%% SB version
%\subsection{The proof of Proposition 3.9} \label{appendix:equivariantization_proof}

%% arXiv version
Here we give a proof of Proposition \ref{prop:bimodule=equivariantization}. To be precise, we reformulate the proposition as follows.
%% SB version
%Here we give a proof of Proposition 3.9. To be precise, we reformulate the proposition as follows.

Let $G$ be a (not necessarily abelian) finite group and $H \subseteq G$ be a subgroup. We equip the quotient set $G/H$ with the left translation $G$-action: $(g,x) \mapsto gx$ for $g \in G$ and $x \in G/H$. It induces a $G$-action on the function algebra $F_H \coloneqq \fun(G/H)$:
\[
(g \cdot f)(x) \coloneqq f(g^{-1}x) , \quad g \in G , \, f \in F_H , \, x \in G/H .
\]
If we choose a basis $\{\delta_x\}_{x \in G/H}$ of $F_H$ consisting of delta functions
\[
\delta_x(y) \coloneqq \delta_{x,y} , \quad x,y \in G/H ,
\]
then the $G$-action on $F_H$ is given by $g \cdot \delta_x = \delta_{gx}$. Moreover, this $G$-action is compatible with the algebra structure so that $\fun(G/H)$ is a condensable algebra in $\rep(G)$.

This $G$-action also induces a $G$-action on the category $\vect_{G/H}$ of finite-dimensional $G/H$-graded vector spaces:
\[
T_g(V)_x \coloneqq V_{g^{-1}x} , 
\]
for $ g \in G $, $x \in G/H$, and $V \in \vect_{G/H} $. It induces a monoidal $G$-action on the multi-fusion category $\Fun(\vect_{G/H},\vect_{G/H})$ defined by conjugation:
\[
g \cdot F \coloneqq T_g \circ F \circ T_{g^{-1}} ,
\]
for $g \in G , \, F \in \Fun(\vect_{G/H},\vect_{G/H}) $. Also we know that $\Fun(\vect_{G/H},\vect_{G/H})$ is monoidally equivalent to the multi-fusion category $\mathrm{Mat}_n(\vect)$ of $n$-by-$n$ matrices valued in $\vect$, where $n \coloneqq \lvert G \rvert / \lvert H \rvert$. An object in $\mathrm{Mat}_n(\vect)$ is a finite-dimensional bi-graded vector space
\[
V = \bigoplus_{x,y \in G/H} V_{x,y} .
\]
The tensor product in $\mathrm{Mat}_n(\vect)$ is defined by
\[
(V \otimes W)_{x,y} = \bigoplus_{z \in G/H} V_{x,z} \otimes W_{z,y} .
\]
Under the monoidal equivalence $\Fun(\vect_{G/H},\vect_{G/H}) \simeq \mathrm{Mat}_n(\vect)$, the monoidal $G$-action on $\mathrm{Mat}_n(\vect)$ is given by
\[
g(V)_{x,y} = V_{g^{-1}x,g^{-1}y} , 
\]
for $ g \in G, \, x,y \in G/H , \, V \in \mathrm{Mat}_n(\vect)$. More generally, let $K \subseteq G$ be another subgroup. Then there is also a conjugation $G$-action on the finite semisimple category $\Fun(\vect_{G/H},\vect_{G/K})$. Under the equivalence $\Fun(\vect_{G/H},\vect_{G/K}) \simeq \mathrm{Mat}_{n \times m}(\vect)$ (where $m \coloneqq \lvert G \rvert / \lvert K \rvert$), the $G$-action is given by
\[
g(V)_{x,y} = V_{g^{-1}x,g^{-1}y} , 
\]
for $g \in G , \, x \in G/K , \, y \in G/H , \, V \in \mathrm{Mat}_{n \times m}(\vect)$.
\begin{prop} \label{prop:bimodule=equivariantization_appendix}
The category ${}_{F_K} \rep(G)_{F_H}$ is equivalent to $\Fun(\vect_{G/H},\vect_{G/K})^G$. When $K = H$, this is an equivalence of fusion categories.
\end{prop}

The main idea is to show that both two categories are equivalent to the third one.

\begin{defn}
Let $G$ be a group and $X$ be a set equipped with a left $G$-action. The \emph{action groupoid} $\CG(X,G)$ is defined by the following data:
\bit
\item $\ob(\CG(X,G)) \coloneqq X$.
\item $\Hom_{\CG(X,G)}(x,y) \coloneqq \{g \in G \mid g \cdot x = y\}$ for $x,y \in X$.
\item The composition and identity morphisms are induced by the multiplication and unit of $G$.
\eit
\end{defn}

Let $\CG$ be a finite groupoid, i.e., a groupoid with finitely many objects and morphisms. Recall that a finite-dimensional $\CG$-representation is a functor $F \colon \CG \to \vect$. More precisely, a finite-dimensional $\CG$-representation is a collection $\{F(x) \in \vect\}_{x \in \CG}$ of finite-dimensional vector spaces, together with linear isomorphisms $F(f) \colon F(x) \to F(y)$ labeled by morphisms $f \colon x \to y$ in $\CG$, such that $F(g) \circ F(f) = F(g \circ f)$ whenever the composition of morphisms $f,g$ in $\CG$ can be defined. All finite-dimensional $\CG$-representations and natural transformations between them form a finite semisimple category $\rep(\CG)$.

\begin{expl} \label{expl:product_groupoid_representation}
Consider the action groupoid $\CG(G/K \times G/H,G)$, where $G$ acts on $G/K \times G/H$ diagonally: $g \cdot (x,y) \coloneqq (gx,gy)$. Then a finite-dimensional $\CG(G/K \times G/H,G)$-representation $F$ is equivalent to a collection of the following data:
\bit
\item finite-dimensional vector spaces $\{F(x,y) \in \vect\}_{x \in G/K,y \in G/H}$;
\item linear isomorphisms $F(g,x,y) \colon F(x,y) \to F(gx,gy)$, such that the following equation holds for every $x \in G/K$, $y \in G/H$ and $g,h \in G$:
\begin{align*}
\bigl( F(x,y) \xrightarrow{F(h,x,y)} F(hx,hy) \xrightarrow{F(g,hx,hy)} F(ghx,ghy) \bigr) \\
= F(gh,x,y) .
\end{align*}

\eit
\end{expl}

\begin{rem} \label{rem:groupoid_abelian}
The connected components of $\CG(G/K \times G/H,G)$ are one-to-one corresponds to double cosets in $K \backslash G / H$. In particular, if $G$ is a finite abelian group and $K = H$, the category $\rep(\CG(G/H \times G/H,G))$ is equivalent to the direct sum of $n = \lvert G \rvert / \lvert H \rvert$ copies of $\rep(H)$. A simple object $F \in \rep(\CG(G/H \times G/H,G))$ only supports on a single connected component and every nonzero $F(x,y)$ is an irreducible $H$-representation. In other words, the simple objects are labeled by pairs $(x,\rho)$ where $x \in G/H$ and $\rho \in \hat H$. More precisely, the simple object corresponding to $(x,\rho)$, denoted by $\mathcal M_{(x,\rho)}$, is the induced representation $\Ind^G_H(\rho)$ equipped with the $G/H \times G/H$-grading
\[
(\mathcal M_{(x,\rho)})_{y,z} = \begin{cases}
(\Ind^G_H(\rho))_z , & y = xz , \\
0 , & \text{otherwise} ,
\end{cases}
\]
which is determined by $x$.
\end{rem}

\begin{lem} \label{lem:bimodule=groupoid}
The category ${}_{F_K} \rep(G)_{F_H}$ is equivalent to $\rep(\CG(G/K \times G/H,G))$ as categories.
\end{lem}

\pf
By definition, an ($F_K$,$F_H$)-bimodule $M \in {}_{F_K} \rep(G)_{F_H}$ in $\rep(G)$ is nothing but
\bit
\item an ($F_K$,$F_H$)-bimodule $M \in {}_{F_K} \vect_{F_H}$ in $\vect$,
\item equipped with a $G$-action, such that the left $F_K$- and right $F_H$-actions on $M$ are $G$-equivariant, i.e.,
\be \label{eq:bimodule_equivariant}
(g \cdot a) \triangleright (g \cdot m) \triangleleft (g \cdot b) = g \cdot (a \triangleright m \triangleleft b)
\ee
for all $g \in G$, $m \in M$ and $a \in F_K$, $b \in F_H$.
\eit
Since $F_H$ and $F_K$ are semisimple algebras, a bimodule $M \in {}_{F_K} \vect_{F_H}$ can be decomposed as
\[
M = \bigoplus_{\substack{x \in G/K \\ y \in G/H}} M_{x,y} ,
\]
where $M_{x,y} \coloneqq \delta_x \triangleright M \triangleleft \delta_y$. Taking an element $m \in M_{x,y}$, the condition \eqref{eq:bimodule_equivariant} implies that
\begin{align*}
\delta_{g z} \triangleright (g \cdot m) \triangleleft \delta_{g w} = (g \cdot \delta_z) \triangleright (g \cdot m) \triangleleft (g \cdot \delta_w) \\
= g \cdot (\delta_z \triangleright m \triangleleft \delta_w) = \delta_{x,z} \delta_{y,w} (g \cdot m) .
\end{align*}

Thus $g \cdot m \in M_{g x,g y}$. Hence an ($F_K$,$F_H$)-bimodule $M \in {}_{F_K} \rep(G)_{F_H}$ in $\rep(G)$ is equivalent to a collection of the following data:
\bit
\item finite-dimensional vector spaces $\{M_{x,y}\}_{x \in G/K,y \in G/H}$;
\item linear isomorphism $\rho_{g,x,y} \colon M_{x,y} \to M_{gx,gy}$, such that the following equation holds for every $x \in G/K$, $y \in G/H$ and $g,h \in G$:
\[
\bigl( M_{x,y} \xrightarrow{\rho_{h,x,y}} M_{hx,hy} \xrightarrow{\rho_{g,hx,hy}} M_{ghx,ghy} \bigr) = \rho_{gh,x,y} .
\]
\eit
By comparing with example \ref{expl:product_groupoid_representation} the lemma is proved.
\epf

\begin{lem} \label{lem:equivariantization=groupoid}
The equivariantization $\Fun(\vect_{G/H},\vect_{G/K})^G$ is equivalent to $\rep(\CG(G/K \times G/H,G))$ as categories.
\end{lem}

\pf
An object in the equivariantization $\Fun(\vect_{G/H},\vect_{G/K})^G$ is a pair $(V,\{v_g\}_{g \in G})$, where $V$ is an object in $\Fun(\vect_{G/H},\vect_{G/K}) \simeq \mathrm{Mat}_{n \times m}(\vect)$ and $v_g \colon g(V) \to V$ is an isomorphism such that the diagram \eqref{diag:equivariant_object_appendix} commutes. By taking the $(x,y)$-component for $x \in G/K$ and $y \in G/H$, we get the following commutative diagram:
\[
\xymatrix{
V_{h^{-1}g^{-1}x,h^{-1}g^{-1}y} \ar@{=}[r] \ar[d]_{g(v_h)_{x,y} = (v_h)_{g^{-1}x,g^{-1}y}} & V_{(gh)^{-1}x,(gh)^{-1}y} \ar[d]^{(v_{gh})_{x,y}} \\
V_{g^{-1}x,g^{-1}y} \ar[r]^{(v_g)_{x,y}} & V_{x,y}
}
\]
By comparing with example \ref{expl:product_groupoid_representation} the lemma is proved.
\epf

\medskip
Now we give the proof.

\pf[Proof of Proposition \ref{prop:bimodule=equivariantization_appendix}]
By Lemma \ref{lem:bimodule=groupoid} and \ref{lem:equivariantization=groupoid} we see that ${}_{F_K} \rep(G)_{F_H}$ is equivalent to the equivariantization $\Fun(\vect_{G/H},\vect_{G/K})^G$ as categories. It suffices to show that this equivalence preserves the tensor product when $K = H$.

First we consider the tensor product of ${}_{F_H} \rep(G)_{F_H}$ (i.e., the relative tensor product $\otimes_{F_H}$). Since $F_H$ is a special Frobenius algebra with $\Delta(\delta_x) \coloneqq \delta_x \otimes \delta_x$ and $\varepsilon(\delta_x) = 1$ for $x \in G/H$, the relative tensor product $M \otimes_{F_H} N$ of two bimodules $M,N \in {}_{F_H} \rep(G)_{F_H}$ is isomorphic to the image of the following idempotent:
\begin{align*}
M \otimes N \xrightarrow{1 \otimes \eta \otimes 1} M \otimes F_H \otimes N \xrightarrow{1 \otimes \Delta \otimes 1} \\
M \otimes F_H \otimes F_H \otimes N \xrightarrow{\mu_M^R \otimes \mu_N^L} M \otimes N ,
\end{align*}

where $\mu_M^R$ and $\mu_N^L$ are right $F_H$-action on $M$ and left $F_H$-action on $N$, respectively. This idempotent maps $m \otimes n \in M \otimes N$ to
\[
\sum_{x \in G/H} (m \triangleleft \delta_x) \otimes (\delta_x \triangleright n) .
\]
Therefore, we have
\[
(M \otimes_{F_H} N)_{x,y} = \bigoplus_{z \in G/H} M_{x,z} \otimes N_{z,y} ,  
\]
for $x,y \in G/H$. The $G$-action on $M \otimes_{F_H} N$ is induced from that of $M \otimes N$, i.e., the diagonal action $g \cdot (m \otimes n) \coloneqq (g \cdot m) \otimes (g \cdot n)$.

On the other hand, the tensor product $(V,\{v_g\}_{g \in G}) \otimes (W,\{w_g\}_{g \in G})$ in $\Fun(\vect_{G/H},\vect_{G/H})^G$ is given by the object $V \otimes W \in \Fun(\vect_{G/H},\vect_{G/H}) \simeq \mathrm{Mat}_n(\vect)$ equipped with the $G$-action $\{v_g \otimes w_g\}_{g \in G}$. This $G$-action on the $(x,y)$-component is
\begin{align*}
(g(V \otimes W))_{x,y} = \bigoplus_{z \in G/H} V_{g^{-1}x,g^{-1}z} \otimes W_{g^{-1}z,g^{-1}y} \\
\xrightarrow{\bigoplus (v_g)_{x,z} \otimes (w_g)_{z,y}} \bigoplus_{z \in G/H} V_{x,z} \otimes W_{z,y} = (V \otimes W)_{x,y} .
\end{align*}

Hence the tensor product of $\Fun(\vect_{G/H},\vect_{G/H})^G$ coincides with that of ${}_{F_H} \rep(G)_{F_H}$. This completes the proof.
\epf

\begin{rem} \label{rem:equivariantization_special_case}
When $K = H = G$, the algebra $F_G = \fun(G/G) = \Cb$ is trivial and Proposition \ref{prop:bimodule=equivariantization_appendix} recovers example \ref{expl:vect_equivariantization_rep_G}. When $K = H = \{e\}$, the algebra $F_{\{e\}} = \fun(G)$ and ${}_{F_{\{e\}}} \rep(G)_{F_{\{e\}}} \simeq \vect_G$ as fusion categories. Thus in this case Proposition \ref{prop:bimodule=equivariantization_appendix} simply says that $\Fun(\vect_G,\vect_G)^G \simeq \vect_G$ as fusion categories.
\end{rem}

\begin{rem}
The representation category $\rep(\CG)$ of a finite groupoid $\CG$ is naturally a multi-fusion category. However, the ``natural'' monoidal structure on $\rep(\CG(G/H \times G/H,G))$ does not coincide with those of ${}_{F_H} \rep(G)_{F_H}$ and $\Fun(\vect_{G/H},\vect_{G/H})^G$.
\end{rem}

\begin{rem} \label{rem:equivariantization_bdy}
Consider the action groupoid $\CG(G/K,H)$, where $H$ acts on $G/K$ by left translation. There is an equivalence of action groupoids $\CG(G/K,H) \simeq \CG(G/K \times G/H,G)$ defined by $x \mapsto (x,H)$. It is straightforward to check that the equivariantization $\vect_{G/K}^H$ is equivalent to $\rep(\CG(G/K,H))$, and hence equivalent to ${}_{F_K} \rep(G)_{F_H}$.
\end{rem}

% \input{appendix_enriched_cat}
% !TeX root = supp.tex
% !TeX program = pdfLaTeX
% !TEX bibfile = Top.bib

\section{Enriched categories} \label{appendix:enriched_categories}

In this appendix we give some examples of enriched (fusion) categories.

\medskip
Let $\CC$ be a fusion category. By the so-called canonical construction \cite{Kel69}, every finite semisimple left $\CC$-module $\CM$ can be promoted to a $\CC$-enriched category $\bc[\CC]{\CM}$:
\bit
\item The objects in $\bc[\CC]{\CM}$ are objects in $\CM$;
\item The hom space $\Hom_{\bc[\CC]{\CM}}(x,y)$ for $x,y \in \CM$ is the internal hom $[x,y]$, which is defined by
\be \label{eq:def_internal_hom}
\Hom_\CC(a,[x,y]) \simeq \Hom_\CM(a\odot x, y), \quad \forall a \in \CC ,\, x,y \in \CM .
\ee
\item The identity morphisms $\one_\CC \to [x,x]$ and composition of morphisms $[y,z] \otimes [x,y] \rightarrow [x,z]$ are induced by the universal property of the internal homs. 
\eit
Moreover, if $\CC$ is braided and $\CM$ is a fusion left $\CC$-module defined by a braided functor $\phi \colon \overline{\CC} \to \FZ_1(\CM)$, then the canonical construction $\bc[\CC]{\CM}$ is a $\CC$-enriched fusion category \cite{MP17,KZ18a}. 

\begin{expl} \label{expl:bimodule_enriched}
Let $\CC$ be a fusion category and $A,B \in \CC$ be separable algebras. Then ${}_A \CC_B$ is a finite semisimple left ${}_{A} \CC_{A}$-module with the module action defined by
\[
X \odot M \coloneqq X \otimes_A M , \quad X \in {}_{A} \CC_{A} , \, M \in {}_A \CC_B .
\]
It induces an enriched category $\bc[{}_{A} \CC_{A}]{{}_A \CC_B}$ by the canonical construction. The internal hom (i.e., the hom space) $[M,N] \in {}_{A} \CC_{A}$ for $M,N \in {}_A \CC_B$ is given by
\be \label{eq:internal_hom_bimodule_enriched}
[M,N] = (M \otimes_B N^R)^L ,
\ee
where both the left and right duals are taken in $\CC$. In the case $A = \one$, this result first appeared in \cite[example 3.19]{EO04} (see also \cite[Lemma 2.1.6]{KZ18} for a proof, which applies to our general case). Similarly, ${}_A \CC_B$ is a finite semisimple right ${}_B \CC_B$-module and can be viewed as a finite semsimple left $({}_B \CC_B)^{rev}$-module. The internal hom $[M,N] \in ({}_B \CC_B)^{rev}$ for $M,N \in {}_A \CC_B$ is given by
\[
[M,N] = (N^L \otimes_A M)^R ,
\]
where both the left and right duals are taken in $\CC$.
\end{expl}

\begin{expl} \label{expl:bimodule_enriched_group}
Let $G$ be a finite group and $H,K \subseteq G$ be subgroups. By taking $\CC = \rep(G)$, $A = F_H$ and $B = F_K$ in example \ref{expl:bimodule_enriched}, we get an enriched category $\bc[{}_{F_H} \rep(G)_{F_H}]{{}_{F_H} \rep(G)_{F_K}}$. Let us compute the internal hom $[M,N] \in {}_{F_H} \rep(G)_{F_H}$ for $M,N \in {}_{F_H} \rep(G)_{F_K}$.

By \eqref{eq:internal_hom_bimodule_enriched}, we have $[M,N] = (M \otimes_{F_K} N^R)^L$. The right dual $N^R$ of $N$ in $\rep(G)$ is given by the dual representation $N^*$, which is automatically bi-graded:
\[
N^* = \bigoplus_{\substack{x \in G/K \\ y \in G/H}} (N^*)_{x,y} \simeq \bigoplus_{\substack{x \in G/K \\ y \in G/H}} (N_{y,x})^* .
\]
The left $F_K$-action on $N^R$ is induced by this $G/K$-grading, i.e., $\delta_x$ acts as a projector onto $\bigoplus_{y \in G/H} (N^*)_{x,y}$ for every $x \in G/K$. The relative tensor product $\otimes_{F_K}$ has been computed in the proof of Proposition \ref{prop:bimodule=equivariantization_appendix}. We have
\begin{align*}
(M \otimes_{F_K} N^R)_{x,y} \simeq \bigoplus_{z \in G/K} M_{x,z} \otimes (N^*)_{z,y} \\
\simeq \bigoplus_{z \in G/K} M_{x,z} \otimes (N_{y,z})^* , \quad x,y \in G/H .
\end{align*}

Then by taking the left dual we have
\[
[M,N] = \bigoplus_{x,y \in G/H} [M,N]_{x,y} \simeq \bigoplus_{x,y \in G/H} \bigoplus_{z \in G/K} N_{x,z} \otimes (M_{y,z})^*
\]
equipped with the diagonal $G$-action on each direct summand.
\end{expl}

\begin{expl} \label{expl:rep_H_enriched_in_bimodule}
Let us consider the special case that $K=G$ in example \ref{expl:bimodule_enriched_group}. Then the bimodule category ${}_{F_H} \rep(G)_{F_G} = {}_{F_H}\rep(G)$ is equivalent to $\rep(H)$, and the equivalence $\rep(H) \to {}_{F_H}\rep(G)$ is given by the induced representation functor $\Ind^G_H$. Thus for two irreducible representations $\rho,\sigma \in \rep(H)$, their internal hom in ${}_{F_H} \rep(G)_{F_H}$ is
\[
[\rho,\sigma] = \Ind^G_H(\sigma) \otimes \Ind^G_H(\rho^*)
\]
equipped with the product $G/H \times G/H$-grading of the $G/H$-gradings on induced representations. By comparing with remark \ref{rem:groupoid_abelian}, we see that
\[
[\rho,\sigma] = \bigoplus_{x \in G/H} \mathcal M_{(x,\sigma \rho^{-1})} .
\]
\end{expl}

\begin{expl} \label{expl:monoidal_module_enriched}
Let $\CC$ be a fusion category and $\CM,\CN$ be finite semisimple left $\CC$-modules. Then the category $\Fun_\CC(\CM,\CN)$ of left $\CC$-module functors is also finite semisimple \cite{ENO05,EO04} and admits a left $\FZ_1(\CC)$-module structure:
\[
(x,\gamma_{-,x}) \odot F \coloneqq x \odot F(-) ,
\]
for $(x,\gamma_{-,x}) \in \FZ_1(\CC) , \, F \in \Fun_\CC(\CM,\CN) $.
The internal hom $[F,G]_{\FZ_1(\CC)}$ of $F,G \in \Fun_\CC(\CM,\CN)$ in $\FZ_1(\CC)$ is the end
\[
\int_{m \in \CM} [F(m),G(m)]_{\CC} \in \CC
\]
equipped with a canonical half-braiding induced by changing the variable in ends, where $[F(m),G(m)]_\CC$ is the internal hom in $\CC$ (see \cite[Proposition 3.5]{DKR15} for more details). It induces an enriched category $\bc[\FZ_1(\CC)]{\Fun_\CC(\CM,\CN)}$ by the canonical construction. If $\CM = \CC_A$ and $\CN = \CC_B$ where $A,B \in \CC$ are separable algebras, then the functor category $\Fun_\CC(\CM,\CN)$ is equivalent to ${}_A \CC_B$.

In particular, when $\CN = \CM$, the functor category $\Fun_\CC(\CM,\CM)$ is a multi-fusion category \cite{ENO05}, which is monoidally equivalent to $({}_{A} \CC_{A})^\rev$ if $\CM = \CC_A$. The above $\FZ_1(\CC)$-module structure can be induced from a central functor $\overline{\FZ_1(\CC)} \to \Fun_\CC(\CM,\CM)$ called the $\alpha$-induction. Moreover, the central structure $\overline{\FZ_1(\CC)} \to \FZ_1(\Fun_\CC(\CM,\CM))$ is a braided equivalence \cite{EO04}. It follows that $\bc[\FZ_1(\CC)]{\Fun_\CC(\CM,\CM)}$ is an enriched fusion category whose $E_1$-center is trivial \cite[Corollary 5.29]{KYZZ21}. Indeed, it is the $E_0$-center of the enriched category $\bc[\CC]{\CM}$ \cite[Corollary 4.41]{KYZZ21}.
\end{expl}

\begin{expl} \label{expl:internal_hom_monoidal_module_enriched}
Let $G$ be a finite abelian group and $H \subseteq G$ be a subgroup. By taking $\CC = \rep(G)$ and $B = A = F_H$ in example \ref{expl:monoidal_module_enriched}, we get an enriched category $\bc[\FZ_1(\rep(G))]{{}_{F_H} \rep(G)_{F_H}}$ (indeed, an enriched fusion category $\bc[\overline{\FZ_1(\rep(G))}]{{}_{F_H} \rep(G)_{F_H}}$). Let us compute the internal hom $[M,N]$ in $\FZ_1(\rep(G))$ for $M,N \in {}_{F_H} \rep(G)_{F_H}$. We do not use the result in example \ref{expl:monoidal_module_enriched}, but compute the internal hom by definition \eqref{eq:def_internal_hom}.

First we focus on the case that $M = N = F_H$. The general cases are discussed later. Since $F_H$ is the tensor unit of ${}_{F_H} \rep(G)_{F_H}$ and ${}_{F_H} \rep(G)_{F_H}$ is a closed left fusion $\FZ_1(\rep(G))$-module, the internal hom $[F_H,F_H] \in \FZ_1(\rep(G))$ is a Lagrangian algebra in $\FZ_1(\rep(G))$ \cite{DMNO13}. Thus it suffices to find its support (i.e., underlying object), which is a Lagrangian subgroup of $(G \times \hat G,q)$ (see example \ref{expl:abelian_center_metric} and \ref{expl:abelian_Lagrangian_center}). Recall that the forgetful functor $F \colon \FZ_1(\rep(G)) \to \rep(G)$ is given by $\mathcal O_{(g,\phi)} \mapsto \phi$, and the half-braiding $\beta_{\psi,(g,\phi)}$ of $\mathcal O_{(g,\phi)}$ is given by
\[
\psi \otimes F(\mathcal O_{(g,\phi)}) = \psi \otimes \phi \xrightarrow{\psi(g)^{-1}} \phi \otimes \psi = F(\mathcal O_{(g,\phi)}) \otimes \psi .
\]

Let us recall some basic facts about $F_H$. As an object in $\rep(G)$,
\[
F_H = \fun(G/H) \simeq \bigoplus_{\psi \in \widehat{G/H}} \psi ,
\]
and this isomorphism (i.e., the Fourier transform) maps $\delta_x \in F_H$ to $\sum_{\psi \in \widehat{G/H}} \psi(x) \cdot 1_\psi$. As an object in ${}_{F_H} \rep(G)_{F_H}$, the tensor unit $F_H$ is described by the following data (see the proof of Lemma \ref{lem:bimodule=groupoid}):
\bit
\item Each $(F_H)_{x,x} = \Cb$ (spanned by the delta function $\delta_x$) and $(F_H)_{x,y} = 0$ for $x \neq y$.
\item The $G$-action $\Cb = (F_H)_{x,x} \to (F_H)_{gx,gx} = \Cb$ is trivial for every $g \in G$ and $x \in G/H$.
\eit

Now we consider the bimodule $\mathcal O_{(g,\phi)} \odot F_H \in {}_{F_H} \rep(G)_{F_H}$ for $g \in G$ and $\phi \in \hat G$.
\bnu[(a)]
\item As an object in $\rep(G)$, $\mathcal O_{(g,\phi)} \odot F_H \in {}_{F_H} \rep(G)_{F_H}$ is the tensor product $\phi \otimes F_H$, which has the same underlying vector space as $F_H$ (i.e., spanned by $\{\delta_x\}_{x \in G/H}$) but equipped with the $G$-action
\[
h \cdot \delta_x \coloneqq \phi(h) \delta_{hx} , \, h \in G , x \in G/H .
\]
\item The right $F_H$-action on $\mathcal O_{(g,\phi)} \odot F_H$ is defined by
\begin{align*}
(\mathcal O_{(g,\phi)} \odot F_H) \otimes F_H = \phi \otimes F_H \otimes F_H \\
\xrightarrow{1 \otimes \mu} \phi \otimes F_H = \mathcal O_{(g,\phi)} \odot F_H ,
\end{align*}

where $\mu$ is the multiplication of $F_H$. Thus we see the right $F_H$-action on $\mathcal O_{(g,\phi)} \odot F_H$ is given by
\[
\delta_x \triangleleft \delta_y = \delta_{x,y} \delta_x .
\]
\item The left $F_H$-action on $\mathcal O_{(g,\phi)} \odot F_H$ is defined by
\begin{align*}
F_H \otimes (\mathcal O_{(g,\phi)} \odot F_H) = F_H \otimes \phi \otimes F_H \xrightarrow{\beta_{F_H,(g,\phi)} \otimes 1} \\
\phi \odot F_H \otimes F_H \xrightarrow{1 \otimes \mu} \phi \otimes F_H = \mathcal O_{(g,\phi)} \odot F_H ,
\end{align*}

where $\beta_{F_H,(g,\phi)}$ is the half-braiding of $\mathcal O_{(g,\phi)}$ with $F_H$:
\begin{align*}
F_H \odot \mathcal O_{(g,\phi)} = \bigoplus_{\psi \in \widehat{G/H}} \psi \otimes \phi \\
\xrightarrow{\bigoplus \psi(g)^{-1}} \bigoplus_{\psi \in \widehat{G/H}} \phi \otimes \psi = \mathcal O_{(g,\phi)} \odot F_H .
\end{align*}

Under the isomorphism $F_H \simeq \bigoplus_{\psi \in \widehat{G/H}} \psi$, the half-braiding $\beta_{F_H,(g,\phi)}$ maps $\delta_y \otimes 1_\phi \in F_H \odot \mathcal O_{(g,\phi)}$ to
\begin{align*}
&\sum_{\psi \in \widehat{G/H}} \psi(g)^{-1} \psi(y) 1_\phi \otimes 1_\psi \\
&= 1_\phi \otimes \sum_{\psi \in \widehat{G/H}} \psi(g^{-1}y) 1_\psi
= 1_\phi \otimes \delta_{g^{-1} y} .
\end{align*}
Thus the left $F_H$-action on $F_H$ is given by
\[
\delta_y \triangleright \delta_x = \delta_{x,g^{-1}y} \delta_x = \delta_{gx,y} \delta_x .
\]
\enu
Therefore, the bimodule $\mathcal O_{(g,\phi)} \odot F_H \in {}_{F_H} \rep(G)_{F_H}$ is described by the following data:
\bit
\item Each $(\mathcal O_{(g,\phi)} \odot F_H)_{gx,x} = \Cb$ and other components are zero.
\item The $G$-action $\Cb = (\mathcal O_{(g,\phi)} \odot F_H)_{gx,x} \to (\mathcal O_{(g,\phi)} \odot F_H)_{hgx,hx} = \Cb$ is given by $\phi(h)$ for $g,h \in G$ and $x \in G/H$.
\eit

It was proved that for an 1d condensable algebra $B$ in a fusion category $\CC$, $A:=[\one_{{}_B \CC_B},\one_{{}_B \CC_B}]$ is a Lagrangian algebra in $\FZ_1(\CC)$ such that $\FZ_1(\CC)_A\simeq {}_B \CC_B$ \cite{KYZ21}.
Now for $\CC=\rep(G)$ and $B=F_H$, $[F_H,F_H]$ is a Lagrangian algebra in $\FZ_1(\rep(G))$.

By the definition of internal homs \eqref{eq:def_internal_hom} and Remark \ref{rem:groupoid_abelian}, we have
\begin{align*}
&\Hom_{\FZ_1(\rep(G))}(\mathcal O_{(g,\phi)},[F_H,F_H]) \simeq\\
&\Hom_{{}_{F_H} \rep(G)_{F_H}}(\mathcal O_{(g,\phi)} \odot F_H,F_H) \simeq\\
&\begin{cases} \Cb , & (g,\phi) \in H \oplus \widehat{G/H} , \\ 0 , & \text{otherwise} . \end{cases}
\end{align*}

Then by Schur's lemma, we have
\[
[F_H,F_H] \simeq \bigoplus_{(g,\phi) \in H \oplus \widehat{G/H}} \mathcal O_{(g,\phi)} .
\]
Hence the support of the Lagrangian algebra $[F_H,F_H]$ is the Lagrangian subgroup $H \oplus \widehat{G/H} \subset G \oplus \hat G$.

Now we discuss the general internal homs. By the basic properties of internal homs \cite{Ost03} we have
\[
P \otimes [F_H, F_H] \otimes Q^L \simeq [Q \odot F_H, P \odot F_H]
\]
for $P,Q \in \FZ_1(\rep(G))$. By the above calculation it is not hard to see that every bimodule $M \in {}_{F_H} \rep(G)_{F_H}$ is isomorphic to $P \odot F_H$ for some $P \in \FZ_1(\rep(G))$. Indeed, every simple bimodule $M$ is isomorphic to $\mathcal O_{(g,\phi)} \odot F_H$ for some $(g,\phi) \in G \times \hat G$. Then the internal hom $[M,N]$ for simple bimodules $M,N \in {}_{F_H} \rep(G)_{F_H}$ can be obtained explicitly by using the above formula, and the support of this internal hom is a coset of the Lagrangian subgroup $H \oplus \widehat{G/H}$.
\end{expl}

\end{appendix}
% The \nocite command causes all entries in a bibliography to be printed out
% whether or not they are actually referenced in the text. This is appropriate
% for the sample file to show the different styles of references, but authors
% most likely will not want to use it.
% \nocite{*}

% \bibliography{apssamp}% Produces the bibliography via BibTeX.
\bibliography{Top}

\end{document}